%% file: main.tex
\RequirePackage{rotating}
\documentclass[fleqn,usenatbib]{mnras}

\usepackage{newtxtext,newtxmath}
\usepackage[T1]{fontenc}
\usepackage{graphicx}	% Including figure files
\usepackage{gensymb}
\usepackage{amsmath}	% Advanced maths commands
\usepackage{siunitx}	% \num{} formatting in claims_macros

\usepackage[normalem]{ulem}
\usepackage{tabularx}
\usepackage{subcaption}
\usepackage{rotating}
\usepackage{adjustbox}
\usepackage[usenames,dvipsnames]{xcolor}
\usepackage{hyperref}
\hypersetup{
    colorlinks=true,
    linkcolor=MidnightBlue,
    filecolor=MidnightBlue,
    citecolor=MidnightBlue,
    urlcolor=MidnightBlue,
}
\usepackage{orcidlink}
\usepackage[switch, modulo]{lineno}
\newcommand{\bs}[1]{\boldsymbol{#1}}
\graphicspath{ {Figures/} }
\linenumbers
\nolinenumbers
\title[UNIONS-3500 Weak Lensing: III. 2D Cosmological constraints in configuration space]{UNIONS-3500 Weak Lensing: III. 2D Cosmological Constraints in Configuration Space}

\author[UNIONS Collaboration: L. W. K. Goh et al.]{
L. W. K. Goh$^{1,2}$\orcidlink{0000-0002-0104-8132}\thanks{E-mail: lgoh@roe.ac.uk},
S. Guerrini$^{3}$\orcidlink{0009-0004-3655-4870},
C. Daley$^{4}$\orcidlink{0000-0002-3760-2086},
F. Hervas-Peters$^{4,5}$\orcidlink{0009-0008-1839-2969},
M. Kilbinger$^{4}$\orcidlink{0000-0001-9513-7138},
A. Wittje$^{6}$\orcidlink{0000-0002-8173-3438},
\newauthor
C. Murray$^{3}$\orcidlink{0000-0002-4668-1273},
S. Fabbro$^{7,8}$\orcidlink{0000-0003-2239-7988},
H. Hildebrandt$^{6}$\orcidlink{0000-0002-9814-3338},
M. J. Hudson$^{9,10,11}$\orcidlink{0000-0002-1437-3786},
L. van Waerbeke$^{12}$\orcidlink{0000-0002-2637-8728},
A. H. Wright$^{6}$\orcidlink{0000-0001-7363-7932},
\newauthor
T. de Boer$^{13}$\orcidlink{0000-0001-5486-2747},
J.-C. Cuillandre$^{4}$\orcidlink{0000-0002-3263-8645},
E. Magnier$^{13}$,
A. W. McConnachie$^{7}$\orcidlink{0000-0003-4666-6564}\\
$^{1}$ Institute for Astronomy, University of Edinburgh, Royal Observatory, Blackford Hill, Edinburgh EH9 3HJ, UK\\
$^{2}$ Higgs Centre for Theoretical Physics, School of Physics and Astronomy, The University of Edinburgh, Edinburgh EH9 3FD, UK\\
$^{3}$ Universit\'e Paris Cit\'e, Universit\'e Paris-Saclay, CEA, CNRS, AIM, F-91191, Gif-sur-Yvette, France\\
$^{4}$ Universit\'e Paris-Saclay, Universit\'e Paris Cit\'e, CEA, CNRS, AIM, F-91191, Gif-sur-Yvette, France\\
$^{5}$ Department of Astronomy, Steward Observatory, University of Arizona, 933 North Cherry Avenue, Tucson, AZ 85721-0065, USA\\
$^{6}$ Ruhr University Bochum, Faculty of Physics and Astronomy, Astronomical Institute (AIRUB), German Centre for Cosmological Lensing, 44780 Bochum, Germany\\
$^{7}$ NRC Herzberg Astronomy and Astrophysics, 5071 West Saanich Road, Victoria, BC V8Z 6M7, Canada\\
$^{8}$ Department of Computer Science, University of British Columbia, 2366 Main Mall, Vancouver, BC V6T 1Z4, Canada\\
$^{9}$ Department of Physics and Astronomy, University of Waterloo, 200 University Avenue West, Waterloo, ON N2L 3G1, Canada\\
$^{10}$ Waterloo Centre for Astrophysics, University of Waterloo, Waterloo, ON N2L 3G1, Canada\\
$^{11}$ Perimeter Institute for Theoretical Physics, 31 Caroline St. North, Waterloo, ON N2L 2Y5, Canada\\
$^{12}$ Department of Physics and Astronomy, University of British Columbia, 6224 Agricultural Road, Vancouver, BC V6T 1Z1, Canada\\
$^{13}$ Institute for Astronomy, University of Hawaii, 2680 Woodlawn Drive, Honolulu HI 96822\\
}

\date{Accepted XXX. Received YYY; in original form ZZZ}

\pubyear{\the\year{}}

\begin{document}
\input{claims_macros}
\label{firstpage}
\pagerange{\pageref{firstpage}--\pageref{lastpage}}
\maketitle

\begin{abstract}
We present the first cosmological constraints from the cosmic shear analysis of the UNIONS-3500 weak lensing galaxy catalogue in configuration space. The Ultraviolet Near Infrared Optical Northern Survey (UNIONS) is the largest and deepest photometric survey of the northern hemisphere to date, with the UNIONS-3500 catalogue using high-quality $r$-band imaging across 3500 deg$^2$ of the sky. We perform a 2D cosmic shear analysis with a single tomographic bin, using the two-point correlation function (2PCF) statistic. Assuming a flat $\Lambda$CDM model, we obtain constraints on the clustering amplitude of $S_8 \equiv\sigma_8\sqrt{\Omega_{\rm m}/0.3}= 0.831^{+0.067}_{-0.078}$, which is consistent with constraints from \textit{Planck} CMB measurements and precedent cosmic shear results within $1\sigma$. We outline the construction of our cosmological inference pipeline, including the estimation of the source redshift distribution, shear calibration, and covariance matrix, and describe methodologies for the mitigation of systematic effects arising from PSF systematics and $B$-modes. We demonstrate that our results are robust to variations in analysis choices, including scale cuts, prior ranges, and nonlinear modelling. This paper is part of a coordinated release which collectively demonstrates the maturity and readiness of UNIONS to deliver competitive cosmological results, positioning it as a key stepping stone towards the forthcoming era of Stage IV weak lensing experiments.
\end{abstract}

\begin{keywords}
Cosmology:observations--gravitational lensing: weak--cosmological parameters
\end{keywords}

%%%%%%%%%%%%%%%%%%%%%%%%%%%%%%%%%%%%%%%%%%%%%%%%%%

%%%%%%%%%%%%%%%%% BODY OF PAPER %%%%%%%%%%%%%%%%%%

\section{Introduction}

Since its first detection two and a half decades ago \citep{Bacon2000, 2000astro.ph..3338K, vanWaerbeke2000, Wittman2000}, weak gravitational lensing has emerged as a powerful tool to probe the Universe, particularly the properties of its most elusive components: dark energy and dark matter. Weak lensing exploits the fact that along the line of sight, the observed shapes of background galaxies are distorted---or `sheared'---as their light is deflected by the intervening gravitational potential of foreground matter distributions. By measuring the shapes of these sheared galaxies, or more specifically, the correlations between them, we are able to trace the underlying dark matter distribution, more commonly known as the large-scale structure (LSS) of the Universe.

An additional advantage of weak lensing as a cosmological probe is that it offers insight into the evolution of the LSS at low redshifts and on relatively small scales, thus making it highly complementary to other cosmological probes such as the cosmic microwave background (CMB) and baryonic acoustic oscillations (BAO). Weak lensing best constrains the combination of $\sigma_8$, the amplitude of clustering at a scale of $8\,h^{-1}$~Mpc, and the present-day total matter density $\Omega_\mathrm{m}$, often expressed as a single combined parameter $S_8\equiv\sigma_8\sqrt{\Omega_\mathrm{m}/0.3}$. 

There have been several galaxy surveys dedicated to weak lensing science, with Stage III surveys \cite[see][for a definition of the different weak-lensing survey `stages']{2006astro.ph..9591A} such as the Kilo-Degree Survey \cite[KiDS;][]{kidslegacy_catalogue} and the Dark Energy Survey \cite[DES;][]{des_y6} publishing their last data releases, as well as the Hyper-Suprime Cam survey releasing their latest data set comprising three years' worth of observations \cite[HSC Y3;][]{hsc_dr3}. In recent years, low-redshift weak lensing and galaxy clustering analyses have consistently favoured a lower value of $S_8$ than that inferred from high-redshift CMB measurements \cite[e.g.][]{planck2018}, with discrepancies at the level of $1$--$3\,\sigma$ \cite[see for example][ for an in-depth review]{cosmoverse}. Whether this tension signals new physics or simply reflects residual systematics, the current cosmological landscape appears poised for an answer. Most recently, the KiDS-Legacy cosmic shear analysis \citep{Wright_kids_2025}, which incorporates improved modelling of systematic effects, together with a re-analysis of HSC Y3 data using clustering-redshift techniques \citep{hsc_clustering_redshift}, seem to suggest a possible resolution.

Looking ahead, Stage IV surveys, including \textit{Euclid} \citep{euclid_overview} and the Vera C.~Rubin Observatory's Legacy Survey of Space and Time \cite[LSST;][]{lsst} have begun operations and are expected to publish their first data releases in the coming years. With this new generation of experiments, sub-percent precision on key cosmological measurements becomes achievable, making the coming decade an exciting one for both the quantity and quality of observational data.

Into this landscape comes the Ultraviolet Near Infrared Optical Northern Survey \citep[UNIONS;][]{gwynUNIONSUltravioletNearInfrared2025}, an ongoing photometric Stage III survey targeting up to 6250~deg$^2$ of the northern sky. In particular, UNIONS $r$-band imaging with the Canada-France Hawai'i Telescope (CFHT) enables high-quality weak lensing measurements through its excellent seeing.

This paper, along with four other accompanying papers, constitutes the first major UNIONS cosmic shear analysis, based on 3500~deg$^2$ of data accumulated over more than five years of coordinated observations from three wide-field telescopes in Hawai'i. We present configuration space cosmological constraints from the largest northern hemisphere weak lensing data set to date, which we name UNIONS-3500, hence offering a fitting complement to preceding Stage III analyses \citep{desy3-cosmo, Kilo-DegreeSurvey:2023gfr, hsc-y3, hsc-y3-2,  decade, Wright_kids_2025}, which have been largely derived from  observations in the southern sky.

This paper is structured as follows: Sect. \ref{sec:unions_data} provides a brief overview of the UNIONS-3500 galaxy catalogue. In Sect. \ref{sec:modelling}, we outline the theoretical modelling of the cosmic shear two-point correlation function (2PCF), the central probe used in this work, along with the additional components required for a complete cosmic shear analysis, including systematic error modelling, the covariance matrix and redshift distribution estimation, and a brief outline of external data sets included with the analysis. In Sect. \ref{sec:inference_pipeline}, we describe our inference pipeline and detail how we take into account systematic effects and define our scale cuts. In Sect. \ref{sec:results} we present our results in terms of the marginalised cosmological parameter constraints, across various analysis setups and data set combinations. Finally,  we offer our conclusions in Sect. \ref{sec:conclusions}. 

This paper forms part of a larger release: Table \ref{tab:unions_papers} provides a summary of the accompanying papers, which detail catalogue construction (Paper~I; \citealt{kilbinger.etal25}); $B$-mode validation (Paper~II; \citealt{daley.etal25}); cosmological constraints in configuration space (Paper~III; this work); cosmological constraints in harmonic space (Paper~IV; \citealt{guerrini.etal25b}), and lastly image simulations and shear calibration (Paper~V; \citealt{hervaspaters.etal25}).

\begin{table*}
\centering
\caption{List of associated publications in this coordinated UNIONS release.}
\label{tab:unions_papers}
\begin{tabular}{l l l}
\hline
\textbf{Paper Index} & \textbf{Author} & \textbf{Title} \\
\hline
I & \cite{kilbinger.etal25} & Weak lensing catalogues \\
% II & \cite{hervaspaters.etal25} & Image simulations and shear calibrations\\
II & \cite{daley.etal25} & $B$-mode validation and comparison\\
III & This work & Cosmological constraints in configuration space\\
IV & \cite{guerrini.etal25b} & Cosmological constraints in harmonic space \\
V & \cite{hervaspaters.etal25} & Simulations and validation \\
\hline
\end{tabular}
\end{table*}

\section{The UNIONS data set}\label{sec:unions_data}

UNIONS combines multi-band photometric images from multiple telescopes. The Canada-France Imaging Survey (CFIS) provides $u$- and $r$-band images from CFHT.  The shape measurement of galaxies relies on high-quality images taken in the $r$ band, which benefit from exquisite seeing of ${\sim}0.7$~arcsec, making it ideal for weak lensing science. The Panoramic Survey Telescope and Rapid Response System (Pan-STARRS) provides imaging in the $i$- and $z$-bands. The Subaru telescope also takes images in the $z$-band within the framework of the WISHES (Wide Imaging with Subaru HSC of the Euclid Sky) programme, and in the $g$-band through the WHIGS (Waterloo-Hawai'i IfA $g$-band Survey) programme. UNIONS provides wide-field multiband photometry over the optical bands, contributing to \textit{Euclid}'s photometric redshifts in the northern sky. An extension to UNIONS is currently collecting data in all five bands down to 15$\degree$ in declination. For a full review of the survey and its strategies, see \cite{gwynUNIONSUltravioletNearInfrared2025}. 

\subsection{UNIONS-3500 weak lensing catalogue}

This paper presents the first cosmological parameter estimation using the UNIONS-3500 2D weak lensing catalogue. Since this galaxy sample comprises images collected specifically in the $r$ band, we do not yet possess colour information to construct multiple tomographic bins; this analysis therefore relies on a single, two-dimensional redshift distribution (see Sect. \ref{sec:nz} for more details). 

The shape measurement was performed with \texttt{ShapePipe} \citep{farrensShapePipeModularWeaklensing2022a, shapepipe_axel}, which also incorporates the galaxy point-spread function (PSF) modelling based on \texttt{PSFex} \citep{bertinAutomatedMorphometrySExtractor2011a}. Our lensing sample gathers data collected until the end of 2022, and is composed of over 61 million galaxies totalling an area of 3500~deg$^2$, corresponding to 2894~deg$^2$ of effective area after masking. The shapes of the galaxies were measured using \texttt{ngmix} \citep{sheldonNGMIXGaussianMixture2015} and the calibration of the galaxy ellipticities was performed using \texttt{Metacalibration} \citep{huffMetacalibrationDirectSelfCalibration2017, sheldonPracticalWeaklensingShear2017}. This gives an effective galaxy number density of 4.96~arcmin$^{-2}$ and per-component shape noise of $\sigma_{e}=0.27$. Calibrated shapes using \texttt{Metacalibration} are saved but undergo a second correction step to remove residual per-component PSF leakage, an empirical process adapted from \cite{liKiDSLegacyCalibrationUnifying2023}. The catalogue contains the spin-2 ellipticities for each galaxy, $e_1$ and $e_2$, corrected for the per-component additive biases $c_1$ and $c_2$. We also include their non-leakage corrected counterparts. Specific details on the PSF fitting and validation, shape measurement and leakage correction methods can be found in \hyperlink{cite.kilbinger.etal25}{Paper I}.

Additionally, different size cuts and masking schemes were explored before converging on the one employed to create the fiducial UNIONS-3500 shear catalogue. In its current version, a size cut of $r_{\rm h, gal}/r_{\rm h, PSF} > 0.7$ was applied, where $r_{\rm h}$ is defined as the half-light radius. The effective mask is an amalgamation of flags generated during MegaPipe image processing (bad pixels, chip defects), at \texttt{ShapePipe} processing time (foreground objects, blended galaxy images), and in post-processing (cosmic rays, exposure coverage); see \hyperlink{cite.kilbinger.etal25}{Paper I} for more details. In total, this nominally excludes objects detected as blends, regions around Messier and NGC objects, bright stars, stellar halos, diffraction spikes, defects, and cosmic rays. We also explored a more comprehensive masking of halo-like emissions around faint and bright stars, but $B$-mode null tests showed this additional masking introduced spurious systematic contamination; ultimately, it was not employed. For an in-depth study of the impact of size cuts and masking schemes, we refer the reader to \hyperlink{cite.kilbinger.etal25}{Paper I} and \hyperlink{cite.daley.etal25}{Paper II}. 

%--------------------------------------------------------------------
\section{Modelling}\label{sec:modelling}
In this section, we detail the modelling of the cosmic shear 2-point correlation function (2PCF), including validation tests for systematic effects, the covariance matrix, calibration of the shear magnification bias, and estimation of the source redshift distribution.

\subsection{Cosmic shear 2-point correlation function}\label{sec:2pcf}

In the regime of weak lensing, the magnitude of distortion of the galaxy images is too small to be detected individually. Hence, an ensemble of galaxies is typically analysed together by computing the correlation of the measured shapes between pairs of galaxies. There exist several summary statistics that capture this information, most notably the two-point correlation functions in configuration space $\xi^{ij}_{\pm}(\theta)$, and the angular power spectra $C^{ij}_\ell$ in Fourier or harmonic space. In principle, these two statistics capture similar information on the sky, albeit with different sensitivities to scales and masking effects. For comprehensive reviews on weak lensing theory and methodologies, see for example \cite{Bartelmann_1999yn} and \cite{Kilbinger_2014cea}. For results on the cosmic shear analysis of UNIONS-3500 in Fourier space, we refer the reader to \hyperlink{cite.guerrini.etal25b}{Paper IV}.

The cosmic shear 2PCF is the expectation value of the product of the tangential and cross components of the galaxy shear signal, $\bs{\gamma}_t$ and $\bs{\gamma}_\times$ respectively \citep{Kaiser1992},
\begin{equation}
\xi_\pm(\theta)=\left<{\gamma}_t{\gamma}_t\right>(\theta)\pm\left<{\gamma}_\times{\gamma}_\times\right>(\theta)\,,
\end{equation}
which is a function of $\theta$, the angle of separation between a pair of galaxies on the sky. The $\xi_\pm(\theta)$ functions can also be derived from their Fourier space counterpart by a Hankel transform, such that, assuming the flat-sky approximations, it is given by \citep{Kaiser:1996tp}
\begin{equation}\label{eq:cell_to_xi}
    \xi_{\pm}(\theta)=\int^{\infty}_{0}\frac{\ell \textrm{d}\ell}{2\pi}\textrm{J}_{n}(\theta \ell)\,C_\ell\,,
\end{equation}
where J$_n$ is the $n$th-order Bessel function of the first kind, with $n=0$ for $\xi_+$ and $n=4$ for $\xi_-$. Note that we have dropped the notation for tomographic indices $i$ and $j$ since we are only considering a 2D analysis. The cosmic shear power spectrum $C_\ell$ is given by the integral of the product of the lensing efficiency as a function of the comoving distance $q(\chi)$, and the nonlinear matter power spectrum $P_\mathrm{NL}(k, z(\chi))$,
\begin{equation} \label{eq:Cell}
C_\ell = 
\int_0^{\chi_\textrm{H}}\textrm{d}\chi \ 
\frac{q^2(\chi)}
{\chi^2} P_\mathrm{NL}(k, z(\chi)) ,
\end{equation}
where under the Limber approximation $k=\frac{\ell+1/2}{\chi}$, $\chi_\textrm{H}$ is the comoving distance to the horizon, and
\begin{equation}\label{eq:lens_eff_cs}
    q(\chi)=\frac{3}{2}\Omega_{\rm m}\frac{H_0}{c^2}\int_\chi^{\chi_\textrm{H}}\textrm{d}\chi' n(\chi')\frac{\chi-\chi'}{\chi'}\,.
\end{equation}
Here $H_0=100h$ km s$^{-1}$Mpc$^{-1}$ is the Hubble constant and $n(\chi)$ is the galaxy redshift distribution.

The cosmic shear observable is the observed ellipticity of a galaxy $\bs{e}^\mathrm{obs}$, which is the sum of its intrinsic ellipticity $\bs{\epsilon}^\mathrm{s}$ and shear $\bs{\gamma}$ (boldface denotes spin-2 quantities):
\begin{equation}
    \bs{e}^\mathrm{obs} = \bs{\epsilon}^\mathrm{s} + \bs{\gamma}\,.
\end{equation}
Following the notation of \cite{Schneider:2002jd}, we define the angular bin width as $\Delta\theta$ and $\Delta_{\theta}(\phi)=1$ for $\theta-\Delta\theta/2 \leq \phi \leq \theta+\Delta\theta/2$ and zero elsewhere. The 2PCF estimator for a galaxy pair $i$, $j$ at separation $|\theta_i-\theta_j|$ is
\begin{equation}\label{eq:xi_estimator}
    \hat{\xi}_{\pm}(\theta)=\frac{\Sigma_{ij}w_iw_j (e_{it}e_{jt}\pm e_{i\times}e_{j\times})\Delta_\theta(|\theta_i-\theta_j|)}{N^{ij}_{\rm{p}}(\theta)}\,,
\end{equation}
where $w$ is the galaxy weight, computed during the shape measurement step, and $N^{ij}_{\rm p}(\theta)$ is the effective number of galaxy pairs in the galaxy bin.

\begin{figure*}
    \centering
    \includegraphics[width=\linewidth]{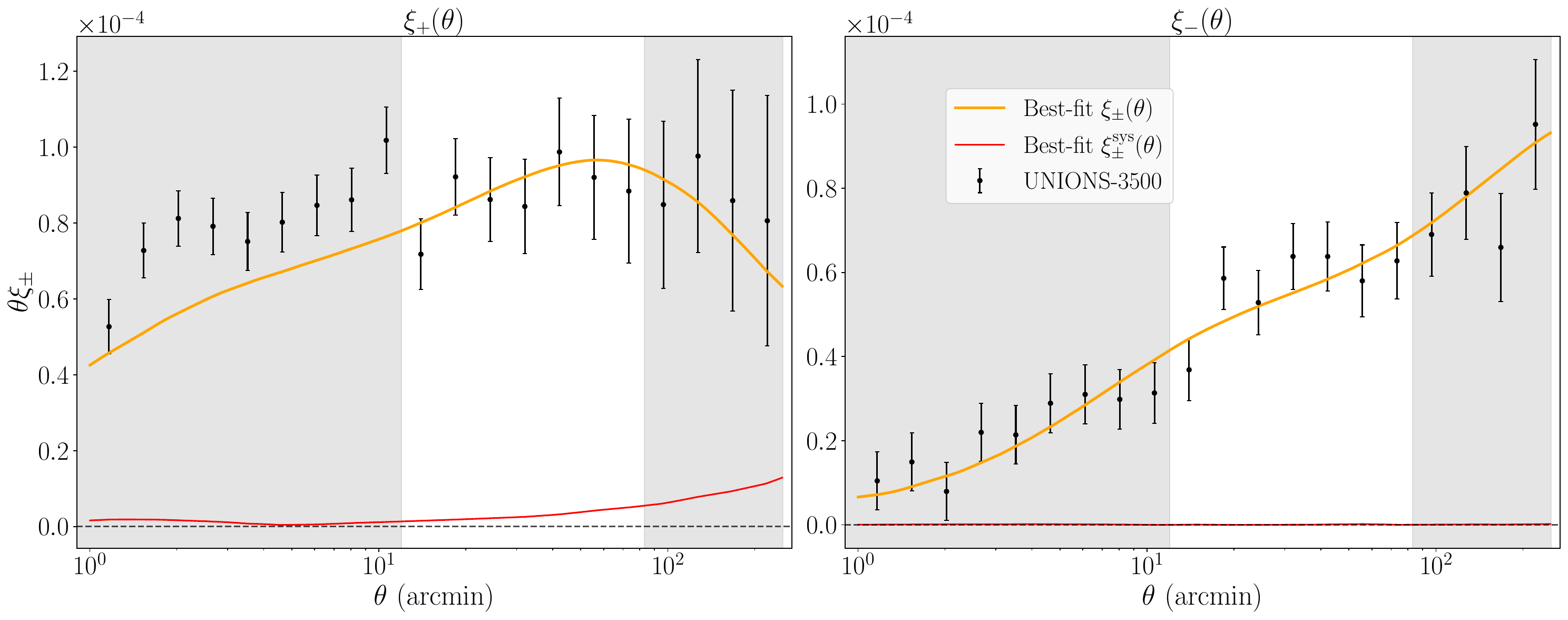}
    \caption{Real space 2PCF data vectors estimated from the data in black (\emph{left:} $\xi_+$, \emph{right:} $\xi_-$) as a function of angular separation $\theta$, with error bars computed from the diagonals of the covariance matrix. The grey sections denote the fiducial scale cuts employed in this analysis. The solid orange line shows the best-fit model to the data obtained in this analysis, which includes both the cosmological and PSF-leakage $\xi_\pm^{\rm{sys}}$ signal. We also plot solely the best-fit $\xi_\pm^{\rm{sys}}$ signal in red, which we have obtained in a joint fit during the inference (see Sect. \ref{sec:psf_inference}).}
    \label{fig:xi_pm}
\end{figure*}

In Fig. \ref{fig:xi_pm} we present the $\hat{\xi}_{\pm}(\theta)$ data vectors computed by \texttt{TreeCorr} \citep{2004MNRAS.352..338J}, where we have binned the separation angle $\theta$ into $20$ logarithmically-spaced bins over $1$--$250$~arcmin. We also include their associated uncertainties given by the diagonals of the covariance matrix, whose calculation is detailed in Sect. \ref{sec:covmat}. 

\begin{figure*}
    \centering
    \includegraphics[width=\linewidth]{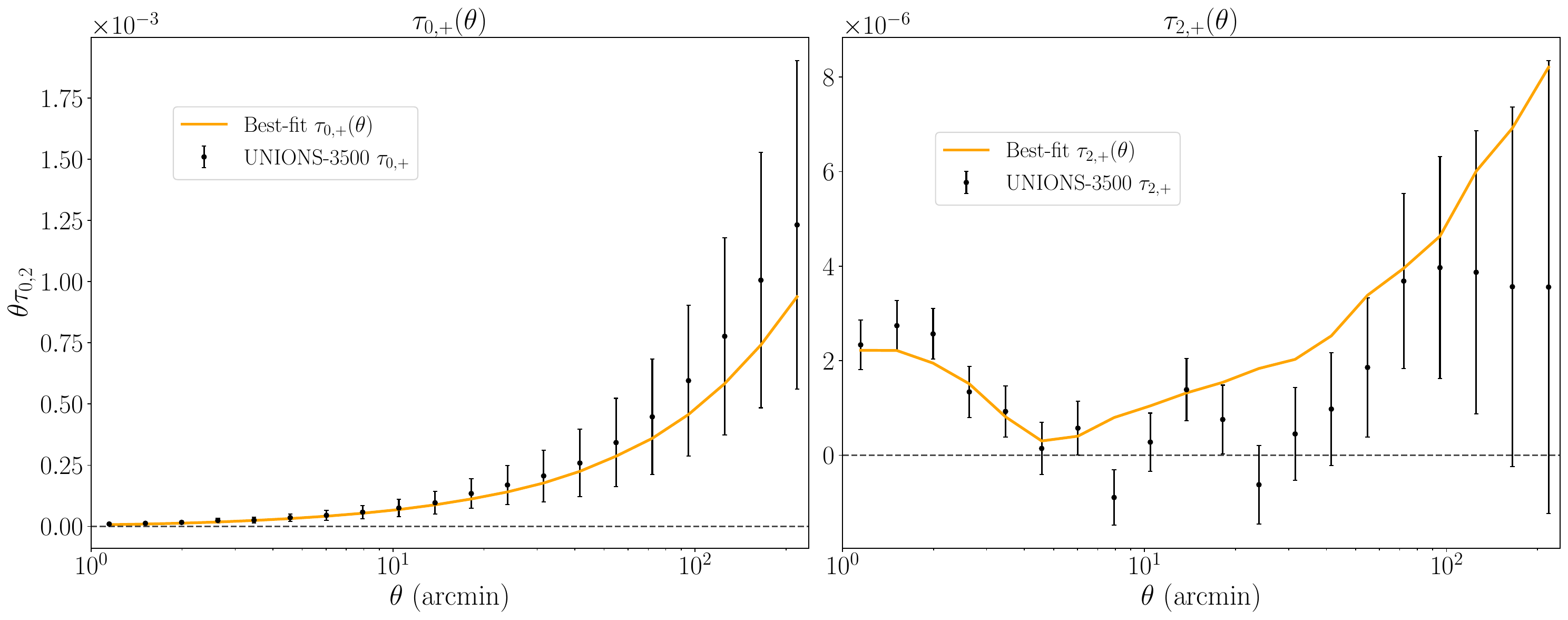}
    \caption{The PSF systematic statistics $\tau_{0,+}(\theta)$ (\emph{left panel}) and $\tau_{2,+}(\theta)$ (\emph{right}) statistics estimated based on Eqs.~\eqref{eq:rho-stats} and \eqref{eq:tau-stats}, as a function of angular separation $\theta$, along with their error bars calculated from the semi-analytic covariance matrix. We also include the $\tau_0$ and $\tau_2$ best-fit obtained from the inference.}
    \label{fig:tau_02}
\end{figure*}
\subsection{Systematic effects and validation tests}\label{sec:sys_tests}
We assess the robustness of our data vectors to systematic measurement effects, focusing on PSF modelling, configuration-space $E$/$B$-mode decomposition, and the Complete Orthogonal Sets of $E$/$B$-mode Integrals (COSEBIs; \citealt{schneider.eifler.krause10, asgari.schneider.simon12}). These tests also drove the evolution of the catalogue version that led to the fiducial sample used here.

\subsubsection{PSF systematic effects}
\label{sec:psf_leakage}
The shear signal of a galaxy is estimated from its observed ellipticity. However, this measurement is often noisy, since it is contaminated by the PSF of the instrument. PSF systematics and mismodelling can contribute to the observed galaxy ellipticity in the form of an additive bias factor, $\bs{e}^{\mathrm{ sys}}$, such that
\begin{align}
    \bs{e}^{\mathrm{obs}} = \bs{\epsilon}^{\mathrm{s}} + \bs{e}^{\mathrm{ sys}} + \bs{\gamma}.
\end{align}
Quantifying this bias is essential for recovering an unbiased shear signal. A commonly used model for PSF leakage is given by \citep{2008A&A...484...67P, 2016MNRAS.460.2245J}
\begin{align}
    \bs{e}^{\mathrm{ sys}} = \alpha_\mathrm{PSF} \bs{e}^{\mathrm p} + \beta_\mathrm{PSF} \delta \bs{e}^{\mathrm p} + \eta_\mathrm{PSF} \delta \bs{T}^{\mathrm p},
\end{align}
where $\alpha_\mathrm{PSF}$, $\beta_\mathrm{PSF}$ and $\eta_\mathrm{PSF}$ are constant free parameters, $\bs{e}^{\mathrm p}$ is the ellipticity of the PSF model, $\delta \bs{e}^{\mathrm p} = \bs{e}^* -\bs{e}^{\mathrm p}$ is the PSF ellipticity residual and $\delta \bs{T}^{\mathrm p} = \bs{e}^* (T^* - T^{\mathrm p})/T^*$ is the PSF size residual. We use the superscript `p' to refer to the properties of the PSF (size or ellipticity), and `$*$' to denote those of point-like sources (also referred to as stars) which, by convention, have zero size before convolution with the PSF. Residual terms can only be evaluated at the positions of the stars.

The PSF-PSF and galaxy-PSF 2PCFs can be used to derive the amplitude of the leakage bias in the measured galaxy-galaxy correlation function. This is done by estimating the $\rho$ statistics \cite[introduced by][]{2010MNRAS.404..350R,2016MNRAS.460.2245J}, representing PSF-PSF correlations, and $\tau$ statistics \cite[introduced by][]{2020PASJ...72...16H,gatti2021shapecatalogue, Giblin21}, quantifying galaxy-PSF correlations, directly from the data. They are respectively given by (where we have dropped the `PSF' subscript in $\alpha, \beta$ and $\eta$ for clarity of notation)
\begin{equation} \label{eq:rho-stats}
\begin{aligned}
\rho_{0}(\theta) &= \langle \bs{e}^\mathrm{p} \bs{e}^\mathrm{p} \rangle(\theta), 
&\qquad \rho_{1}(\theta) &= \langle \delta \bs{e}^\mathrm{p} \, \delta \bs{e}^\mathrm{p} \rangle(\theta), \\
\rho_{2}(\theta) &= \langle \bs{e}^\mathrm{p} \delta \bs{e}^\mathrm{p} \rangle(\theta), 
&\qquad \rho_{3}(\theta) &= \langle \delta \bs{T}^\mathrm{p} \, \delta \bs{T}^\mathrm{p} \rangle(\theta), \\
\rho_{4}(\theta) &= \langle \delta \bs{e}^\mathrm{p} \, \delta \bs{T}^\mathrm{p} \rangle(\theta), 
&\qquad \rho_{5}(\theta) &= \langle \bs{e}^\mathrm{p} \delta \bs{T}^\mathrm{p} \rangle(\theta)\,,
\end{aligned}
\end{equation}
and
\begin{equation}\label{eq:tau-stats}
\begin{aligned}
\tau_{0}(\theta) &= \langle \bs{e} \, \bs{e}^\mathrm{p} \rangle(\theta) &= \alpha\rho_0(\theta)+\beta\rho_2(\theta)+\eta\rho_5(\theta), \\
\tau_{2}(\theta) &= \langle \bs{e} \, \delta \bs{e}^\mathrm{p} \rangle(\theta) &= \alpha\rho_2(\theta)+\beta\rho_1(\theta)+\eta\rho_4(\theta), \\
\tau_{5}(\theta) &= \langle \bs{e} \, \delta \bs{T}^\mathrm{p} \rangle(\theta) &= \alpha\rho_5(\theta)+\beta\rho_4(\theta)+\eta\rho_3(\theta)\,.
\end{aligned}
\end{equation}
where we only consider the `$+$' component of each field, since the `$-$' component tends to be noise-dominated and does not significantly increase cosmological information.

The second equality in Eq.~\eqref{eq:tau-stats} shows how $\alpha$, $\beta$, and $\eta$ are estimated from the $\rho(\theta)$ and $\tau(\theta)$ statistics. Further details can be found in \cite{guerriniGalaxyPointSpread2025}. In Fig. \ref{fig:tau_02}, we plot the $\tau_{0,+}(\theta)$ and $\tau_{2,+}(\theta)$ statistics estimated from the catalogue, along with their best-fit signal recovered from the inference. 

Subsequently, we can compute the 2PCF estimator of the PSF systematics as
\begin{align}\label{eq:xi_sys}
\begin{split}
    \xi_\pm^{\mathrm{sys}}(\theta) &= \alpha^2 \rho_0(\theta) + \beta^2 \rho_1(\theta) + \eta^2 \rho_3(\theta)\\
    &+ 2\alpha\beta\,\rho_2(\theta) + 2\alpha \eta \,\rho_5(\theta) + 2 \beta \eta \,\rho_4(\theta)\;,
\end{split}
\end{align}
such that the total estimated 2PCF signal is a sum of the true shear signal and a contribution from the PSF systematics, 
\begin{align}\label{eq: xi with leakage bias}
    \xi_\pm^{\mathrm{obs}}(\theta; \alpha, \beta, \eta) = \xi_\pm^{\bs{\gamma} \bs{\gamma}}(\theta) + \xi_\pm^{\mathrm{ sys}}(\theta; \alpha, \beta, \eta)\,.
\end{align}
Section 4 of \hyperlink{cite.kilbinger.etal25}{Paper I} details the measurement of the $\xi_\pm^{\mathrm{sys}}(\theta)$ signal, where a non-negligible leakage bias at large scales was detected for both the $\xi_\pm$ data vectors, despite a noticeable improvement after applying an empirical leakage correction. We therefore modelled PSF systematics by additionally sampling the $\alpha$ and $\beta$ parameters at the inference step (see Sect. \ref{sec:psf_inference} for details), fixing $\eta =0$ since \cite{guerriniGalaxyPointSpread2025} demonstrated that it is strongly correlated with $\alpha$, and that fixing $\eta$ does not significantly modify the estimate of the leakage bias. We also used this systematic test to inform our scale cuts: requiring that PSF systematics contribute less than $10\%$ of the total signal gave an upper scale cut of $83~$arcmin for both $\xi_\pm$. 

\subsubsection{E/B mode validation tests}\label{sec:eb_modes}

Gravitational lensing produces a shear field that is curl-free to leading order; higher-order effects and intrinsic alignments can produce $B$-modes but remain below Stage III sensitivity \citep{schneider.etal22}. Significant $B$ modes are therefore evidence of residual systematics, and we use them as one of the determining factors informing our scale cuts. For a detailed comparison of configuration- and harmonic-space $B$ modes across catalogue variants (changes in PSF size cuts, masking, etc.), see \hyperlink{cite.daley.etal25}{Paper II}; here we summarise the results for the fiducial catalogue used for the cosmic shear analysis.

We computed two $B$-mode statistics: pure-mode correlation functions $\xi_\pm^{E/B}(\theta)$ in configuration space \citep{schneider.etal22}; and the first six COSEBIs modes $B_n$. For each statistic, we calculate a probability to exceed (PTE) and require PTE $>0.05$ to pass the null test. Preliminary blinded analyses with the initial version of the catalogue passed configuration-space $B$-mode null tests but failed in harmonic space.
We therefore introduced a more conservative size cut, removing approximately 25~per~cent of galaxies and reducing $n_{\rm eff}$ from $6.48$ to $4.96$~arcmin$^{-2}$ and the per-component shape noise from $0.28$ to $0.27$.
The stricter cut improved PSF systematics statistics, and the harmonic-space $B$-mode null tests passed.

Over the full angular range, COSEBIs  show significant $B$-modes with a PTE of $\configPteSixThreeCosebisFull$, while $\xi_\pm^B$ passes with a PTE of $\configPteSixThreeCombinedFull$. After restricting to scales of $12-83$~arcmin for both $\xi_\pm$ signals, both null tests pass with PTEs of $\configPteSixThreeCosebis$ and $\configPteSixThreeCombined$ respectively. These results then informed our fiducial scale cuts (see Sect.~\ref{sec:scale_cuts}).

\subsection{Redshift distribution}\label{sec:nz}

We estimate the redshift distribution of our weak lensing source sample based on the colour-redshift relation. Our shear catalogue used for this analysis, however, is not yet fully covered by UNIONS multi-band photometry. In order to still employ the colour-based approach for the redshift calibration, we exploit the spatial overlap between the UNIONS $r$-band data and the Canada–France–Hawai'i Telescope Lensing Survey \citep[CFHTLenS;][]{heymans2012cfhtlens,erben2013cfhtlens} W3 field, which covers $44.2~\mathrm{deg}^2$. CFHTLenS provides significantly deeper $ugriz$ photometry \citep{hildebrandt2012cfhtlens} than UNIONS, such that essentially all detected sources have CFHTLenS counterparts. We therefore cross-match these sources to CFHTLenS and adopt the associated $ugriz$ magnitudes. The matched sample is assumed to trace the same underlying colour–redshift distribution as the full UNIONS population.

To calibrate the redshift distribution, we assemble a spectroscopic sample ($\sim$ 65\,500 galaxies) that occupies the same photometric space as the matched UNIONS-CFHTLenS sources. This sample combines data from three deep spectroscopic surveys: the DEEP2 Galaxy Redshift Survey \citep{newman2013deep2}, the VIMOS VLT Deep Survey \citep[VVDS;][]{lefevre2005vvds}, and the Vimos Public Redshift Survey \citep[VIPERS;][]{scodeggio2018vipers}. All of these surveys were observed with the CFHTLenS $ugriz$ filters, enabling direct comparisons in colour–magnitude space. After applying standard survey-specific quality cuts, we obtain a clean and representative set of galaxies for the calibration of the redshift distribution.

Using the multi-band photometry of the spectroscopic sample, we train a self-organising map (SOM; \citealt{kohonen1982som}), which arranges galaxies based on their positions in the multi-dimensional magnitude space \citep{masters2015c3r2,wright2020som}. The initial SOM is defined on a $101 \times 101$ cell grid, providing a fine-grained tiling of colour–magnitude space. For robust statistical sampling, we subsequently hierarchically cluster the SOM into approximately 5000 effective resolution elements. This preserves the structure of the photometric manifold while ensuring that each region contains a sufficient number of spectroscopic objects for more robust mean statistics.

We then populate the SOM with the UNIONS sources by assigning each galaxy to its best-matching SOM cell based on its $ugriz$ photometry. Each UNIONS galaxy carries two lensing-related weights: (i) the galaxy weight $w$ (as explained in \hyperlink{cite.kilbinger.etal25}{Paper I}); and (ii) an additional weighting based on the average shear response $\langle R_\gamma \rangle$ .%\citep{myles2021redshiftcalibration}.

%The shear-response weight $w^{R}$ is constructed as follows. 
For the average shear response, we use the values that result from binning the whole UNIONS-3500 sample in a two-dimensional space defined by the signal-to-noise ratio (SNR) and the galaxy size parameter $r_{\rm h}/r_{\rm PSF}$ (shown in Figure 3 of \hyperlink{cite.kilbinger.etal25}{Paper I}). %is similar to the binning scheme presented in \citet{gatti2021shapecatalogue} and is used to compute the mean shear response $\langle R \rangle$ in each bin. 
Each galaxy $j$ is assigned the average shear response value of its bin $\langle R_\gamma \rangle_{\mathrm{bin}(j)}$, which we use as an additional multiplicative weight.%,
%\begin{equation}
%    w^{R}_j = \langle R \rangle_{\mathrm{bin}(j)}.
%\end{equation}

To account for potential selection biases in the spectroscopic sample, we adopt the ``prior volume weighting'' scheme described in \citet{wright2025legacynz}. This correction addresses the fact that the probability distribution of redshift at a fixed colour may differ between the spectroscopic and wide-field samples due to differing selection functions or spectroscopic success rates (e.g., colour-redshift degeneracies, \citealt{hartley2020}).
Following the methodology of \citet{wright2025legacynz}, we define a corrective weight for each spectroscopic galaxy $k$ as the ratio of the redshift probability density functions (PDFs) of a simulated wide-field sample ($P_{\rm w}$) and the calibration sample ($P_{\rm c}$):
\begin{equation}
w_k^{\rm prior} = \frac{P_{\rm w}(z_k)}{P_{\rm c}(z_k)}.
\end{equation}
The wide-field PDF, $P_{\rm w}(z)$, is derived from an analytic model for a magnitude-limited sample, $N(z, m)$, which was calibrated using SURFS-Shark light cones \citep{Elahi/etal:2018,lagos/etal:2018}. For our UNIONS sample, we approximate this selection using an $r$-band magnitude-limited model ($20 \leq r \leq 23.75$) to mimic the effective depth and the impact of the lensing shape weights. This prior volume weight ensures that even before the SOM cell assignment, the spectroscopic galaxies are adjusted to better represent the expected underlying redshift baseline of the UNIONS survey.

Combining these weighting schemes, the SOM weight for cell $i$ is defined as
\begin{equation}
    w_i^{\mathrm{SOM}} =
    \frac{\displaystyle \sum_{j} w_{j}\, \langle R_\gamma \rangle_{\mathrm{bin}(j)}}
    {\displaystyle \sum_{k} w_{k}^{\rm prior}},
\end{equation}
where the numerator sums the weighted counts of UNIONS galaxies $j$ assigned to cell $i$, and the denominator is the sum of the prior volume weights $w_k^{\rm prior}$ of spectroscopic calibration galaxies $k$ in that cell \citep{wright2025legacynz}. This definition ensures that the spectroscopic sample is reweighted to match the effective distribution of sources contributing to the shear signal.

\begin{figure}
    \centering
    \includegraphics[width=\linewidth]{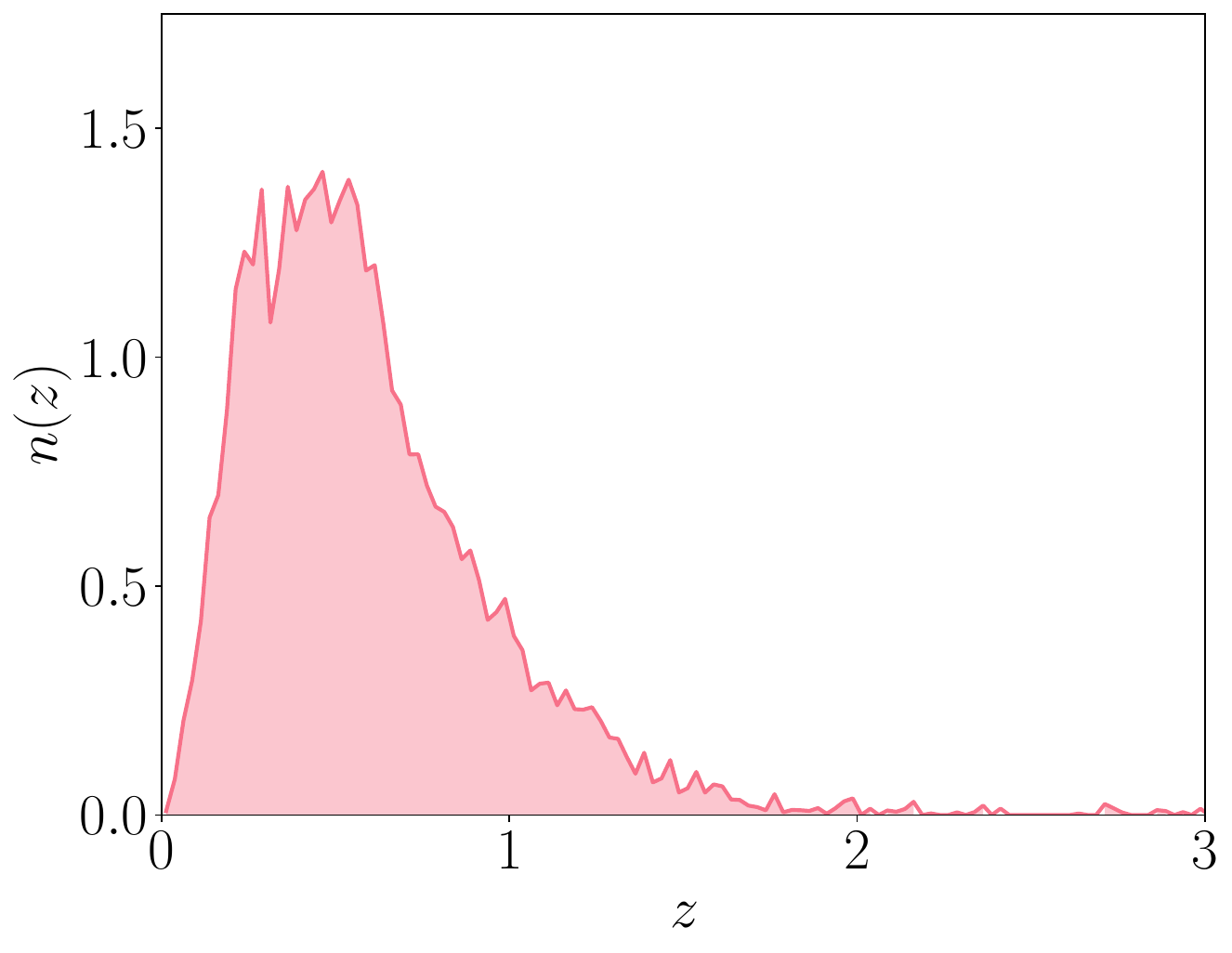}
    \caption{Normalised redshift distribution, $n(z)$.}
    \label{fig:nz}
\end{figure}

Finally, the redshift distribution of the UNIONS weak-lensing sample is obtained by applying the SOM weights to the spectroscopic redshift distributions in each cell:
\begin{equation}
    n(z)
    = \sum_i w_i^{\mathrm{SOM}}\, n_i^{\mathrm{spec}}(z),
\end{equation}
where $n_i^{\mathrm{spec}}(z)$ is the redshift distribution of spectroscopic galaxies located in SOM cell $i$. To obtain a statistically robust estimate of the final redshift distribution, we conduct a bootstrap resampling over the spatially binned spectroscopic calibration sample. We generate 1000 realisations of $n(z)$ and adopt the mean of these realisations as our final estimate. The final calibrated $n(z)$ distribution 
% and the underlying spectroscopic distribution $n^{\rm spec}(z)$
is shown in Fig.~\ref{fig:nz}.

\subsection{Shear multiplicative bias}

Errors in the shear measurement step potentially introduce systematic biases, such that the true ellipticity of a galaxy is given by
\begin{equation}
    \bs{e}^{\rm obs} = (1+m)\,\bs{e}^{\rm true} + \bs{c}
\end{equation}
where $\bs{c}$ is the additive component, corrected for during catalogue processing, and $m$ is the multiplicative bias, for which we detail its method of estimation in this section.

In Stage III ground-based surveys, blended objects, which are galaxies that receive flux from a close angular neighbour and cannot be spatially resolved, have been determined to be the leading contributor of multiplicative bias \citep{maccrann_dark_2021,liKiDSLegacyCalibrationUnifying2023}. In \hyperlink{cite.hervaspaters.etal25}{Paper V}, the image simulations and shear calibration efforts are detailed in full. In short, two suites of images were run: one with galaxies placed on a grid and one with realistic galaxy placements obtained from an $N$-body simulation. Each is run with four shears ($\gamma_1 = \pm 0.025$, $\gamma_2 = \pm 0.025$) to test how well the input shear is recovered.

The image simulations revealed three important sources of multiplicative bias that proved challenging to disentangle. Firstly, a bug in the shear measurement pipeline led to the WCS offsets between different epochs not being correctly propagated, giving rise to a $\Delta m$ contribution of approximately $-0.02$. Secondly, a selection effect was introduced when \texttt{SExtractor} Flags $= 0$ were selected, since objects with Flags $1$ and $2$ are, on average, more elliptical. While most selection effects are calibrated through \texttt{Metacalibration}, this one was not captured, as the detection was not repeated on the differently sheared branches. This also led to a multiplicative bias $\Delta m \approx -0.02$. Finally, physically blended objects estimated from simulated images with galaxies placed at realistic positions based on an $N$-body simulation gave a multiplicative bias $\Delta m \approx -0.017$.

Jointly, all effects were cumulatively responsible for a multiplicative bias $m = -0.057 \pm 0.0047$. We conservatively multiply the uncertainty by a factor of 3 to account for the fact that these strong effects might not be exactly captured by the simulated images. This leads to a total multiplicative bias $m = -0.057 \pm 0.014$.

\subsection{Covariance modelling}\label{sec:covmat}

\begin{figure}
    \centering
    \includegraphics[width=\linewidth]{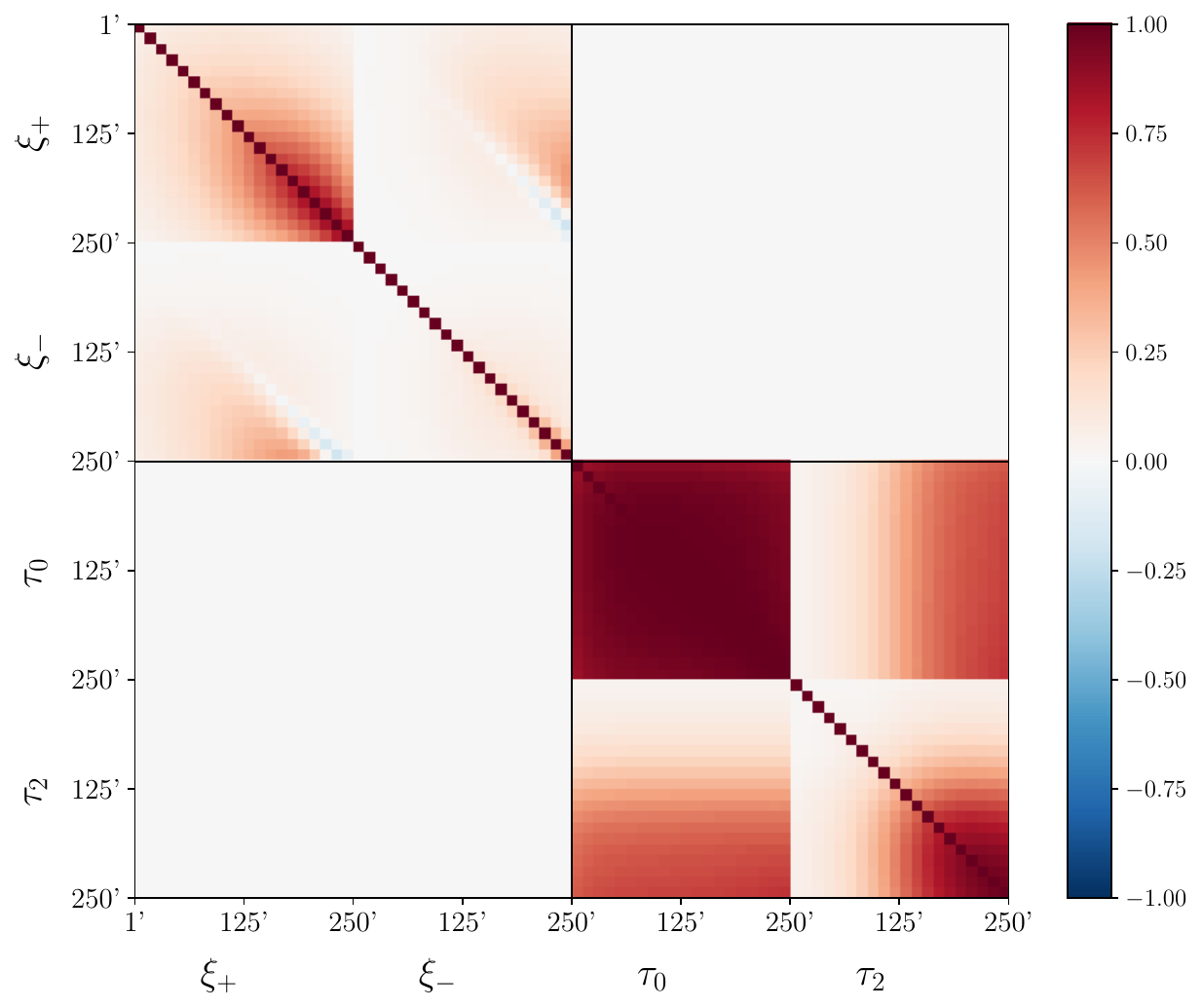}
    \caption{Correlation matrix of the two-point correlation functions, $\xi_\pm(\theta)$ and the $\tau_{0,2}(\theta)$ PSF systematics statistics (see Sect. \ref{sec:psf_inference}).}
    \label{fig:full_cov_matrix}
    \end{figure}

We calculate the covariance matrix between angular bins by taking into account the Gaussian (G), connected non-Gaussian (nG), and super sample covariance (SSC) contributions. In a tomographic analysis, the total covariance matrix of two angular power spectra $C^{ij}_{\ell_1}$ and $C^{kl}_{\ell_2}$ in Fourier space is given by
\begin{align}
        {\mathrm{Cov}}\left[C^{ij}_{\ell_1},C^{kl}_{\ell_2}\right]\,  = &
        \,{\mathrm{Cov}_{\mathrm G}}\left[C^{ij}_{\ell_1},C^{kl}_{\ell_2}\right]\,+ \nonumber {\mathrm{Cov}_{\mathrm{nG}}}\left[C^{ij}_{\ell_1},C^{kl}_{\ell_2}\right] + \nonumber \\
        &\,{\mathrm{Cov}_{\mathrm{SSC}}}\left[C^{ij}_{\ell_1},C^{kl}_{\ell_2}\right].
\end{align}

Following the notation of \cite{PhysRevD.70.043009}, the Gaussian covariance term of any galaxy $g$ or shear $\gamma$ field can be expressed as
\begin{align}
\label{eq: covgauss}
{\mathrm {Cov}_{\mathrm G}}= &
\left[(2\ell_1 + 1)\,f_{\mathrm{sky}} \, \Delta \ell \right]^{-1}
\delta_{\ell_1 \ell_2}^{\mathrm K} 
\, \times \nonumber \\ & 4\;
 \Bigg\{ \left[C^{AC,ik}_{\ell_1} + {N}^{AC,ik}_{\ell_1}\right]
\left[C^{BD,jl}_{\ell_2} + N^{BD,jl}_{\ell_2} \right]
 + \nonumber \\ & 
 \left[C^{AD,il}_{\ell_1} + N^{AD,il}_{\ell_1} \right]
\left[C^{BC,jk}_{\ell_2} + N^{BC,jk}_{\ell_2} \right]
\Bigg\}
\end{align}
where $A, B, C, D$ represents either $g$ or $\gamma$ for the observed fraction of the sky $f_{\mathrm{sky}}$, and noise power spectra $N_{ij}$. In the case of cosmic shear, $A=B=C=D=\gamma$, hence it is simply $N^{\gamma\gamma,ij}_\ell=\frac{\sigma_{\bs{e}}^2}{2\bar{n}^{i}}\,\delta_{ij}^{\mathrm K}$. Here $\bar{n}^i$ is the mean galaxy number density of bin $i$ and $\sigma_{\bs{e}}$ the shape noise, while $\delta_{ij}^{\mathrm K}$ denotes the Kronecker delta. However, transitioning to small scales, the cosmic shear field becomes significantly non-Gaussian, leading to extra contributions arising from the connected four-point function of these fields. For the full expression of the connected non-Gaussian covariance terms, see for example \cite{Cooray:2002dia} and \cite{takada_jain_2009}.

Lastly, the SSC term, which captures the uncertainty due to a change in the background density at modes larger than the survey area, can be modelled following \cite{Takada2013}:
\begin{multline}\label{eq: covSSCintermediate}
   {\mathrm {Cov}_{\mathrm{SSC}}}\simeq \frac{1}{f_{\mathrm{sky}}}
   \int d\chi\,\frac{q^{A,i}(\chi)q^{B,j}(\chi)q^{C,k}(\chi)q^{D,l}(\chi)}{\chi^4}\times\\
   \frac{\partial P_{AB}(k_{\ell_1} , z)}{\partial \delta_{\mathrm b}}\,
   \frac{\partial P_{CD}(k_{\ell_2} , z)}{\partial \delta_{\mathrm b}}\,
   \sigma_{\mathrm b}^2(z) \; ,
\end{multline}
where $\sigma_{\mathrm b}^2$ is the covariance of the background density field $\delta_\mathrm{b}$ within the survey window. In the case of a cosmic shear-only analysis, $q^{A,B,C,D}_i(\chi)$ is the cosmic shear lensing efficiency given in Eq. \eqref{eq:lens_eff_cs}, and $\sigma_{\mathrm b}$ is given by \citep{Lacasa_2016}
\begin{equation}\label{eq: sigma}
   \sigma_{\mathrm b}^2(z) = \frac{1}{2 \pi^{2}} \int dk \; k^{2} \,P_\mathrm{L}\left(k , z\right)\, [\textrm{j}_{0}(kr)]^2 \,,
\end{equation}
where j$_0$ is the spherical Bessel function of zeroth order, and  $P_\mathrm{L}(k,z)$ refers to the linear matter power spectrum.

The covariance matrix for the 2PCF in configuration space follows from a Hankel transformation,
\begin{align}
    {\mathrm{Cov}} \left(\xi^{ij}(\theta_1),\xi^{kl}(\theta_2)\right)&=\frac{1}{4\pi^2} \nonumber\\&
    \times\int \frac{d\ell_1}{\ell_1}
    \int \frac{d\ell_2}{\ell_2}\,\ell_1^2\,\ell_2^2\, \textrm{J}_n(\ell_1\theta_1)\, \textrm{J}_n(\ell_2\theta_2) \nonumber\\& 
    \times\left[{\mathrm{Cov}}\left(C^{ij}_{\ell_1},C^{kl}_{\ell_2}\right)\right],
\end{align}
where once again $n=0$ for $\xi_+$ and $n=4$ for $\xi_-$. We calculate our covariance matrix based on the above equations, using the \texttt{CosmoCov} software \citep{2017MNRAS.470.2100K,Fang__2020},  where we have adopted the fiducial input cosmology following table 1 of \cite{onecovariance}. We plot the full correlation matrix (where $\textup{Corr}_{ij}=\textup{Cov}_{ij}/\sqrt{\textup{Cov}_{ii}\times\textup{Cov}_{jj}}$) in Fig. \ref{fig:full_cov_matrix}.

\subsubsection{Effect of masking on covariance estimation}
As was investigated in \cite{troxel2018maskingkids} and \cite{friedrich2018maskingdes}, masking can result in a noncontiguous and patchy survey footprint, thus inducing additional uncertainties in the covariance matrix of the correlation functions. Following \cite{Schneider:2002jd},  the pure shape noise term of the covariance matrix in configuration space (referred to as $N^{\gamma\gamma}_\ell$ in the previous section) is given by
\begin{equation}\label{eq:noise_cov}
    {\mathrm{Cov}}^\mathrm{SN} \left(\xi_{\pm}^{ij}(\theta_1),\xi_{\pm}^{kl}(\theta_2)\right)=\frac{(\sigma_{\bs{e}}^i\sigma_{\bs{e}}^j)^2}{N^{ij}_{\mathrm p}(\theta_1)}\delta_{\theta_1\theta_2}^{\mathrm K}\left(\delta_{ik}^{\mathrm K}\delta_{jl}^{\mathrm K}+\delta_{il}^{\mathrm K}\delta_{jk}^{\mathrm K}\right),
\end{equation}
where $N_{\mathrm p}^{ij}(\theta)$ is as defined in Eq. \eqref{eq:xi_estimator}. When boundary and masking effects are neglected, it can be approximated by
\begin{equation}
    N_{\mathrm{p}}^{ij}(\theta)=2\pi A\theta\Delta_\theta\bar{n}_i\bar{n}_j\;,
\end{equation}
where $A$ is the survey area. On the other hand, a more accurate estimation of $N_{\mathrm p}^{ij}$ can be obtained by taking into account the survey mask, such that 
\begin{equation}
        N_{\mathrm{p}}^{ij}(\theta)=2\pi A\theta\Delta_\theta\bar{n}_i\bar{n}_jw_{\rm{mask}}(\theta)\;, 
\end{equation}
where the mask power spectrum $w_{\rm{mask}}(\theta)$, normalised to the survey area, modifies $N_{\mathrm{p}}^{ij}$.

\begin{figure}
    \centering
    \begin{subfigure}{\linewidth}
        \centering
        \includegraphics[width=\linewidth]{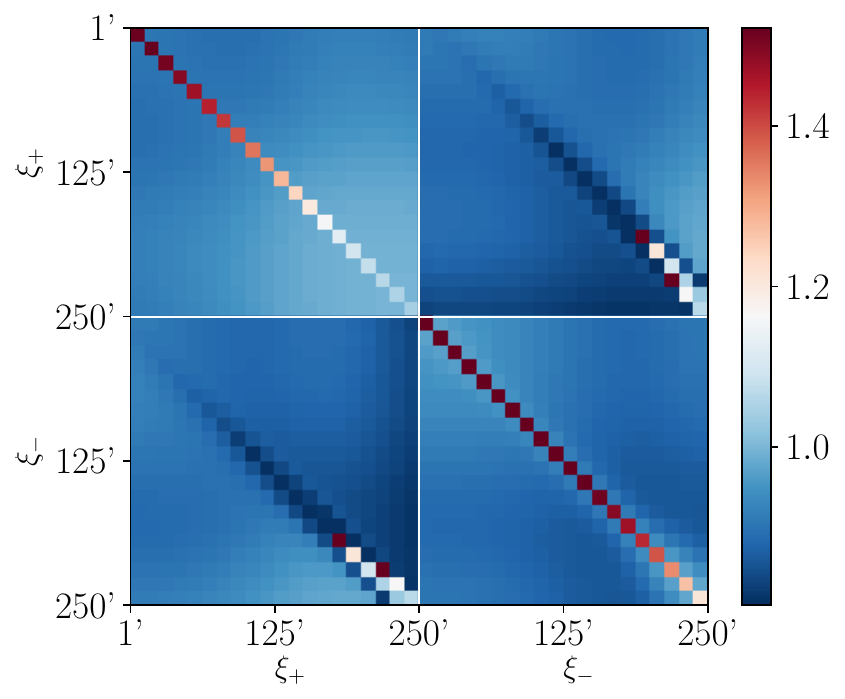}
    \end{subfigure}
    \vspace{1em} % optional vertical space
    \begin{subfigure}{\linewidth}
        \centering
        \includegraphics[width=\linewidth]{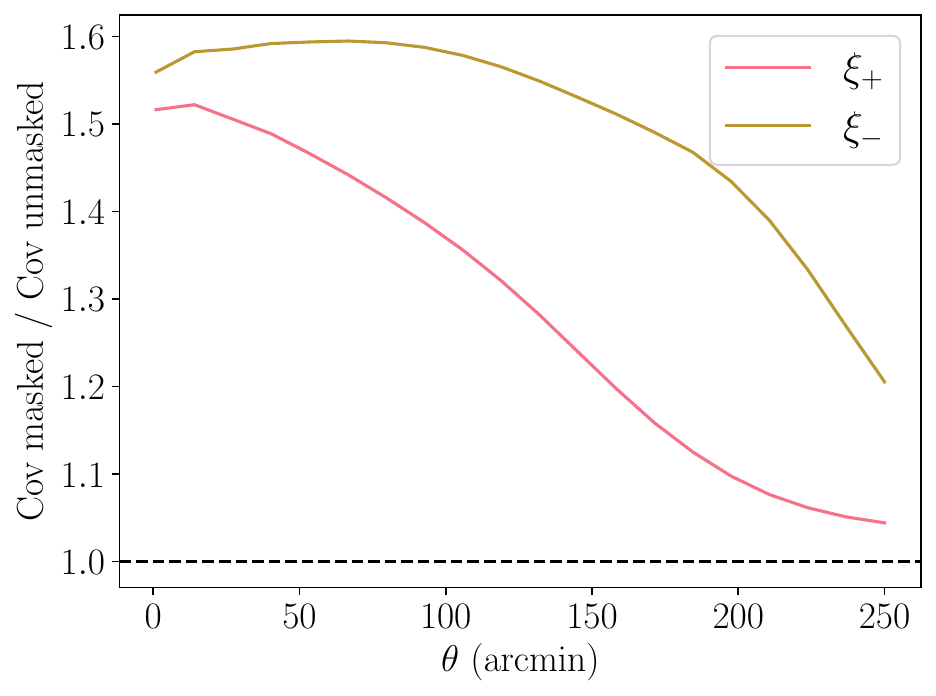}
    \end{subfigure}
    \caption{Ratio of the total covariance matrix with and without the survey mask, for both the auto-covariance and cross-covariance of the $\xi_\pm(\theta)$ correlation functions (top) and their diagonals (bottom), over $\theta\in[1,250]~$arcmin.}
    \label{fig:masked_covmat}
\end{figure}

We examine the effect of the UNIONS mask on the covariance matrix, and present the full ratio of the masked to unmasked covariance matrix, including the cross-covariances and non-Gaussian and SSC contributions, in Fig. \ref{fig:masked_covmat}. We find that overall, including masking effects increases the diagonal terms of the auto-covariance (i.e. the noise terms as defined in Eq. \eqref{eq:noise_cov}), with a peak in the ratio occurring at approximately $1$~arcmin for $\xi_+$ and $20$~arcmin for $\xi_-$.  This ratio decreases with increasing $\theta$, as the angular separation of the measurement becomes larger than the masking scale. The overall impact of masking is larger for the uncertainties in $\xi_-$ than $\xi_+$. Our results are consistent with the trend and values found in \cite{troxel2018maskingkids}, as shown in the bottom panel of their Fig. 1. On the other hand, the ratios of the off-diagonal and cross-covariance terms are reduced (except at large scales). This is because the off-diagonal terms include an SSC contribution, which is suppressed under masking. This is apparent in the integrand for $\sigma_{\mathrm b}^2$, where the survey window is rescaled by the mask power spectrum, thereby downweighting the contribution of modes suppressed by the survey mask to the overall background variance.

We adopt the masked covariance matrix as our fiducial choice, and also present results with the unmasked covariance to quantify its impact on our analysis.

%--------------------------------------------------------------------
\subsection{External data sets}
Here we summarise the external data sets incorporated into our inference runs. These include highly complementary measurements from the CMB, which provide information at high redshifts, as well as BAO data, which offer information on the expansion history at low redshifts. Combined with cosmic shear, they give a more complete description of the Universe at both the background and perturbation levels, tightening overall cosmological constraints. 

\subsubsection{Cosmic Microwave Background}
The CMB, relic radiation from free-streaming photons that decoupled from baryons at redshifts $z\simeq1100$, provides highly precise constraints on early-Universe physics. We include data from the 2018 data release of the \textit{Planck} mission \citep{planck_data}, specifically those of the \textit{TT}, \textit{TE} and \textit{EE} auto- and cross-spectra, spanning a multipole range of $2\leq \ell \leq 2508$ for \textit{TT}, and $2 \leq \ell \leq 1996$ for \textit{TE} and \textit{EE}. To avoid having to sample over the large number of CMB nuisance parameters, we employ instead the \texttt{plik\_lite} likelihood \citep{plike_lite}, which assumes a Gaussian likelihood and only marginalises over the nuisance parameter $A_{\mathrm{lens}}$, the correction amplitude due to lensing effects. Furthermore, we do not include the lensing power spectrum likelihood, as we currently lack a prescription for the cross-correlation in the covariance matrix between cosmic shear and lensing data.

\subsubsection{Baryonic Acoustic Oscillations}
We include observations of the BAO scale from the second data release of the Dark Energy Spectroscopic Instrument \cite[DESI DR2;][]{desi_dr2_1, desi_dr2_2}, in the form of measurements of the transverse and comoving BAO distance $(D_\mathrm{M}$ and $D_\textrm{H}$ respectively). They span a redshift range of $0.1 < z< 3$, and represent the most comprehensive and precise measurement of the BAO scale to date. These results attracted significant attention, as they favour a cosmological model with a varying dark energy equation of state over a cosmological constant $\Lambda$. However, we shall only consider a $\Lambda$CDM model in all our analyses, for which BAO can lend competitive constraining power on $\Omega_{\rm m}$. We employ the full range of BAO measurements taken from the Bright Galaxy Survey (BGS), Luminous Red Galaxies (LRG), Emission Line Galaxies (ELG), quasars (QSO) and Lyman-$\alpha$ forest (Ly$\alpha$) samples.

%--------------------------------------------------------------------

\section{Inference pipeline setup}\label{sec:inference_pipeline}

We use Bayesian inference to derive constraints on the cosmological parameters of the concordance $\Lambda$CDM model from the data vectors, covariance matrix, and redshift distribution described above. 

Bayesian inference relies on Bayes' Theorem, which gives the probability distribution of the parameters $\theta$ given a model $M$ and the observed data $d$. This probability distribution $P(\theta|d, M)$, known as the posterior distribution, is defined as 
\begin{equation}\label{eq:BayesTh}
P(\theta|d, M)=\frac{\mathcal{L}(d|\theta, M) \Pi(\theta|M)}{\mathcal{Z}(d|M)},
\end{equation}
where $\Pi(\theta|M)$ is the prior distribution, quantifying our initial knowledge of the distribution of $\theta$, $\mathcal{Z}({d}|M)$ is the evidence, which gives the probability of observing the data given $M$, and $\mathcal{L}({d}|\theta, M)$ is known as the likelihood function, which is the probability of observing the data $d$ given the model $M$ with parameter values $\theta$.

We adopt a Gaussian likelihood:
\begin{equation}\label{eq:likelihood_gaussian}
-2\log{\mathcal{L}} \propto [d-T(\theta)]^{\mathrm{T}}\hspace{0.4mm} \mathbf{C}^{-1} [d-T(\theta)],
\end{equation}
where $d$ is the data vector, $T(\theta)$ is the theory vector derived from the model, and $\mathbf{C}$ is the covariance matrix of the data. 

\subsection{Likelihood inference}

We use the Einstein--Boltzmann solver \texttt{CAMB} \citep{Lewis:1999bs} to calculate cosmological quantities and the linear matter power spectrum, and the \textsc{CosmoSIS} \citep{Zuntz:2014csq} pipeline for parameter inference. Our baseline sampler is \texttt{Polychord} \citep{Polychord, Polychord2}, following many other Stage III survey analyses \citep{desy3-cosmo, hsc-y3, hsc-y3-2, decade}. \texttt{Polychord} produces reliable posterior estimates at reasonable computational cost \citep{polychord-test}, though it becomes expensive as the parameter space grows. We therefore also run \texttt{Nautilus} \citep{nautilus}, which leverages deep learning to improve sampling efficiency \citep{euclid-cloe,Wright_kids_2025}, as a consistency check.

We adopt prior ranges for the cosmological parameters as presented in Table 2 of \cite{Kilo-DegreeSurvey:2023gfr}, except that we additionally impose a prior on the baryonic energy density parameter $\omega_{\mathrm b}\equiv\Omega_{\mathrm b}h^2$, which was derived from Big Bang nucleosynthesis (BBN) constraints. Computed from the \texttt{PRyModial} code \citep{prymordial, Schoneberg:2024ifp}, it gives $\omega_\mathrm{b}=0.02218\pm0.00055$ assuming a $\Lambda$CDM model. In total, the cosmological parameters being sampled are $\{\omega_{\mathrm c},\omega_{\mathrm b}, H_0, n_{\mathrm s},S_8\}$: the energy density of cold dark matter; the energy density of baryons; the present-day value of the Hubble parameter; the spectral index; and the amplitude of clustering $S_8\equiv\sigma_8\sqrt{\Omega_\mathrm{m}/0.3}$ \footnote{\cite{2021A&A...646A.129J} argued that sampling $S_8$ gives a more uninformative prior volume as opposed to sampling $A_{\mathrm{s}}$, the amplitude of the primordial power spectrum. This methodology has since been adopted in \cite{asgari_kids-1000_2021} and \cite{Wright_kids_2025}.}. Since the optical depth of reionisation $\tau_{\mathrm{reio}}$ is not well constrained with cosmic shear data, we fix it at the \textit{Planck} 2018 best-fit value of $\tau_{\mathrm{reio}}=0.0544$ \citep{planck2018}. However, when including CMB data, we do not impose the BBN prior, and additionally sample $\tau_{\mathrm{reio}}$. We also assume a flat Universe and one massive and two massless neutrino species, with a total mass of $\Sigma m_{\nu}=0.06$eV. The following subsections describe the modelling of intrinsic alignment, the nonlinear matter power spectrum, and PSF systematics. Table \ref{tab:inference_priors} lists the full set of priors for the cosmological and nuisance parameters that we sample in our analysis. 

When reporting the best-fit parameter values for all parameters except $S_8$ and $\Omega_{\rm m}$, we quote their 1D marginalised mode, or marginalised maximum a posteriori (MAP), and their 68\% confidence interval (CI), as in recent analyses such as \cite{hsc-y3, hsc-y3-2}. This is obtained by evaluating the peak of the 1D posterior distribution after marginalising over all remaining parameters, which we estimate using the in-built Gaussian KDE smoothing kernel of \texttt{getdist} \citep{getdist}. Other commonly adopted statistics include the weighted mean of the marginalised posterior \citep[e.g.][]{des_y6_cosmo}, while others report the MAP value with a 68\% confidence region defined by the projected joint highest posterior density \citep[PJ-HPD;][]{2021A&A...646A.129J}, as in \cite{Wright_kids_2025}. Each approach has distinct advantages and limitations \citep[see][for a detailed discussion]{Kilo-DegreeSurvey:2023gfr}. In particular, the MAP and its associated PJ-HPD interval can be challenging to estimate robustly from nested sampling chains, due to the relatively sparse sampling of high-posterior regions \citep{des-muir}, and can be unstable in the presence of strong parameter degeneracies. Conversely, statistics based on marginalisation, such as the weighted mean and marginalised mode, are susceptible to volume projection effects, especially for non-Gaussian posteriors. We thus attempt to balance a trade-off by choosing instead to report the 2D marginalised mode for $S_8$ and $\Omega_{\rm m}$, the two parameters which are the most important and also most sensitive to projection effects in cosmic shear analyses (see Sect. \ref{sec:glass_mocks}). By marginalising over both parameters simultaneously, we mitigate the worst of the impact due to projection effects when marginalising in one dimension. For completeness, we also quote the weighted mean of $S_8$ in Table \ref{tab:best_fit_chi}. 

\subsection{Intrinsic alignment} \label{sec:ia}
Multiple strategies to disentangle intrinsic alignment (IA) from cosmic shear require accurate estimates of galaxy redshifts \citep{Joachimi_nulling_2010}. Recent cosmic shear measurements have been somewhat successful in directly estimating the intrinsic alignment parameters when jointly fitting the shear-shear ($\bs{\gamma\gamma}$), shear-intrinsic ($\bs{\gamma}$I) and intrinsic-intrinsic (II) correlations to the signal \citep{asgari_kids-1000_2021,Secco_DES_2022}. This success can be traced back to the very different ways these effects act with respect to galaxy separation: small angular bins are more affected by II contributions, while bins with larger separation have a strong $\bs{\gamma}$I contribution. 

In Fourier space, the noiseless cosmic shear angular power spectrum decomposes as
\begin{equation}
C^{\bs{\epsilon}\bs{\epsilon}}_\ell=C^{\bs{\gamma}\bs{\gamma}}_\ell+C^{\bs{\gamma} \textrm{I}}_\ell+C^{\textrm{II}}_\ell\,.
\end{equation}
Models exist which give an approximate theoretical prescription for the latter two terms, with one of the widely employed models being the redshift-dependent nonlinear linear alignment model \citep[NLA;][]{hirata_intrinsic_2004,bridle_dark_2007}. Here, the $3$D power spectra $P^{\bs{\gamma} \textrm{I}}(k,z)$ and $P^{\textrm{II}}(k,z)$ are expressed as a rescaling of the nonlinear matter power spectrum
\begin{align}
P^{\bs{\gamma} \textrm{I}}(k,z) &=
 - A_{\rm IA} C_1 \rho_{\rm crit}
 \frac{\Omega_{\mathrm{m}}}{D(z)}
 P_{\rm NL}(k,z)\,, \\
P^{\textrm{II}}(k,z) &=
 \left(
 A_{\rm IA} C_1 \rho_{\rm crit}
 \frac{\Omega_{\mathrm{m}}}{D(z)}
 \right)^2
 P_{\rm NL}(k,z)\,,
\end{align}
where $A_{\rm{IA}}$ is a dimensionless IA amplitude parameter to be marginalised over, $\rho_{\rm{crit}}$ is the critical energy density, $D(z)$ is the growth factor and $C_1=5\times10^{-14}h^{-2}M^{-1}_\odot$Mpc$^3$.

In this analysis, the use of a single redshift bin prohibits us from efficiently constraining $A_{\mathrm{IA}}$, necessitating the imposition of a well-motivated prior. Since multiple direct intrinsic alignment measurement models have been proposed \citep{joachimi_constraints_2011, mandelbaum_wigglez_2011, singh_intrinsic_2015,johnston_kidsgama_2019, fortuna_kids-1000_2021,samuroff_dark_2023,Navarro-Girones_pau_2025}, we choose to rely on these, further ignoring the impact of any higher-order terms, such as those proposed in the tidal alignment and tidal torquing model \citep[TATT;][]{blazek_beyond_2019}. This strategy has also been adopted in previous works \citep{fortuna_halo_2021,Li_Kids_2023,Wright_kids_2025}.

We use a data-driven approach to estimate $A_{\rm{IA}}$, making use of the same W3 CFHTLenS \texttt{BPZ} \citep{Benitez_BPZ_2000} catalogue from which the $n(z)$ was derived. 
% Since our knowledge of intrinsic alignment is already limited by external data, we do not 
% employ the first-order NLA model \citep{hirata_intrinsic_2004,bridle_dark_2007}, nor 
We then adopt the commonly employed method of splitting intrinsic alignment contributions coming from red and blue galaxies \citep{Krause_IA_2015}, motivated by the distinction between pressure-supported and rotationally supported galaxies, respectively. The parametrisation is given by
\begin{equation}
    A_{\mathrm{IA}}=f_{\mathrm r} \, A_{\mathrm {IA,r}}+f_{\mathrm b} \,A_{\mathrm{IA,b}} \ .
\end{equation}
where $f_{\mathrm r}$ indicates the fraction of red galaxies, and $f_{\mathrm b}$ indicates the fraction of blue galaxies within our sample. For simplicity, we do not introduce any luminosity or redshift dependence, hence $A_{\mathrm{IA}}$ is a constant parameter. To separate blue from red galaxies, we choose to assign all galaxies with $T_B < 1.9$ as red galaxies and treat the rest as blue. The determination of the $T_B$ index is done through \texttt{BPZ}, where it utilises six model templates, arranged by increasing star formation activity, and performs linear interpolation between neighbouring templates to find the best-matching spectral energy distribution. The output parameter, $T_B$, represents the selected best-fit template, or more precisely the blend of adjacent templates, in steps of 0.1. From there, we obtain estimates for $f_{\mathrm r}=0.245$ and $f_{\mathrm b}=0.745$.

For the values $A_{\mathrm{IA,r}}$ and $A_{\mathrm{IA,b}}$, we use prior values derived from direct measurements. These rely on accurate redshift estimation, either photometric or spectroscopic, to compute the projected shape-density correlation function $w_{g+}$ with limited separation. For the blue galaxy sample, we adopt the result from \cite{johnston_kidsgama_2019}, which fitted the available measurements to obtain $A_{\mathrm{IA,b}}=0.21\pm0.37$. While other works using this model have set the intrinsic alignment of blue galaxies to 0 (consistent with the data), we prefer a more conservative approach and adopt this observationally grounded prior with a non-negligible Gaussian width. For an estimate of $A_{\mathrm {IA,r}}$, we sample from the posterior distribution of the double power-law fit done on $A_{\mathrm{IA}}(L/L_0)$ at the mean luminosity value of red galaxies measured on W3, following \cite{Hervas_Peters_IA_2024}. The luminosities are obtained from \texttt{LePhare} \citep{Lephare_2011}, where the mean value for UNIONS is found to be $(L/L_0)=-0.77$. Sampling the posterior gives us $A_{\mathrm{IA,r}}=2.75\pm0.49$, a value in accordance with most KiDS estimates, but not with fiducial DES values \citep{Kilo-DegreeSurvey:2023gfr}. We then add the Gaussian errors in quadrature to arrive at a prior of $A_{\mathrm{IA}}=0.83 \pm 0.39$. We have tested the robustness of this prior against changes in $T_B$, finding only small differences. 

This method is mainly limited by two factors. First, the galaxies for which reliable spectroscopic or photometric redshifts are available are not representative of either the blue or red population, since they mostly need to be bright to have reliable photometric redshifts or be targeted by spectroscopic surveys. Second, these various methods are all obtained with different lensing surveys. It has been shown that the shape algorithm used \citep{singh_intrinsic_2016} and the band the galaxy has been measured in \cite{georgiou_dependence_2019} impact the measured amplitude of the intrinsic alignment signal. These two reasons motivate the doubling of the width of our Gaussian prior to $\pm0.78$ instead. For comparison, we also conduct a run assuming an uninformative flat prior on $A_{\mathrm{IA}}$, and additionally assume the absence of any IA effects (setting $A_{\rm IA}=0$).

\subsection{PSF systematics}\label{sec:psf_inference}
To mitigate systematic effects from PSF mismodelling, we employ two safeguards: firstly, we apply scale cuts to ensure that the additive contribution from the leakage signal, $\xi_\pm^{\mathrm{sys}}$, remains below $10\%$ of the total measured signal. Additionally, we have chosen to jointly fit the observed correlation function signal $\xi_\pm^{\mathrm{obs}}$ with the $\tau$ statistics as introduced in Sect. \ref{sec:psf_leakage}. Specifically, we sample $\alpha$ and $\beta$ as nuisance parameters and compute the theoretical $\tau$ statistics according to Eq. \eqref{eq:tau-stats}, given the $\rho(\theta)$'s estimated from the data. Since we set $\eta=0$, this would require only the computation of $\tau_0(\theta)$ and $\tau_2(\theta)$, where we only consider the $\tau_+$ signal. The $\tau$ statistic likelihood is thus calculated as
\begin{equation}
    \chi^2_{\tau_{0,2}} = [\tau_{0,2,\mathrm{d}}(\theta)\,-\,\tau_{0,2,\mathrm{t}}(\theta)]^{\mathrm{T}}\mathbf{C}_{\tau}^{-1}[\tau_{0,2,\mathrm{d}}(\theta)\,-\,\tau_{0,2,\mathrm{t}}(\theta)]\,,
\end{equation}
where the subscripts `d' and `t' denote the data and theory vectors respectively. $\mathbf{C}_{\tau}$ is a semi-analytical covariance, presented in Fig.~\ref{fig:full_cov_matrix} concatenated with the shear-shear covariance; see \cite{guerriniGalaxyPointSpread2025} for its full derivation. Cross-correlations between the shear-shear 2PCF and the $\tau$ statistics are neglected for simplicity.  

Additionally, we include the PSF systematic contribution as an additive bias to the cosmological signal, by calculating the theoretical $\xi_\pm^{\mathrm{sys}}$ following Eq. \eqref{eq:xi_sys}, given the sampled values of $\alpha$ and $\beta$. The total log-likelihood being minimised is thus
\begin{equation}
\chi^2_{\mathrm{tot}}=\chi^2_{\xi_\pm^{\mathrm{obs}}}+\chi^2_{\tau_{0,2,+}}\,.
\end{equation}
This joint fit accounts for the additive leakage bias and marginalises over PSF systematic uncertainty. Other Stage III surveys employ a similar approach \cite[see for example][]{hsc-y3-2,HSC-PSF}.

For the fiducial analysis, we use Gaussian priors on $\alpha$ and $\beta$ rather than flat priors. For data vectors with significant PSF systematics, unconstrained $\alpha$ and $\beta$ might risk biased posteriors. Specifically, if these parameters are left unconstrained (i.e., with wide priors) during the main cosmological inference step, they may fit the cosmological signal $\xi_\pm^{\gamma\gamma}$ rather than the systematic $\xi_\pm^{\mathrm{sys}}$ (Eq.~\eqref{eq: xi with leakage bias}), since non-trivial degeneracies exist between leakage and cosmological parameters. Consequently, the resulting posteriors on $\alpha$ and $\beta$ would be inconsistent with those obtained instead from a direct $\rho$/$\tau$ fit.

To obtain the priors on $\alpha$ and $\beta$, we first run a separate `PSF systematics inference step' whereby we solely fit $\alpha$ and $\beta$ based on the $\rho$ and $\tau$ statistics estimated from the data, per Eq. \eqref{eq:tau-stats}. This is the methodology described above. We use a vanilla $\texttt{emcee}$ sampler to minimise $\chi^2_{\tau_{0,2}}$, and the resulting posteriors of $\alpha$ and $\beta$ serve as priors in the cosmological inference step (see Table \ref{tab:inference_priors} for their values). We refer the reader to \cite{guerriniGalaxyPointSpread2025} for a detailed description of the methodology of this PSF systematics inference step. 

\begin{figure}
    \centering
    \includegraphics[width=\linewidth]{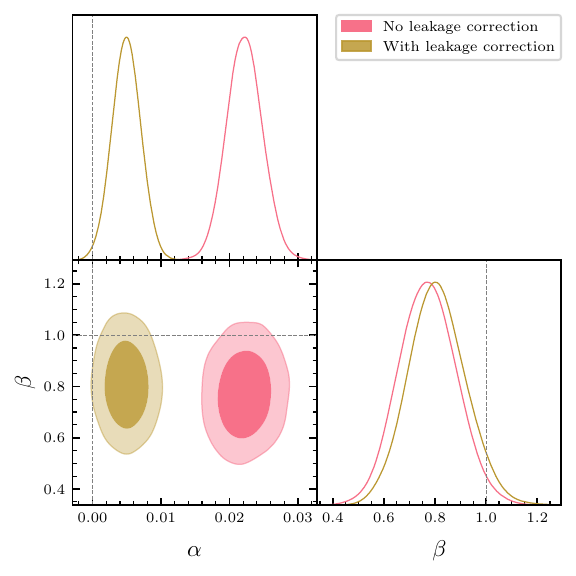}
    \caption{Marginalised posteriors of $\alpha_\mathrm{PSF}$ and $\beta_\mathrm{PSF}$ obtained from the PSF inference step (see Sect. \ref{sec:psf_inference}), which are subsequently adopted as priors when they are sampled in the cosmological inference analysis. We include the results for both sets of ellipticities: the object-wise leakage corrected one (gold) and the non-leakage corrected one (pink). The catalogue-level object-wise leakage correction effectively removes the leakage component of the additive bias ($\alpha$ is close to 0 for the gold contours).}
    \label{fig: psf_leakage_prior}
\end{figure}

We also conduct a PSF inference run where we use the catalogue containing the set of galaxy ellipticities that have not been corrected for PSF leakage, i.e., not corrected object-wise (see \hyperlink{cite.kilbinger.etal25}{Paper I} for details). These ellipticities give different values of $\tau(\theta)$, $\rho(\theta)$, and subsequently different posteriors for $\alpha$ and $\beta$. Figure~\ref{fig: psf_leakage_prior} shows the marginalised posteriors of $\alpha$ and $\beta$ for both the object-wise leakage-corrected and uncorrected ellipticities. The non-leakage-corrected case (in pink) shows a larger mean $\alpha$, as expected from the greater residual PSF leakage when this correction is not applied. On the other hand, $\beta$ remains largely constant, showing that it mainly accounts for the PSF model ellipticity, $\bs{e}^{\mathrm{p}}$. Nonetheless, leakage not corrected at the object level can be absorbed into a larger $\alpha$, and thus a larger $\xi_\pm^{\mathrm{sys}}$ signal, giving consistent cosmological posteriors with the leakage-corrected case.

For comparison, we shall also conduct a cosmological inference run where we adopt flat priors on $\alpha$ and $\beta$ (where $\alpha\in[-0.1,0.1]$ and $\beta\in[-2.0,2.0]$), as well as a setup where we use the non-leakage corrected set of ellipticities, to assess the impact of the PSF inference step on the resultant estimate of $\xi_{\pm}^{\rm{sys}}$.  

\begin{table}
	\centering

	\caption{Sampled cosmological and nuisance parameters with their adopted priors. Uniform and Gaussian priors are listed in the table, with CMB-specific ranges indicated where relevant.}
	\label{tab:inference_priors}
	\begin{tabularx} {\linewidth}{XX}% four columns, alignment for each
		\hline
		\textbf{Parameter} & \textbf{Prior}\\
        \hline
        \multicolumn{2}{c}{\textbf{Cosmology}} \\
		\hline
        $\omega_{\mathrm b}$ [no CMB] & $\mathcal{N}(0.02218,0.00055)$\\
        $\omega_{\mathrm b}$ [with CMB] & $\mathcal{U}(0.0190, 0.0260)$\\
        $\omega_{\mathrm c}$ & $\mathcal U(0.051, 0.255)$\\
        $h$ & $\mathcal U(0.64, 0.82)$ \\
        $n_{\mathrm s}$ & $\mathcal U(0.84, 1.10)$ \\
        $S_8$  & $\mathcal U(0.1,1.3)$ \\
        $\log_{10}(T_{\mathrm{AGN}}/\mathrm{K})$ &  $\mathcal U(7.3, 8.0)$ \\
        $\tau_{\mathrm{reio}}$ [with CMB] & $\mathcal{U}(0.01, 0.80)$ \\
        \hline
        \multicolumn{2}{c}{\textbf{Nuisance}} \\
        \hline
        $A_{\mathrm{IA}}$ & $\mathcal N(0.83, 0.78)$\\
        $\alpha_\mathrm{PSF}$ (leakage corrected)& $\mathcal N(0.005, 0.0022)$\\
        $\beta_\mathrm{PSF}$ (leakage corrected)& $\mathcal N(0.810, 0.1148)$\\
        $\alpha_\mathrm{PSF}$ (non-leakage corrected)& $\mathcal N(0.022, 0.0026)$\\
        $\beta_\mathrm{PSF}$ (non-leakage corrected)& $\mathcal N(0.798, 0.1150)$\\
        $\Delta z$& $\mathcal{N}(-0.003,0.018)$\\
        $m$& $\mathcal{N}(-0.057, 0.014)$\\
		\hline
	\end{tabularx}
\end{table}

\subsection{Redshift calibration bias estimation} \label{sec:nz_bias}

Biases in the calibration of the redshift distribution $n(z)$ propagate into cosmological parameter inference. We therefore marginalise over $\Delta z$, the bias in the mean redshift. Here we describe how we obtain an informed prior on this parameter.

Because the true $n(z)$ of the data is unknown, this bias can only be assessed using simulations, where the true redshifts are available. By constructing mock catalogues that mimic the photometric properties, selection functions, and redshift-calibration steps of the real UNIONS analysis, we directly compare the recovered SOM-based $n(z)$ to the true underlying distribution, obtaining estimates of the systematic shift $\Delta z$.

We construct our mock catalogues from the MICE2 galaxy catalogue, generated from the MICE $N$-body simulation \citep{fosalba2015mice2a}, which adopts a flat $\Lambda$CDM cosmology and evolves a large-volume dark-matter distribution with sufficient resolution for weak-lensing applications. Dark-matter haloes are identified using a friends-of-friends algorithm \citep{crocce2015mice2}, and galaxies are populated up to $z=1.4$ through a hybrid scheme combining halo abundance matching with a halo occupation distribution model, calibrated to reproduce the observed luminosity function and galaxy clustering \citep{carretero2015mice2}. To account for empirical luminosity evolution, we apply the redshift-dependent magnitude-evolution correction of \citet{fosalba2015mice2b}.

To perform the same redshift calibration analysis as in the real UNIONS data, we require that the MICE2 galaxies have realistic multi-band photometry. In the observational data, the UNIONS $r$-band sources are matched to CFHTLenS to obtain five-band $ugriz$ magnitudes \citep{hildebrandt2012cfhtlens}, which are used for the SOM-based redshift calibration. For the mock catalogues, we therefore replicate this photometric setup by generating CFHTLenS-like noisy fluxes for every simulated galaxy. We adopt a noise-modelling framework inspired by \citet{vandenbusch2020kidsredshiftcalibration}, adapted to the depth and filter properties of the CFHTLenS imaging \citep{erben2013cfhtlens}. Through this procedure, we obtain noisy fluxes depending on galaxy size, seeing variations and the intrinsic brightness of each source. 

In a further step, we employ a kNN-based matching procedure following the approach presented in \citet{wright2025legacynz} to ensure that the simulated catalogues replicate the photometric and spectroscopic selection functions of the UNIONS analysis. The UNIONS weak-lensing sample is matched to deeper CFHTLenS $ugriz$ photometry in the W3 field. To reproduce this in the mocks, we match the noisy MICE2 galaxies to the real UNIONS--CFHTLenS sources using kNN-matching in colour--magnitude space. For each real galaxy, we select the nearest simulated object and transfer its multi-band properties to construct a photometric sample that follows the joint distributions of colour and magnitude observed in the data. We also transfer the weights $w$ and the average shear response $\langle R_\gamma\rangle$ from the real UNIONS shear catalogue to the simulations, since we require these in the SOM-based redshift calibration (as described in Sect.\,\ref{sec:nz}). 

To replicate the spectroscopic calibration set consisting of DEEP2, VVDS, and VIPERS, we create a second mock sample using kNN-matching between these surveys and the noisy MICE2 galaxies. Matching in slices of redshift in  colour--magnitude space ensures that the mock calibration sample follows the same selection function as the combined real spectroscopic data set. 

The mock photometric and spectroscopic samples are processed with the same SOM-based redshift calibration pipeline that is used for the real UNIONS data (Sect.~\ref{sec:nz}). The SOM is trained on the spectroscopic-like mock sample, and the photometric mocks are projected onto the trained map to yield a recovered $n(z)$, weighted by both the shape and shear response weights. From the bootstrap resampling, we recover a redshift distribution, which we compare to the true redshift distribution of the simulated sources. This provides a direct measure of the redshift-calibration bias, which we find to be $\Delta z=-0.003$.

The width of the prior on $\Delta z$ is derived directly from the data: we use the standard deviation of the bootstrap realisations of the SOM-based $n(z)$ obtained by resampling the spectroscopic calibration sample. This bootstrap-based estimate of the uncertainty is fully data-driven, and we verified with the mock catalogues that it is consistent with the scatter expected from the true redshift-calibration error budget. Moreover, we include a systematic error from using several mock patches and by changing the parameters of the SOM (e.g., hierarchical clustering). 
We therefore estimate the total uncertainty to be $\pm 0.018$. For completeness, we also consider two additional setups where we impose a flat, uninformative prior on $\Delta z\in[-0.1,0.1]$, and assume no uncertainty (fixing $\Delta z =0$).

\subsection{Nonlinear matter power spectrum}\label{sec:nonlinear}

\begin{figure}
    \centering
    \includegraphics[width=\linewidth]{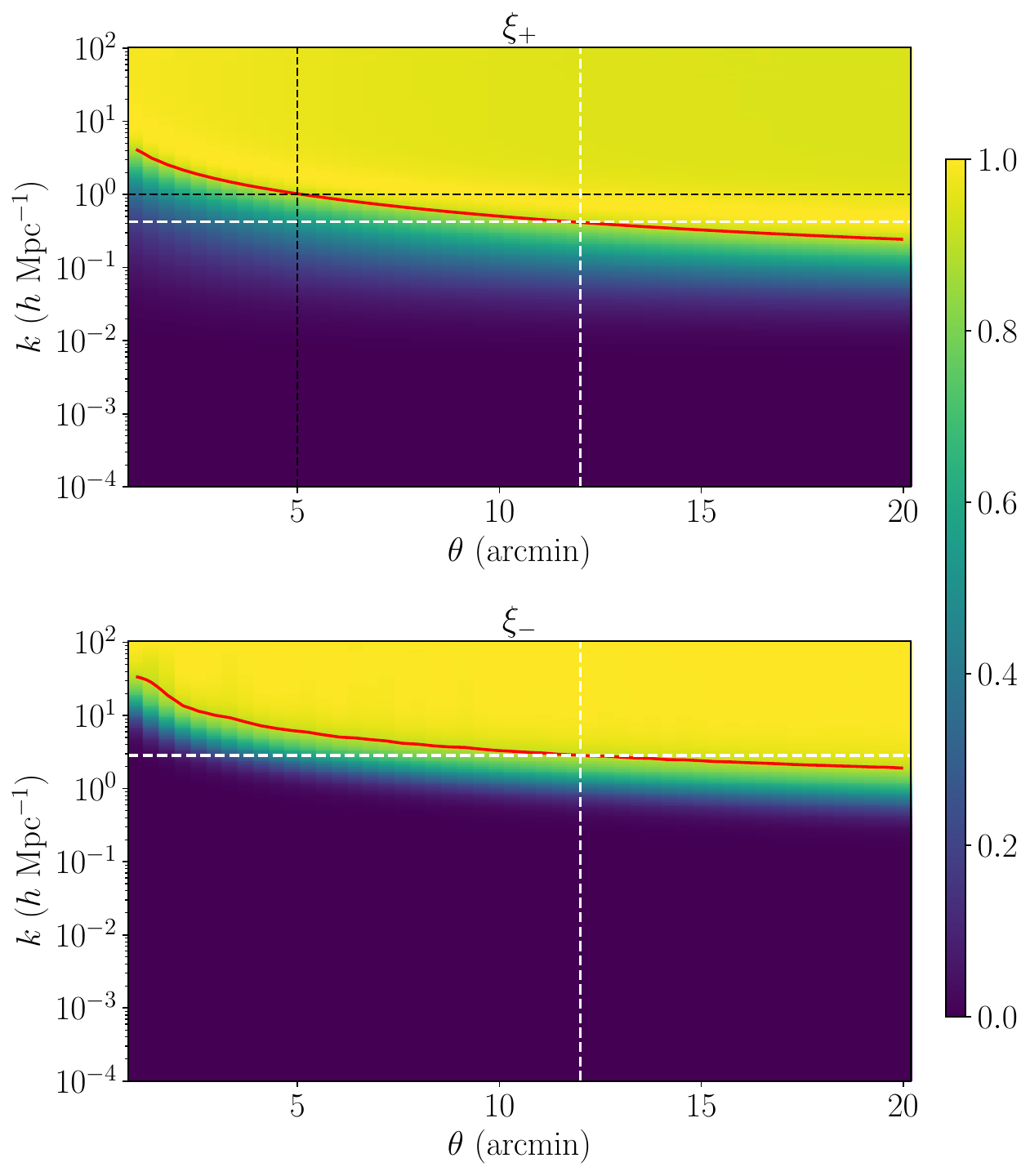}
    \caption{2D heatmap of the ratio of the 2PCF signal as a function of wavenumber $k$ and angular separation $\theta$ (upper: $\xi_+$; lower: $\xi_-$). The red contours mark the boundary beyond which smaller scales (i.e. larger $k$ values) contribute less than 10\% of the total signal. White vertical dashed lines mark the fiducial scale cuts: a 12~arcmin lower bound for $\xi_+$ would signify that wavenumbers no larger than   $k_{\mathrm{max}}=0.43 \,h\,\mathrm{Mpc}^{-1}$ contribute 90\% of the signal, while $k_{\mathrm{max}}=2.85 \,h\,\mathrm{Mpc}^{-1}$ for $\xi_-$. Black lines mark the 5~arcmin lower scale cut for $\xi_+$ explored in one inference run (Sect.~\ref{sec:scale_cuts}), which then corresponds to a $k_{\mathrm{max}}=1 \,h\,\mathrm{Mpc}^{-1}$.}
    \label{fig:k_contributions}
\end{figure}

Cosmic shear probes small scales of the matter power spectrum, requiring accurate modelling of the nonlinear regime. Following the methodology of \cite{DES:2020daw}, we derive the contribution of the nonlinear matter power spectrum at each wavenumber $k$ to the 2PCF signal at different $\theta$ angles, by calculating the integral given in Eq. \eqref{eq:cell_to_xi}. We adopt the \textit{Planck} 2018 best-fit cosmology \citep{planck2018} as our fiducial model, and employ the Limber approximation $k=(\ell+1/2)/\chi(z)$, to compute, for a given $k_{\mathrm{max}}$, the corresponding radial comoving distance up to which we should integrate Eq. \eqref{eq:Cell}. Figure \ref{fig:k_contributions} presents a heatmap of the ratio of the total 2PCF signal as a function of $k$ and $\theta$ scale, for both $\xi_+$ and $\xi_-$. With the conservative criterion that scales beyond $k_{\mathrm{max}}=3\,h\,\mathrm{Mpc}^{-1}$ contribute less than 10\% of the signal, our $n(z)$ implies minimum angular scales of 1.5~arcmin for $\xi_+$ and 11~arcmin for $\xi_-$. However, as described in Sect. \ref{sec:scale_cuts}, by considering the effects of $B$-mode contamination, we instead adopt a fiducial angular lower bound of $\theta=12$~arcmin, which would then imply that scales beyond $k_{\mathrm{max}}=0.43\,h\,\mathrm{Mpc}^{-1}$ do not contribute to more than 10\% of the $\xi_+$ signal, and scales beyond  $k_{\mathrm{max}}=2.85\,h\,\mathrm{Mpc}^{-1}$ for the $\xi_-$ signal. These scale cuts are safely within the range of accuracy of the nonlinear prescriptions adopted below. We mark out these boundaries in Fig. \ref{fig:k_contributions} as well.

For our fiducial analysis, we use the most updated version of \texttt{HMCode2020} \citep{Mead:2020vgs}, which employs the halo model formalism as well as baryonic feedback modelling to calculate the nonlinear matter power spectrum, with up to 2.5\% accuracy at scales $k\leq 10\,h\,\mathrm{Mpc}^{-1}$ when compared to $N$-body simulations. It has also been widely adopted in other Stage III cosmic shear analyses \citep{2019PASJ...71...43H, gatti2021shapecatalogue,Wright_kids_2025}.

When modelling the small-scale baryonic feedback arising from active galactic nuclei (AGN), an additional parameter $\log(T_{\mathrm{AGN}})$ is introduced, which quantifies the `temperature' or strength of this feedback. Calibrations with $N$-body simulations estimate it to be around $7.6$--$8.0$. In our fiducial analysis, we marginalise over this parameter, while also including a test where we do not account for baryonic feedback to quantify its impact on the clustering cosmological parameters $\sigma_8$ and $S_8$. We also run a test without baryonic feedback (disabling the `feedback' option in \texttt{HMCode2020}) and one using the nonlinear prescription of \texttt{Halofit} \citep{Takahashi:2012em} instead, which is based on a halo-fitting model, to quantify the sensitivity of our data to a difference in nonlinear modelling.

\subsection{Scale cuts} \label{sec:scale_cuts}

Our baseline scale cuts were determined by three criteria: PSF systematics, $B$-mode analyses, and potential mismodelling of the nonlinear power spectrum at small scales, as described in Sects.~\ref{sec:psf_leakage}, \ref{sec:eb_modes}, and \ref{sec:nonlinear}, respectively. By setting the requirements that the real space and COSEBIs $B$-mode null tests are passed with a PTE of at least 0.05, and that the PSF systematics signal $\xi_\pm^{\mathrm{sys}}$ does not contribute more than 10\% of the total signal, we obtain $\theta\in[5, 83]$~arcmin for $\xi_+$ and $\theta\in[12,83]$ for $\xi_-$.

To assess the sensitivity of our scale cut choices to cosmology, we conducted blinded inference analyses by successively increasing the values of the lower scale cut of $\xi_+$, with $\theta_{\mathrm{min}}$ varying from $3.98$ to $12.0$~arcmin. The results showed a nontrivial $S_8$ dependence on scale, with a systematic downward drift in $S_8$ and an improved $\chi^2$ value with increasing $\theta_{\mathrm{min}}$. We therefore remove angular bins where the induced shift in the mean $S_8$ exceeds $0.2~\sigma$. This behaviour eventually stabilises, going from the ninth to the tenth angular bin (representing a $\theta_{\mathrm{min}}$ increase from $9.10$~ arcmin to $12.0$~ arcmin), where we also find the largest $\chi^2$ improvement of 0.7. This could be correlated with the exclusion of the localised feature around $10$~arcmin of the $\xi_+$ function (see Fig. \ref{fig:xi_pm}), coincident with the $9.4$~arcmin size of the MegaCam CCD. This feature persisted with similar amplitude across different variations of PSF size cuts and masking schemes, appearing in both the cosmological and systematic $\xi_+^B$ signals. Additionally, we found that scale cut combinations of $\theta\in[5,83]$~arcmin and $\theta\in[12,83]$~arcmin pass the COSEBIs $B$-mode null test, but not for the particular case of $\theta\in[9,83]$~arcmin (see \hyperlink{cite.daley.etal25}{Paper II} for details). The fiducial scale cut we therefore adopt is $\theta\in[12,83]$~arcmin for both the $\xi_{\pm}$ correlation functions. On the other hand, we utilise the full angular range to fit the $\tau(\theta)$ statistics. This amounts to a total of 54 data points that we fit to: 7 each from $\xi_{\pm}$, and 20 each from $\tau_{0,2}$.

For completeness, we also present inference results where we allow for small-scale contributions in $\xi_+$, i.e. when $\theta\in[5,83]$~arcmin. 

\subsection{Pipeline validation with mock catalogues}\label{sec:mocks}

\begin{figure}
    \centering
    \includegraphics[width=\linewidth]{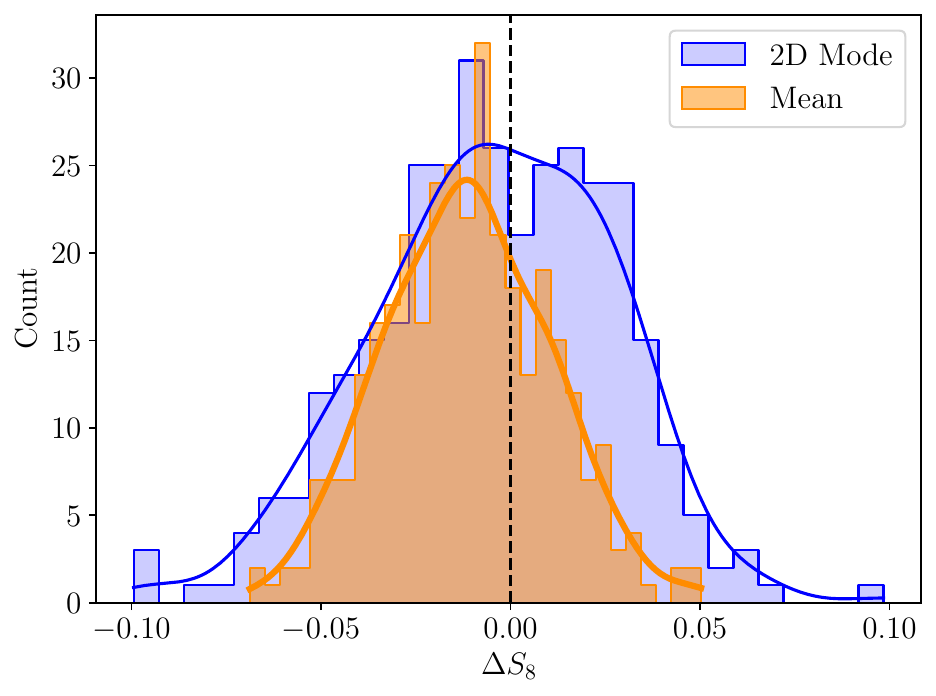}
    \caption{Histogram of the difference in best-fit $S_8$ values obtained from the 341 mocks, when run with the configuration space versus harmonic space pipelines (harmonic $-$ configuration). The orange histogram and corresponding fitted KDE distribution are that of the weighted means, while the blue histogram represents the 2D marginalised mode values. We mark out the null value with the black dashed vertical line.}
    \label{fig:s8_mocks}
\end{figure}

With our inference pipeline set up, we validate it by running it on 341 random galaxy mock catalogues generated with \texttt{GLASS} \citep{glass}. The catalogues were created using the same mask, effective number density and $\sigma_{\bs{e}}$ as the data, with a \textit{Planck} 2018 best-fit fiducial cosmology. In addition, one of the blinded $n(z)$ distributions was chosen at random. We refer the reader to Appendix \ref{sec:glass_mocks} for a validation of the mock galaxy survey properties. The mock catalogues are not contaminated by PSF systematics, therefore the $\tau$ statistics are consistent with zero. We nonetheless account for PSF systematics uncertainty by sampling $\alpha$ and $\beta$ with priors obtained from the $\rho$ and $\tau$ statistics of the mock data vectors. Both parameters recover a zero mean to within $1\,\sigma$.

We measure the cosmic shear and PSF correlation functions, compute their respective covariance matrices, and run both the harmonic and configuration space inference pipelines on each mock catalogue, employing the parameters and priors as in Table \ref{tab:inference_priors}, and the same fiducial scale cuts as detailed in Sect. \ref{sec:scale_cuts}. This test acts as a criterion for unblinding (see Sect. \ref{sec:unblinding}). First, we assess whether we recover the fiducial cosmology within a reasonable level of uncertainty. Additionally, we run both our pipelines on the same set of mocks to ascertain that the average $S_8$ mean obtained from both analyses differs by less than 0.5~$\sigma$. 

We plot in Fig. \ref{fig:s8_mocks} the histogram of the difference in the 341 best-fit values of $S_8$ obtained by both pipelines. We see that when the weighted mean is used to determine the best-fit, we obtain an offset from 0 (albeit still within $1\sigma$), while this discrepancy vanishes when using the 2D marginalised mode to estimate the best-fit instead (albeit with slightly more spread). This demonstrates the robustness of both our pipelines which, despite having been developed independently, are able to deliver internally consistent results. We refer the reader to Appendix \ref{sec:glass_mocks} for further details.

\subsection{Blinding Strategy}\label{sec:unblinding}

We conduct a blinded analysis to remain agnostic to the results and to mitigate potential confirmation biases in our analysis choices. The blinding procedure was implemented by an external collaborator, who generated three versions of the $n(z)$ distributions by applying random shifts to two of them. These shifts induce corresponding offsets in the marginalised $S_8$ posterior distributions. We then computed covariance matrices and performed cosmological inference for all three $n(z)$ variants.

Because the data vector itself was not blinded, we were able to directly apply the tests described in Sect.~\ref{sec:sys_tests} as diagnostics of potential systematic contamination. While the covariance matrix used to compute the $B$-mode PTEs varies with each $n(z)$ shift, we verified that the relative differences between them remain below $10\%$. The resulting variations in PTE do not affect our conclusions regarding unblinding criteria or scale cuts. We therefore computed PTEs using all three covariance matrices and adopted the most conservative values when defining our scale cuts.

The majority of this manuscript was prepared prior to unblinding and underwent internal review within the broader UNIONS collaboration, including the external blinding coordinator. Cosmological inference was carried out for all three blinded data vectors, alongside variations in modelling (including PSF systematics, nuisance parameter priors, and nonlinear prescriptions) and analysis setup, as described in the preceding sections. We verified that differences in the inferred cosmological posteriors arising from these choices were consistent and exhibited similar trends across all three blinded analyses.

In addition, as described in the previous section, we imposed a further validation criterion requiring consistency between the Fourier- and configuration-space pipelines when applied independently to mock catalogues (Sect.~\ref{sec:mocks}), specifically in terms of the recovered $S_8$ constraints. Three preliminary versions of the results section and figures were sketched out in advance, such that unblinding required only selecting the appropriate version and finalising the manuscript.

For transparency, we note that the primary authors of the cosmology analysis papers were inadvertently unblinded during the final stages of the project. However, this knowledge did not influence any analysis choices or changes in the inference pipeline. Furthermore, a bug in the code to estimate the mean $\Delta z$ was discovered only after unblinding, which necessitated a rerun of the inference with the correct $\Delta z$ prior. That being said, the post-unblinding rerun did not affect the pipeline validation, since the mock catalogues were run with $\Delta z=0$. Furthermore, since no analysis-setup changes were made after unblinding, we ensured that the rerun did not introduce any confirmation bias.
%--------------------------------------------------------------------
\section{Results}\label{sec:results}

\subsection{Fiducial analysis}

\begin{figure*}
    \centering
    \includegraphics[width=\textwidth]{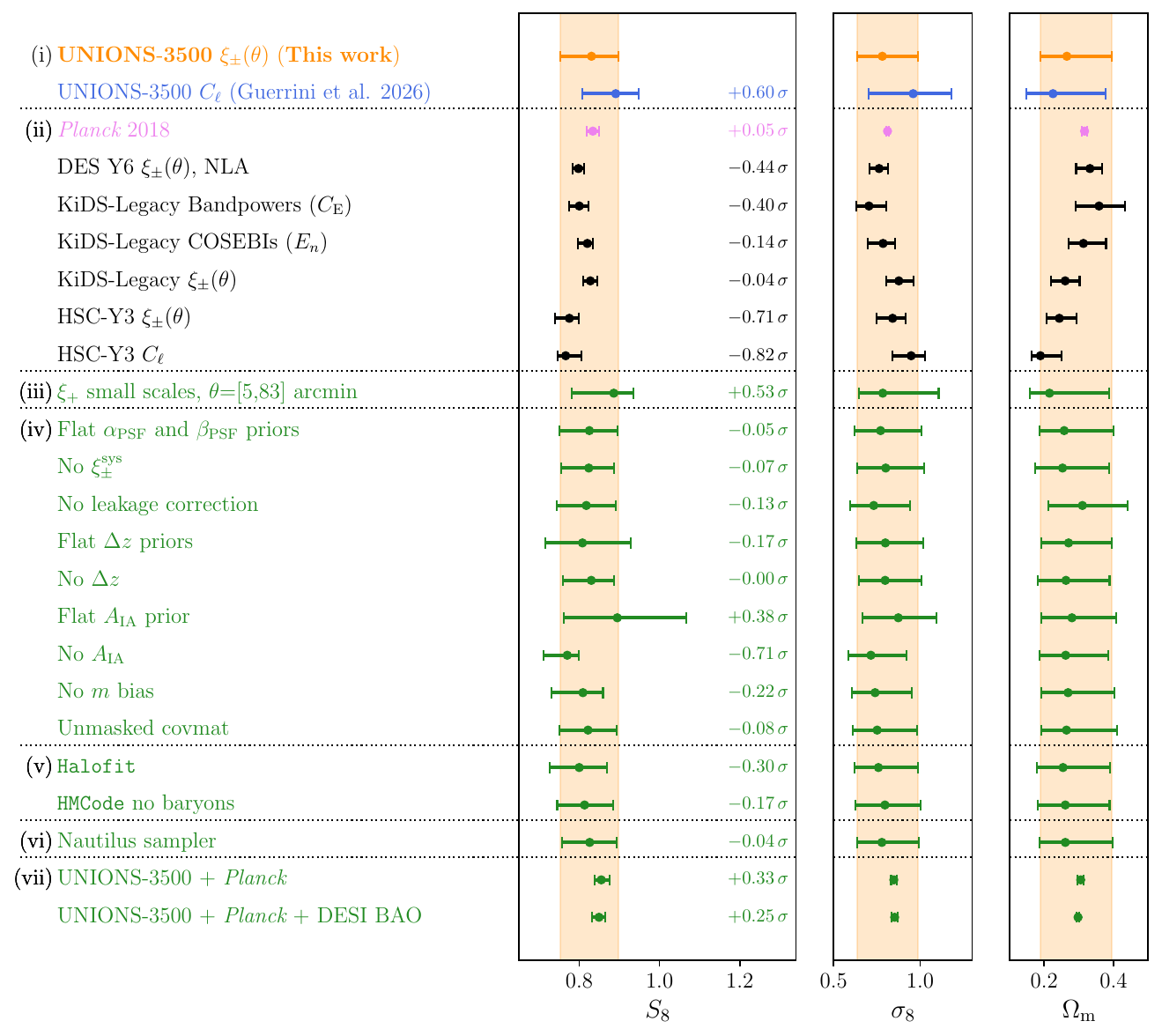}
    \caption{Whisker plot of the best-fit $S_8$, $\sigma_8$, and $\Omega_{\rm m}$ parameters, along with their 68\% confidence interval (CI), for the various observables (orange for configuration space and blue for harmonic space as presented in \protect\hyperlink{cite.guerrini.etal25b}{Paper IV}), comparison with external surveys (written in pink for \protect\textit{Planck} and black for the rest of the weak lensing surveys), and changes in analysis setup (in green). For $S_8$ and $\Omega_{\rm m}$, we report the 2D marginalised mode as the best-fit value, and the 1D marginalised mode for $\sigma_8$, except in the case of the DES Y6 results, where we directly quote the marginalised mean as reported in \protect\cite{des_y6_cs}. The vertical orange bands denote the fiducial 68\% CI of the fiducial result. We also quote the tension in $S_8$ between the fiducial and the various analysis setups, in units of $\sigma$. }
    \label{fig:fid_whisker}
\end{figure*}
\begin{figure*}
    \centering
    \includegraphics[width=0.85\linewidth]{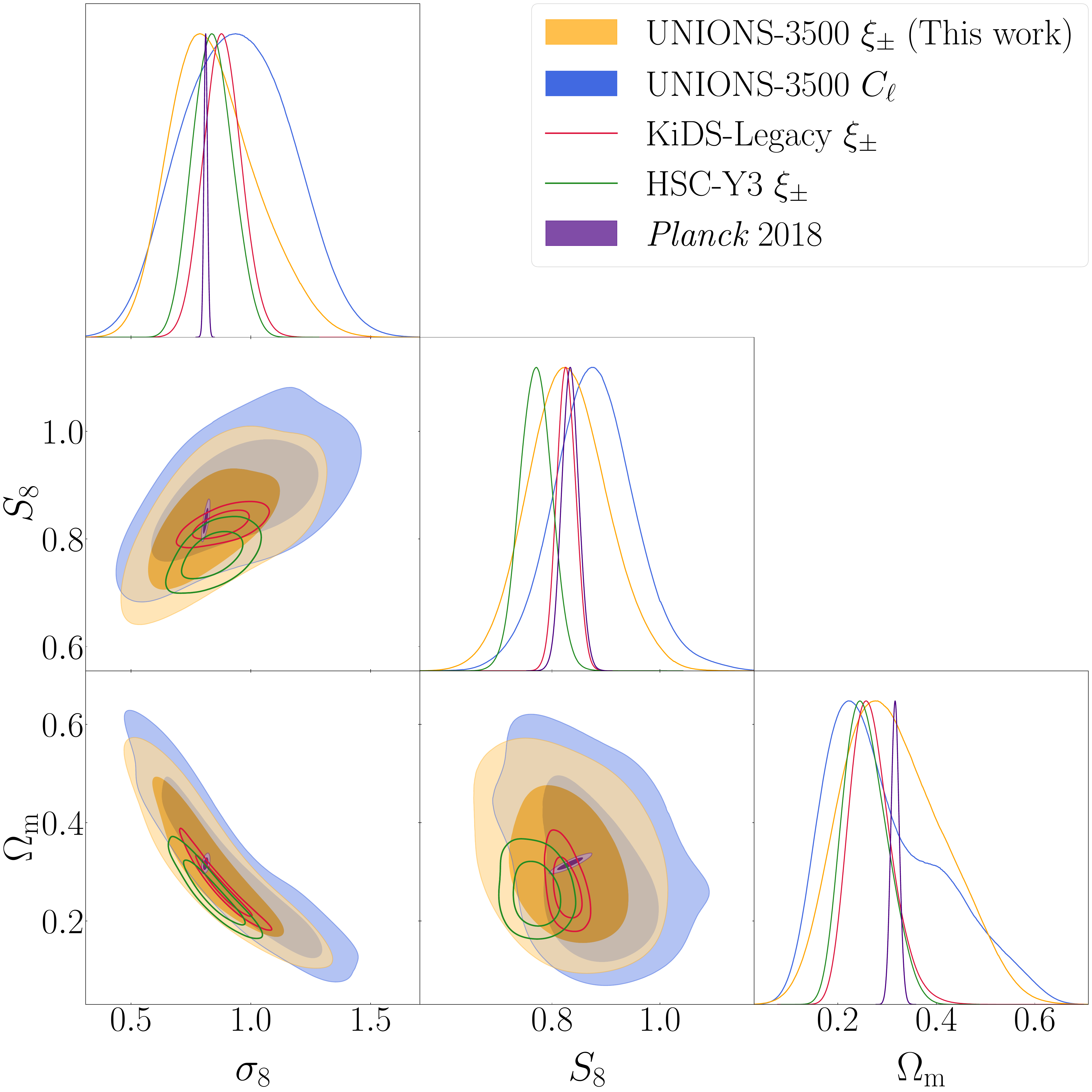}
    \caption{2D marginalised contour plot of $S_8$, $\sigma_8$ and $\Omega_{\rm m}$ of our fiducial results (orange), including those obtained in harmonic space (blue, see \protect\hyperlink{cite.guerrini.etal25b}{Paper IV}), as well as results from external surveys (red unfilled: KiDS-Legacy $\xi_{\pm}$; green unfilled: HSC-Y3 $\xi_{\pm}$; purple filled: \textit{Planck} CMB).}
    \label{fig:fid_contour}
\end{figure*}

We report our best-fit 2D marginalised mode for the $S_8$ parameter, and the matter density $\Omega_{\rm m}$, as well as the 1D marginalised mode of the amplitude of clustering $\sigma_8$, derived from the UNIONS-3500 catalogue in configuration space, which are given by
\begin{gather*}
    S_8=0.831^{+0.067}_{-0.078},\\
    \Omega_{\rm m} = 0.265^{+0.130}_{-0.075},  \\ 
    \sigma_8= 0.782^{+0.206}_{-0.144}.\\
\end{gather*}
Fig. \ref{fig:fid_whisker} presents the whisker plot of the three parameters for all of our analysis runs, as well as the fiducial results obtained in \hyperlink{cite.guerrini.etal25b}{Paper IV}, and comparison to external survey results. In Fig. \ref{fig:fid_contour} we plot the 2D marginalised contours of our fiducial results in configuration and harmonic space, and compare it with those reported in external surveys. 

Additionally, our results are consistent at the $1\sigma$ level with cosmic shear analyses from other Stage III galaxy surveys, such as DES Y6 \citep{des_y6_cosmo}, KiDS-Legacy \citep{Wright_kids_2025}, and HSC-Y3 \citep{hsc-y3,hsc-y3-2}, albeit with a tendency toward systematically higher values. Our measurement provides an independent constraint from the northern hemisphere, among the first of its kind, thus offering a valuable cross-check of other results derived primarily from the southern hemisphere. Our $S_8$ constraint is also fully consistent with that reported by \cite{planck2018}, and therefore does not indicate any significant tension between early- and late-time measurements of the clustering amplitude. We anticipate that future tomographic analyses will substantially improve the constraining power of our catalogue, enabling us to become competitive with other Stage III surveys.

\begin{figure}
    \centering
    \includegraphics[width=\linewidth]{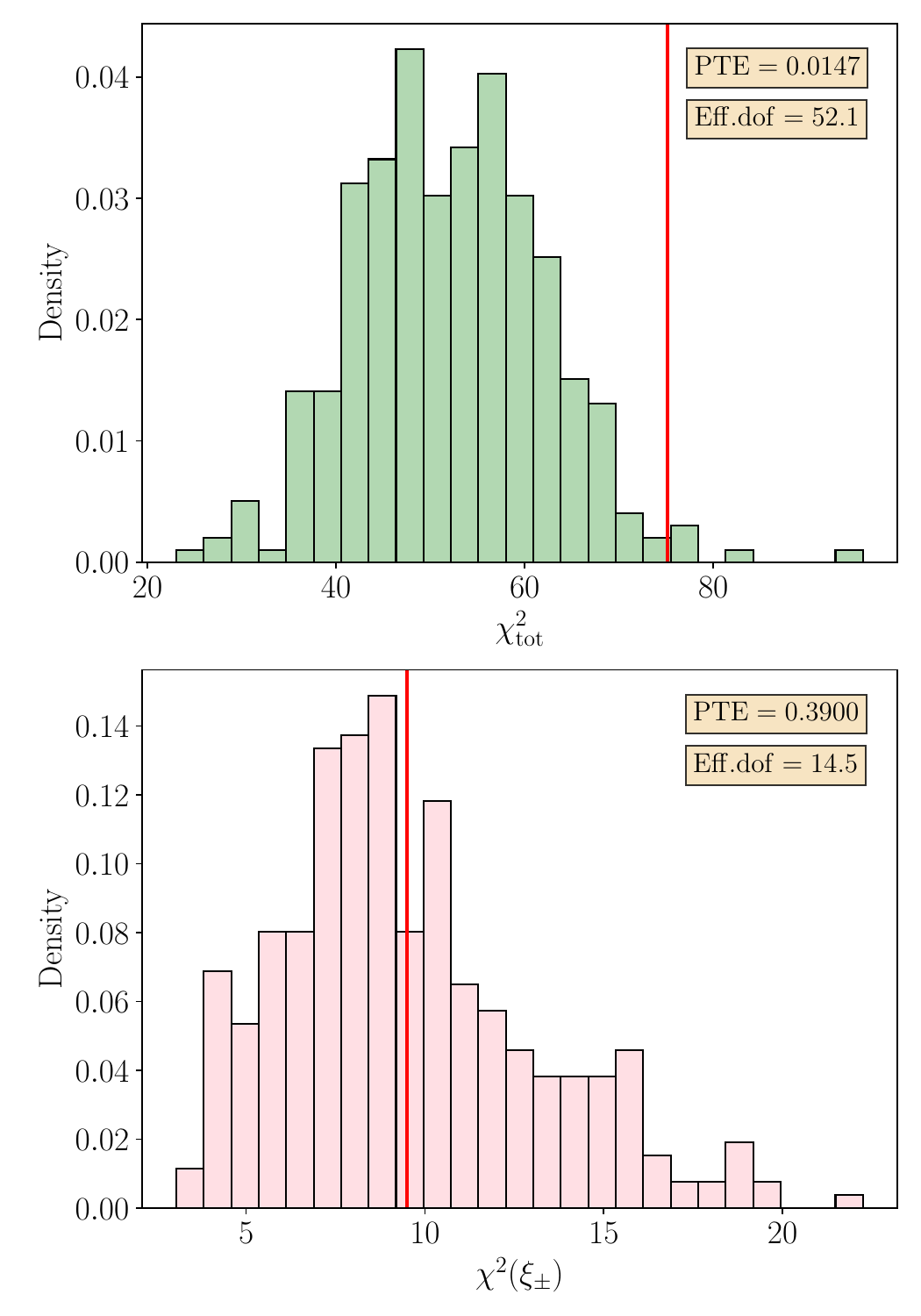}
    \caption{Histogram of $\chi^2$ values when considering the combined cosmological and systematic signals (top, in green), and when only considering the cosmological signal (bottom, pink). The red vertical line denotes the fiducial $\chi^2$. We find an effective degree of freedom of 14.4, and PTE of 0.39 for the $\xi_\pm$ signal, and an effective degree of freedom of 40.4 with a PTE of $5.9\times 10^{-3}$ for the $\tau_{0,2}$ statistics.}
    \label{fig:chi2_tau}
\end{figure}

We estimate the goodness-of-fit of our result by comparing it to those previously derived from our mocks. In Fig. \ref{fig:chi2_tau} we plot the histogram of mock $\chi^2$ values and fit a $\chi^2$ distribution, where we find that the effective degree of freedom is 51.9, while the $\chi^2$ value of our fiducial analysis is 75.33. This is a more robust method of obtaining the degrees of freedom than simply taking the difference between the number of data points and the number of sampled parameters, since it is often the case in cosmic shear experiments that parameters are prior-dominated and highly degenerate. 

We obtain a PTE of $0.0147$, which is mainly driven by the fit to the $\tau_{0,2}$ statistics. When focusing on the pure $\xi_\pm$ data vector, we find a good fit of the $\xi_\pm$ signal with a PTE of $0.39$. In Table \ref{tab:best_fit_chi}, we present the best-fit values of $S_8$ and their 68\% CI, quoting both the 2D marginalised mode and the weighted mean, together with the $\chi^2$ and PTE for each inference setup considered.

\subsubsection{Consistency with harmonic space analysis}\label{sec:consistency_probes}

We find that our results are internally consistent between both summary statistics, where we obtain an $S_8$ value that is $\sim0.6\sigma$ higher in harmonic space. This difference arises from the distinct range of scales probed by the two approaches.

When expressed in $k$-space, we see that a substantial range of the $\ell$ scales employed in the harmonic space analysis correspond to angular scales of $\lesssim 5$ arcmin, which were excluded from the configuration space analysis based on systematics-motivated scale cuts (see Sect.~\ref{sec:scale_cuts}). Notably, our data exhibit a trend of stronger clustering on these small scales, evident from Fig. \ref{fig:xi_pm}, which then leads to a higher $S_8$ best-fit found in the pseudo-$C_\ell$ analysis. We investigate this in more detail in Sect. \ref{sec:scale_cut_mcmc}, where we include more data points at small scales by varying scale cuts.

\subsection{Robustness to modelling choices}
We now consider the impact of various systematic effects and analysis choices, as described in Sect.~\ref{sec:inference_pipeline}. In particular, we highlight the most important ones, namely the impact of prior choices on nuisance parameters describing intrinsic alignments, PSF systematics, and redshift calibration uncertainties, as well as choices related to scale cuts and nonlinear modelling.

\subsubsection{Intrinsic alignment modelling}
\begin{figure}
    \centering
    \includegraphics[width=\linewidth]{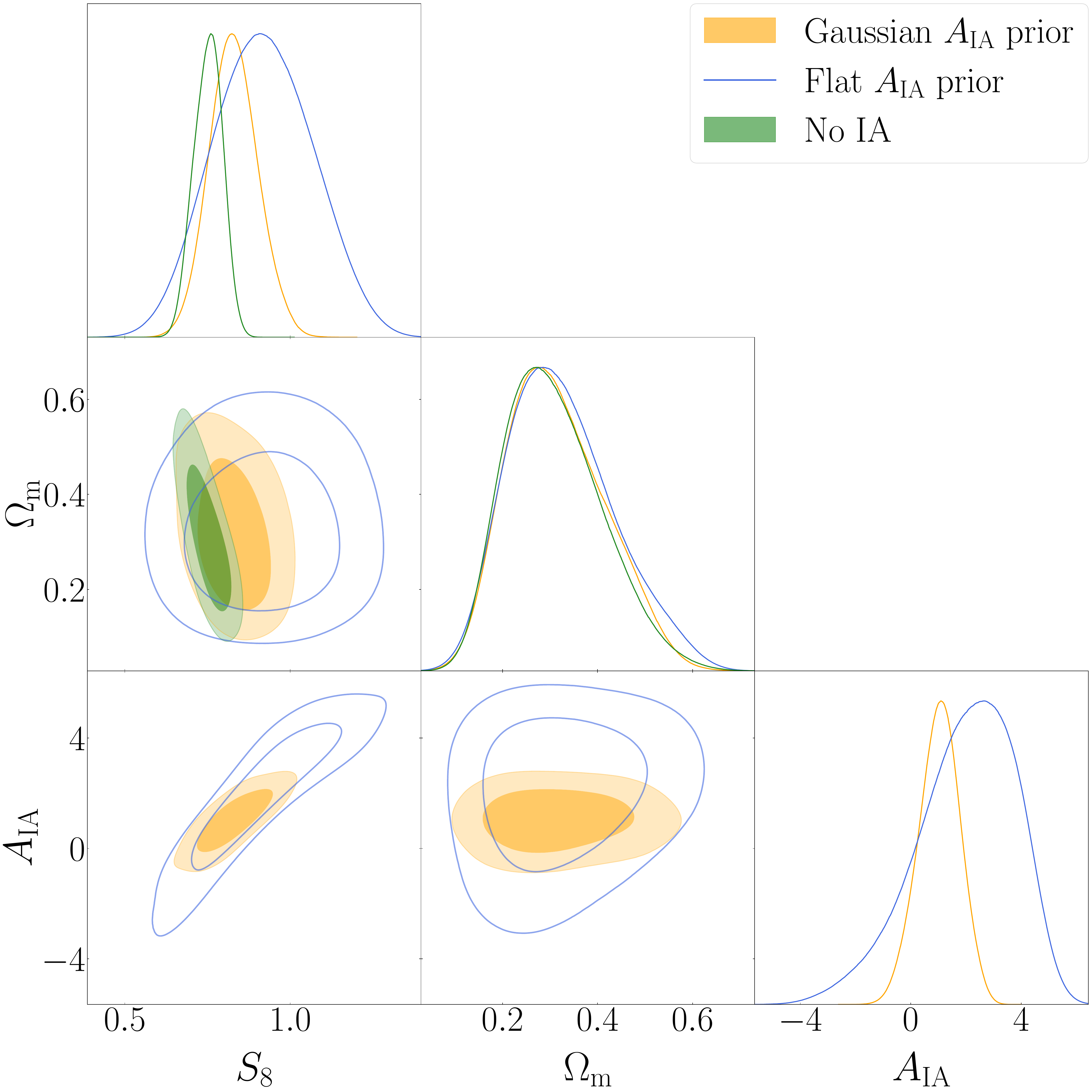}
    \caption{2D marginalised contour plot of $S_8$, $\Omega_{\rm m}$ and $A_{\rm IA}$ comparing the effects of modelling intrinsic alignment: using a Gaussian prior on the amplitude parameter $A_{\rm IA}$ (orange), using a flat prior (blue unfilled) and not including intrinsic alignment (green).}
    \label{fig:ia_contour}
\end{figure}

In Fig. \ref{fig:ia_contour} we present the 2D marginalised contours of $S_8$, $\Omega_{\rm m}$ and $A_{\rm IA}$ for different IA setups: our fiducial choice of imposing a Gaussian prior on its amplitude $A_{\rm IA}$, using a flat prior, and not including IA effects entirely (i.e. setting $A_{\rm IA}=0$). We see that employing an uninformative flat prior has the largest impact on $S_8$ and $A_{\rm IA}$ whereby their constraints are greatly broadened, although their best-fit values remain consistent: we obtain $S_8=0.895^{+0.172}_{-0.133}$ and $A_{\rm IA}=2.61^{+1.55}_{-1.96}$, in the case of a flat $A_{\rm IA}$ prior, compared to $S_8=0.831^{+0.067}_{-0.078}$ and $A_{\rm IA}=1.13^{+0.688}_{-0.798}$ in the fiducial setup with a Gaussian prior. 

This is due to the strong degeneracy between $A_{\rm IA}$ and $S_8$, a well-known phenomenon in cosmic shear analyses whereby both parameters work in opposing manners to impact the shear signal. Since an increase in $A_{\rm IA}$ and thus the IA power spectra $C_\ell^{\gamma I}$ and $C_\ell^{II}$ contribute negatively to the overall shear signal, an enhanced amplitude of clustering would then be required to compensate for it, leading to a strong linear relationship between these two parameters. This is evident in the contour plot of Fig. \ref{fig:ia_contour}. Nonetheless, we are able to recover consistent results from both flat and Gaussian priors, demonstrating the robustness of our technique in estimating the uncertainty of $A_{\rm IA}$ based on galaxy populations (Sect. \ref{sec:ia}).

On the other hand, not accounting for IA naturally gives the tightest constraints, at the expense of a slight bias: we obtain a lower value of $S_8=0.770^{+0.029}_{-0.059}$, which is the largest shift in $S_8$ from the fiducial amongst all the inference tests conducted. Moreover, from Table \ref{tab:best_fit_chi} we see that this modelling choice returns the largest $\chi^2$ and a very low PTE value, thus signifying that the model is a sub-optimal fit to the data. We anticipate that with future tomographic analyses, additional redshift information can break the degeneracy between $S_8$ and $A_{\rm IA}$, which would also imply larger biases should the effects of IA not be modelled accurately.

\subsubsection{Redshift calibration}

\begin{figure}
    \centering
    \includegraphics[width=\linewidth]{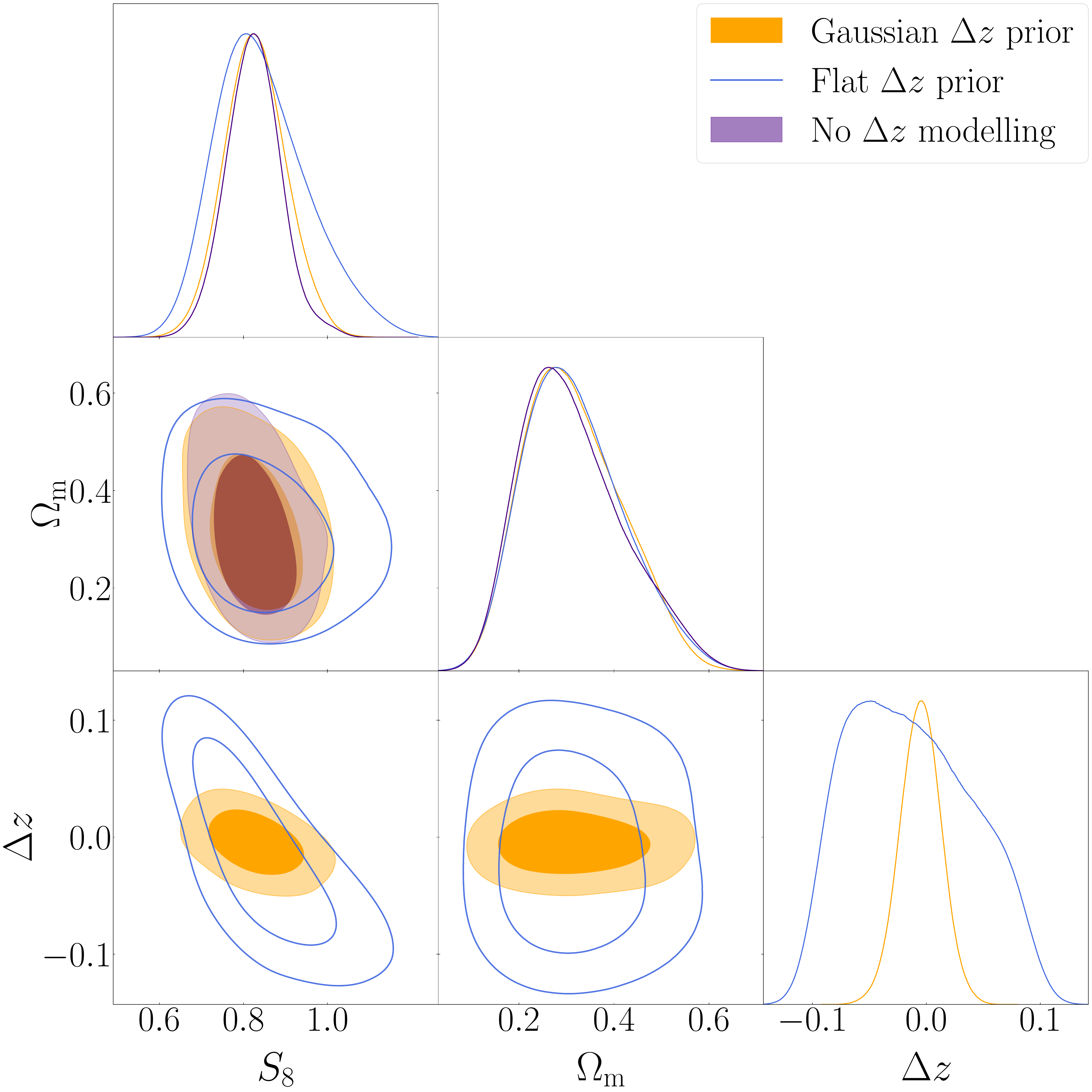}
    \caption{2D marginalised contour plot of $S_8$, $\Omega_{\rm m}$ and $\Delta z$ comparing the effects of modelling redshift uncertainty parameter $\Delta z$: using a Gaussian prior (orange), using a flat prior (blue unfilled), and not accounting for redshift uncertainties (pink).}
    \label{fig:dz_contour}
\end{figure}

We subsequently consider the impact of our choice of priors for $\Delta z$, the uncertainty of the shift in $n(z)$ obtained during redshift estimation and calibration. We present in Fig. \ref{fig:dz_contour} the 2D marginalised contours of $S_8$, $\Omega_{\rm m}$ and $\Delta z$ where, analogous to the previous section, we compare the results from three different analysis choices: the fiducial case of imposing a Gaussian prior, using a flat prior, and fixing $\Delta z = 0$. We see that our results on $S_8$ remain largely consistent in all cases, save for a significant widening of the $\Delta z$ posterior when an uninformative prior is employed: $\Delta z=-0.0450^{+0.077}_{-0.036}$ compared to the fiducial result of $\Delta z=-0.0043^{+0.016}_{-0.018}$, which equates to a three times increase in the size of error bars. Since the estimated mean value of $z$ is already very small (on the order of $10^{-3}$), fixing it at zero does not produce an obvious shift in $S_8$ results. However, this would not be the case should the magnitude of $\Delta z$ be larger.

A degeneracy between $S_8$ and $\Delta z$ is also evident from the contour plot, although in a different direction to that of $A_{\rm IA}$; once again, this has to do with how they impact the amplitude of the $\xi_\pm$ signal. A larger $\Delta z$ would imply a galaxy source distribution with higher $n(z)$ and thus stronger lensing efficiency, as evident from Eq. \ref{eq:lens_eff_cs}, which would then require a lower $S_8$ to counteract. Inclusion of tomographic information would then be able to tighten constraints on these parameters. Furthermore, we see that even when an uncertainty in the mean redshift shift is not taken into account, we recover consistent results, reaffirming the accuracy and robustness of our SOM calibration methodology. 

\subsubsection{PSF systematics}

\begin{figure}
    \centering
    \includegraphics[width=\linewidth]{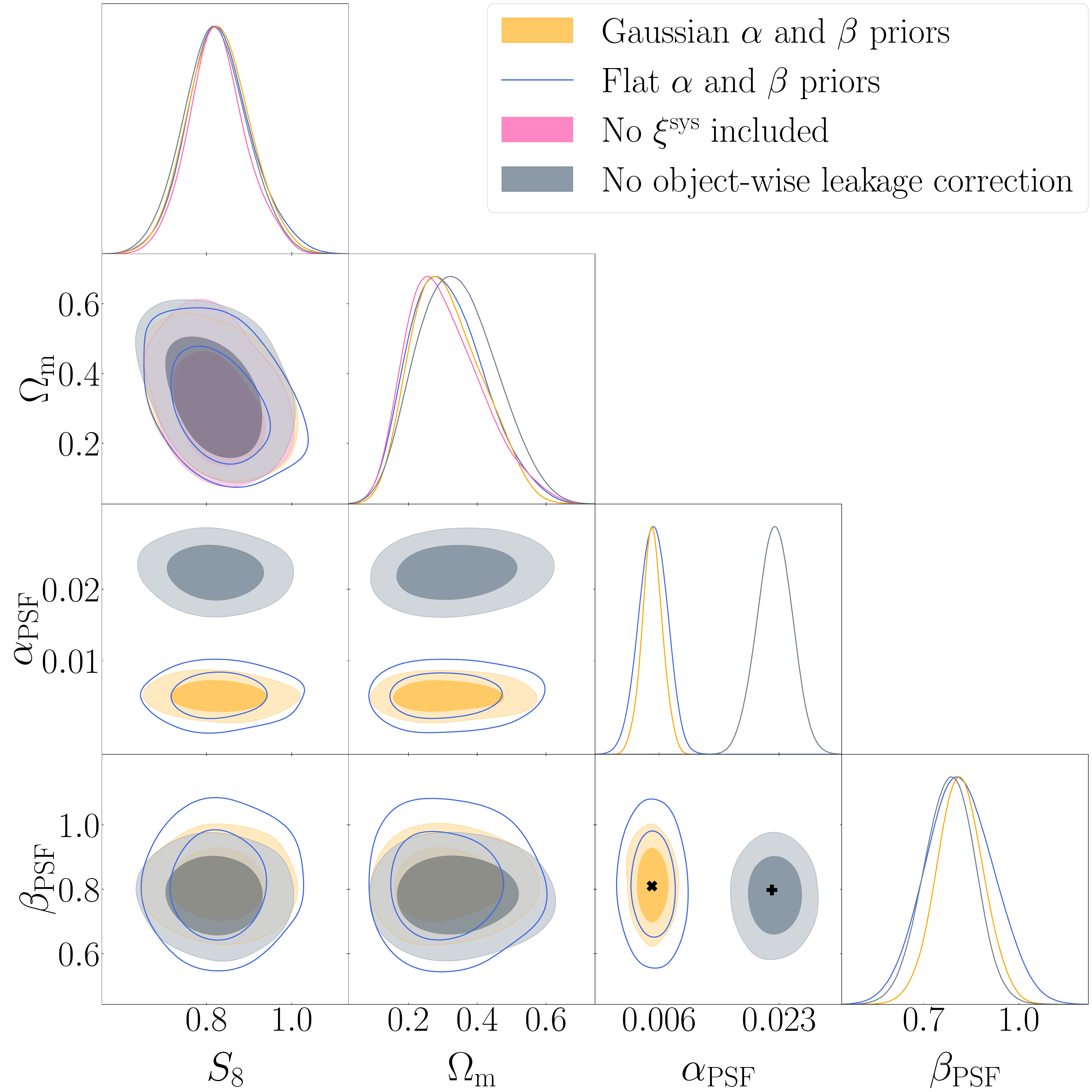}
    \caption{2D marginalised contour plot of $S_8$, $\Omega_{\rm m}$, $\alpha_{\rm PSF}$ and $\beta_{\rm PSF}$ comparing the effects of modelling PSF: using Gaussian priors on the leakage parameters $\alpha_{\rm PSF}$ and $\beta_{\rm PSF}$ (orange), using flat priors (blue unfilled), not including $\xi_{\rm sys}$ in the data vector during inference (pink), and using the set of objects in the catalogue whose ellipticities were not leakage-corrected (grey). The black `x' marks the best-fit values of $\alpha$ and $\beta$ obtained during the PSF inference step (Sect. \ref{sec:psf_inference}) for the object-wise leakage corrected ellipticities, while the black `+' marks those in the non-leakage corrected case.}
    \label{fig:psf_contour}
\end{figure}

We quantify the impact of PSF systematics within our data set, and assess our ability to mitigate it. In Fig. \ref{fig:psf_contour} we plot the 2D marginalised distributions of $S_8$, $\Omega_{\rm m}$, and the leakage parameters $\alpha_{\rm PSF}$ and $\beta_{\rm PSF}$ for four different cases: when imposing Gaussian priors on $\alpha_{\rm PSF}$ and $\beta_{\rm PSF}$, when using flat priors instead, when not accounting for $\xi_\pm^{\rm sys}$ (i.e. not doing a joint fit with the $\tau_{0,2}$ statistics), and when using instead the catalogue derived from the set of ellipticities that were not corrected for per-object PSF leakage. We see that our results for the cosmological parameters are almost perfectly consistent. Employing flat priors on $\alpha$ and $\beta$ works only to slightly enlarge their resultant posterior distributions, demonstrating that even without the additional step of first fitting for $\alpha$ and $\beta$ (as outlined in Sect. \ref{sec:psf_inference}), we are still able to accurately recover their posteriors.

Moreover, comparing the pink and orange contours in the $\Omega_{\rm m}-S_8$ plane of Fig.~\ref{fig:psf_contour}, we find that the impact of $\xi^{\rm sys}$ is negligible, as evidenced by the recovery of consistent $S_8$ posteriors even when it is not jointly modelled in the inference step. This indicates that, at the current level of data precision, adopting a threshold of 10\% for the maximum allowed contribution of $\xi^{\rm sys}$ to the total signal, as was used to define our scale cuts, is sufficient to mitigate any bias arising from PSF contamination. Furthermore, even when an object-wise leakage correction is not applied (grey contours), it can still be accounted for by using the corresponding set of priors for $\alpha$ and $\beta$ instead. This exercise highlights how PSF leakage correction is not just assumed to work at catalogue level, but is also propagated into the cosmological inference with a non-negligible impact.

\subsubsection{Scale cuts}\label{sec:scale_cut_mcmc}

\begin{figure}
    \centering
    \includegraphics[width=\linewidth]{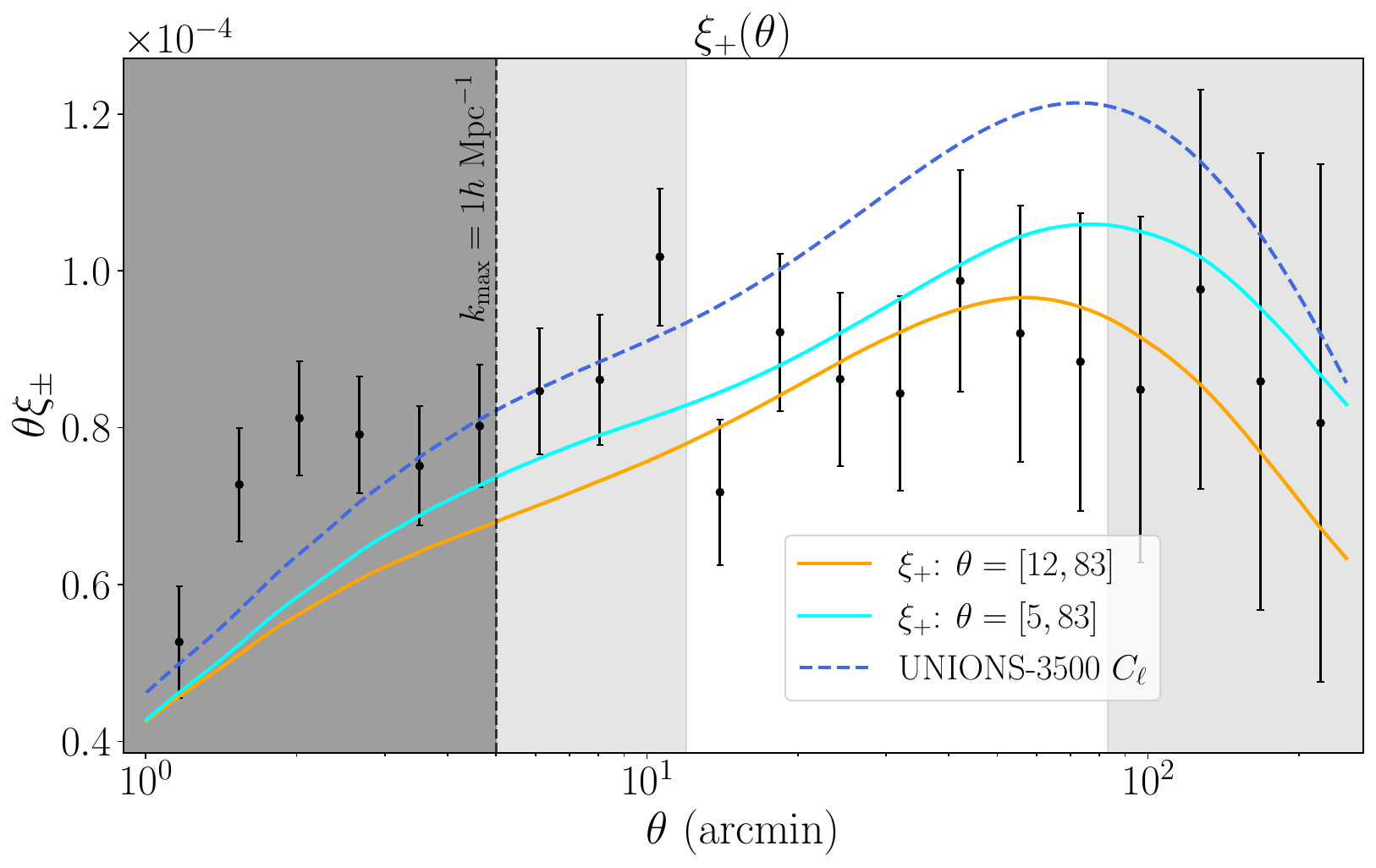}
    \caption{Plot of the best-fit $\xi_+$ data vector for the two scale cuts considered: in the fiducial analysis when $\theta\in[12,83]$ arcmin (orange), and when $\theta\in[5,83]$ arcmin (cyan). We also include the $\xi_+$ vector computed from the best-fit values of the harmonic space analysis in dotted blue. The light grey vertical bands denote the fiducial scale cuts, while the darker grey band on the left demarcates the 5 arcmin cut, which coincides with the boundary where 90\% of the signal receives contributions from $k$ scales up to  $k_{\mathrm{max}}=1\,h\,$Mpc$^{-1}$ (see Fig. \ref{fig:k_contributions}).}
    \label{fig:best-fit-scale-cut-xipm}
\end{figure}

We consider a setup where we include more contribution from small scales, imposing a scale cut of $\theta\in[5,83]$ arcmin for the $\xi_+$ data vector. This gives an increased $S_8=0.886^{+0.049}_{-0.104}$, which nonetheless remains consistent with the fiducial result, and does not seem to significantly tighten the contours in the $\Omega_{\rm m}-S_8$ plane (see Fig. \ref{fig:scale_contour}). 

Not surprisingly, we also find that the $S_8$ results obtained when including smaller angular scales become more consistent with the fiducial harmonic space results (where $S_8=0.891^{+0.057}_{-0.084}$), and the small-scale best-fit $\xi_+$ vector in Fig. \ref{fig:best-fit-scale-cut-xipm} moves towards that derived from the harmonic space analysis. This further reinforces the observation made in Sect. \ref{sec:consistency_probes}: our data exhibit enhanced clustering on small scales, which drives the higher best-fit value of $S_8$ as the lower scale cut is reduced from 12 arcmin to 5 arcmin. When expressed in terms of $k$-space, this smaller scale cut combination has a larger overlap with the $\ell$ range probed by the pseudo-$C_\ell$ estimator, leading to more consistent $S_8$ values between the two approaches. 

However, when naively computing the reduced $\chi^2$ of the pure $\xi_\pm$ data vector, where $\chi^2_{\rm red}=\chi^2/n$ and $n$ is the number of data points, we obtain $\chi^2_{\rm red}=1.34$, which gives a worse fit than in the fiducial case (where $\chi^2_{\rm red}=0.68$), most likely owing to the anomalously high data point at 10 arcmin.

\subsubsection{Nonlinear scales} 

Finally, we consider the impact of modelling choice for the nonlinear power spectrum. We present the best-fit data vectors for all three cases considered in Fig. \ref{fig:nonlin_xipm}: the fiducial case of employing \texttt{HMCode2020} and including the effects of baryonic feedback, \texttt{HMCode2020} without baryonic feedback, and using $\texttt{Halofit}$ instead. We see that not including the effects of baryons leads to the lowest amplitude of clustering, while \texttt{Halofit} gives largely consistent results with our fiducial analysis. 

From Fig. \ref{fig:fid_whisker}, we do not see appreciable changes in the best-fit values of $S_8$, $\sigma_8$ or $\Omega_{\rm m}$. This is generally to be expected, since our adopted scale cuts exclude most of the small scales where nonlinear modelling makes a significant impact (in our case, 90\% of the signal receives contributions from $k$-scales larger than $0.43\,h\,\mathrm{Mpc}^{-1}$). Additionally, using a single tomographic bin results in a broad projection along the line of sight, mixing contributions from a wide range of redshifts and effectively diluting sensitivity to small-scale information. We thus expect to unlock more constraining power from the small scales, and better constrain the amplitude of baryonic feedback $\log(T_{\rm AGN})$ with future tomographic analyses and more precise data in future releases.

\subsection{Combination with external data sets}

\begin{figure}
    \centering
    \includegraphics[width=\linewidth]{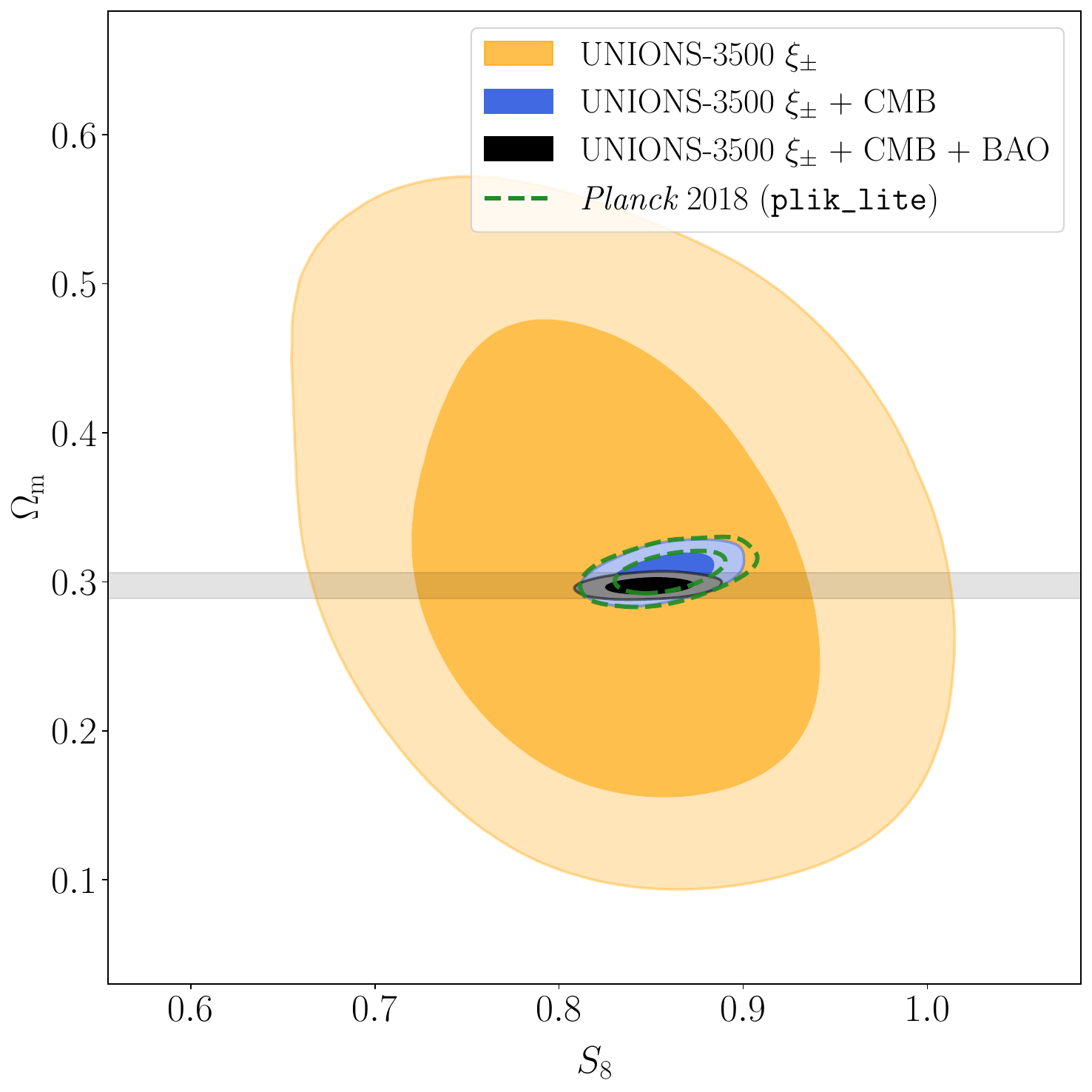}
    \caption{2D marginalised contour of $S_8$ and $\Omega_{\rm m}$ derived from UNIONS-3500 $\xi_\pm$ (orange), UNIONS-3500~+~\protect\textit{Planck} (blue), and UNIONS-3500~+~\protect\textit{Planck}~+~DESI DR2 BAO (black) data sets, comparing them to the \protect\textit{Planck} 2018 results that we have run using the $\texttt{plik\_lite}$ likelihood (green dashed unfilled). The grey band marks the fiducial $1\sigma$ CI of $\Omega_{\rm m}$ from DESI BAO, as reported in \protect\cite{desi_dr2_2}.}
    \label{fig:ext_contour}
\end{figure}

Finally, we consider combined constraints by incorporating external data, including CMB from \textit{Planck} 2018 and BAO from DESI DR2, which we present in Fig. \ref{fig:ext_contour}. We find that, for the UNIONS-3500~+~\textit{Planck} combination, the constraints on $S_8$ are largely driven by CMB data. In particular, the orange contour reduces to the blue one, which are themselves nearly identical to the green contour obtained using \textit{Planck}-only data (with the \texttt{plik\_lite} likelihood). We note that our \textit{Planck}-only posteriors do not exactly match the fiducial results published in \cite{planck2018}, due to differences in sampler, likelihood, parameters and priors employed in our inference pipeline. Including BAO then reduces constraints on $\Omega_{\rm m}$ to that favoured by the data, while $\sigma_8$ stays approximately constant since BAO in itself is insensitive towards the amplitude of clustering. 

%--------------------------------------------------------------------

\section{Conclusion}\label{sec:conclusions}
We have conducted the first 2D cosmic shear analysis and presented cosmological constraints in configuration space with the UNIONS-3500 photometric weak lensing catalogue, the largest of its kind in the northern hemisphere to date. We report a fiducial result of $S_8 = 0.831^{+0.067}_{-0.078}$ and $\Omega_{\rm m} = 0.265^{+0.130}_{-0.075}$ with a flat $\Lambda$CDM model. This is $\sim0.6\,\sigma$ lower compared to the analysis in harmonic space. This difference is primarily driven by the difference in $k$-space sampled by the two summary statistics, which then responded differently to the scale-dependent features in our data. Furthermore, our results are consistent with those from previous Stage III weak lensing surveys, as well as with early-time constraints from \textit{Planck} CMB data. 

We have also outlined our methodology to derive these constraints, ensuring their robustness and ability to mitigate contamination due to systematic effects. We adopted $B-$mode-informed scale cuts, and ensured that PSF systematics is adequately accounted for by jointly fitting the systematic and cosmological signals, on top of employing an additional safeguard of first sampling the leakage parameters $\alpha_{\rm PSF}$ and $\beta_{\rm PSF}$ to obtain more data-informed priors in the subsequent inference step. We estimated the redshift distribution using SOMs and validate our catalogue with dedicated simulations, which are also used to assess shear calibration biases. We then derived data-driven priors on the nuisance parameters quantifying the magnification bias, the redshift uncertainty and the amplitude of intrinsic alignment, thereby maximising the constraining power of our analysis. Finally, both harmonic and configuration space pipelines were simultaneously validated using mock galaxy catalogues, enabling a direct assessment of internal consistency before unblinding; to our knowledge, this represents a novel approach in cosmic shear analyses.

Our final results demonstrate the reliability of our analytical methodology across a wide range of modelling assumptions and analysis configurations. In particular, we find that the inferred cosmological constraints are most sensitive to intrinsic alignment modelling and uncertainties in the source redshift distribution. Adopting a flat prior on $A_{\rm IA}$ increases the uncertainty on $S_8$ by nearly a factor of two, while fixing $A_{\rm IA}=0$ produces a shift of $-0.71\sigma$ in the recovered $S_8$ value, reflecting the well-known degeneracy between intrinsic alignments and the clustering amplitude. Similarly, adopting a flat prior on the redshift shift parameter $\Delta z$ produces the second-largest increase in the $S_8$ uncertainty among the systematic tests considered. Taken together, these findings might suggest that  the current constraining power of our data is driven not only by the statistical precision of the shear measurements, but also by nuisance parameter degeneracies and conservative systematic treatments. 

Yet it is important to note that the priors adopted for these nuisance parameters were intentionally chosen to be conservative relative to their statistical uncertainties, in order to avoid underestimating the impact of systematic effects. This then opens up possibilities for future improvement going forward, both in terms of strengthening the systematic characterisations of redshift distributions, shear biases, PSF systematics and intrinsic alignment models, as well as including more data with future releases and from external surveys.

In this regard, we have already shown that combining UNIONS with external probes such as \textit{Planck} CMB measurements and DESI BAO data substantially tightens constraints on both $S_8$ and $\Omega_{\rm m}$. A natural next step will therefore be to conduct a full tomographic $3\times2$ point analysis incorporating spectroscopic galaxy clustering and galaxy--galaxy lensing measurements using DESI lens galaxies, enabling us to fully exploit the substantial overlap between the two surveys across the northern sky.

As one of the final Stage III weak lensing surveys, UNIONS provides a valuable complement to preceding cosmic shear measurements, offering an independent validation of results with comparable robustness in methodology. Looking ahead, the techniques and validation strategies developed in this work can help pave the way for upcoming Stage IV surveys, whose unprecedented statistical power and improved control of systematics promise to usher in a new and exciting era of weak lensing science.
%--------------------------------------------------------------------

\section*{Acknowledgements}
We would like to thank our external blinding coordinator, Koen Kuijken. We would also like to thank Axel Guinot and Douglas Scott for their comments and helpful feedback. LWKG thanks the University of Edinburgh School of Physics and Astronomy for a postdoctoral Fellowship.  HH is supported by a DFG Heisenberg grant (Hi 1495/5-1), the DFG Collaborative Research Center SFB1491, an ERC Consolidator Grant (No. 770935), and the DLR project 50QE2305. MJH and LVW acknowledge support from NSERC through a Discovery Grant. AHW is supported by the Deutsches Zentrum für Luft- und Raumfahrt (DLR), under project 50QE2305, made possible by the Bundesministerium für Wirtschaft und Klimaschutz, and acknowledges funding from the German Science Foundation DFG, via the Collaborative Research Center SFB1491 "Cosmic Interacting Matters - From Source to Signal". This work was made possible by utilising the CANDIDE cluster at the Institut d’Astrophysique de Paris. The cluster was funded through grants from the PNCG, CNES, DIM-ACAV, the Euclid Consortium, and the Danish National Research Foundation Cosmic Dawn Center (DNRF140). The authors acknowledge the use of the Canadian Advanced Network for Astronomy Research (CANFAR) Science Platform operated by the Canadian Astronomy Data Centre (CADC) and the Digital Research Alliance of Canada (DRAC), with support from the National Research Council of Canada (NRC), the Canadian Space Agency (CSA), CANARIE, and the Canada Foundation for Innovation (CFI). It is maintained by Stephane Rouberol. We are honoured and grateful for the opportunity of observing the Universe from Maunakea and Haleakala, which both have cultural, historical and natural significance in Hawai'i. This work is based on data obtained as part of the Canada-France Imaging Survey, a CFHT large program of the National Research Council of Canada and the French Centre National de la Recherche Scientifique. Based on observations obtained with MegaPrime/MegaCam, a joint project of CFHT and CEA Saclay, at the Canada-France-Hawai'i Telescope (CFHT) which is operated by the National Research Council (NRC) of Canada, the Institut National des Science de l’Univers (INSU) of the Centre National de la Recherche Scientifique (CNRS) of France, and the University of Hawaii. This research is based in part on data collected at Subaru Telescope, which is operated by the National Astronomical Observatory of Japan. Pan-STARRS is a project of the Institute for Astronomy of the University of Hawai'i, and is supported by the NASA SSO Near Earth Observation Program under grants 80NSSC18K0971, NNX14AM74G, NNX12AR65G, NNX13AQ47G, NNX08AR22G, 80NSSC21K1572 and by the State of Hawai'i.
%%%%%%%%%%%%%%%%%%%%%%%%%%%%%%%%%%%%%%%%%%%%%%%%%%
\section*{Data Availability}

A subset of the raw data underlying this article is publicly available via the Canadian Astronomical Data Centre at \url{http://www.cadc-ccda.hia-iha.nrc-cnrc.gc.ca/en/megapipe/}. The remaining raw data and all processed data are available to members of the Canadian and French communities via reasonable requests to the principal investigators of the Canada-France Imaging Survey, Alan McConnachie and Jean-Charles Cuillandre. All inference chains and software used to produce the results will be publicly available to the international community upon paper acceptance.



\bibliographystyle{mnras}
\bibliography{biblio}

%%%%%%%%%%%%%%%%% APPENDICES %%%%%%%%%%%%%%%%%%%%%

\appendix

\section{Validation with GLASS mocks}\label{sec:glass_mocks}

\begin{figure}
    \centering
    \includegraphics[width=\linewidth]{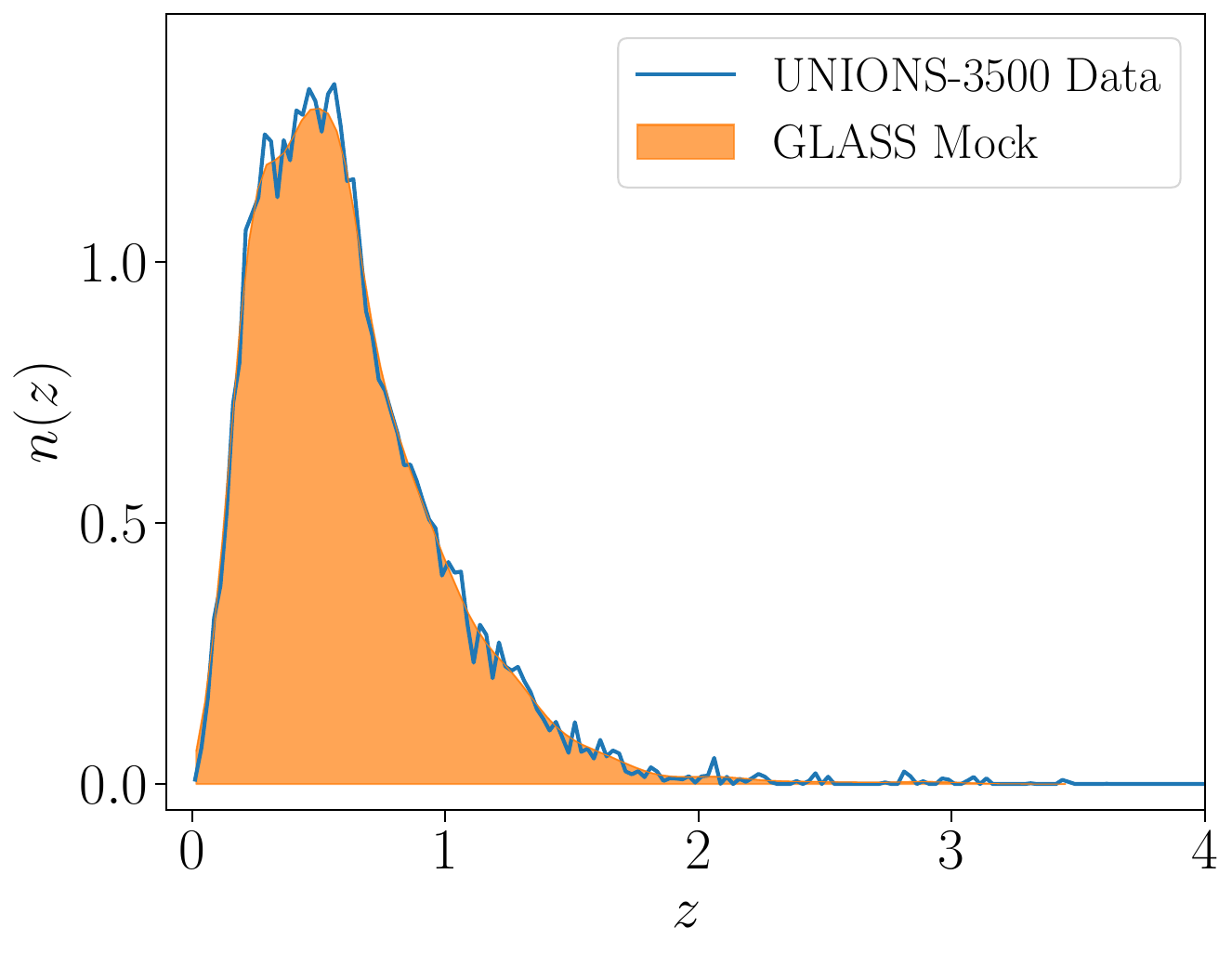}
    \caption{Redshift distribution $n(z)$ of the UNIONS data (blue) and the mock catalogues (orange).}
    \label{fig:nz_mocks}
\end{figure}

We present validation tests using the \texttt{GLASS} galaxy mock catalogues. The redshift distribution of the mocks agrees with the data (Fig. \ref{fig:nz_mocks}), as expected. We do not expect systematic effects to contribute to the mock signal, such that the nuisance parameters listed in Table \ref{tab:inference_priors} should be consistent with zero. Nonetheless, we sample over them, keeping their Gaussian priors with the same standard deviations but centred over a zero mean. 

\begin{figure}
    \centering
    \includegraphics[width=\linewidth]{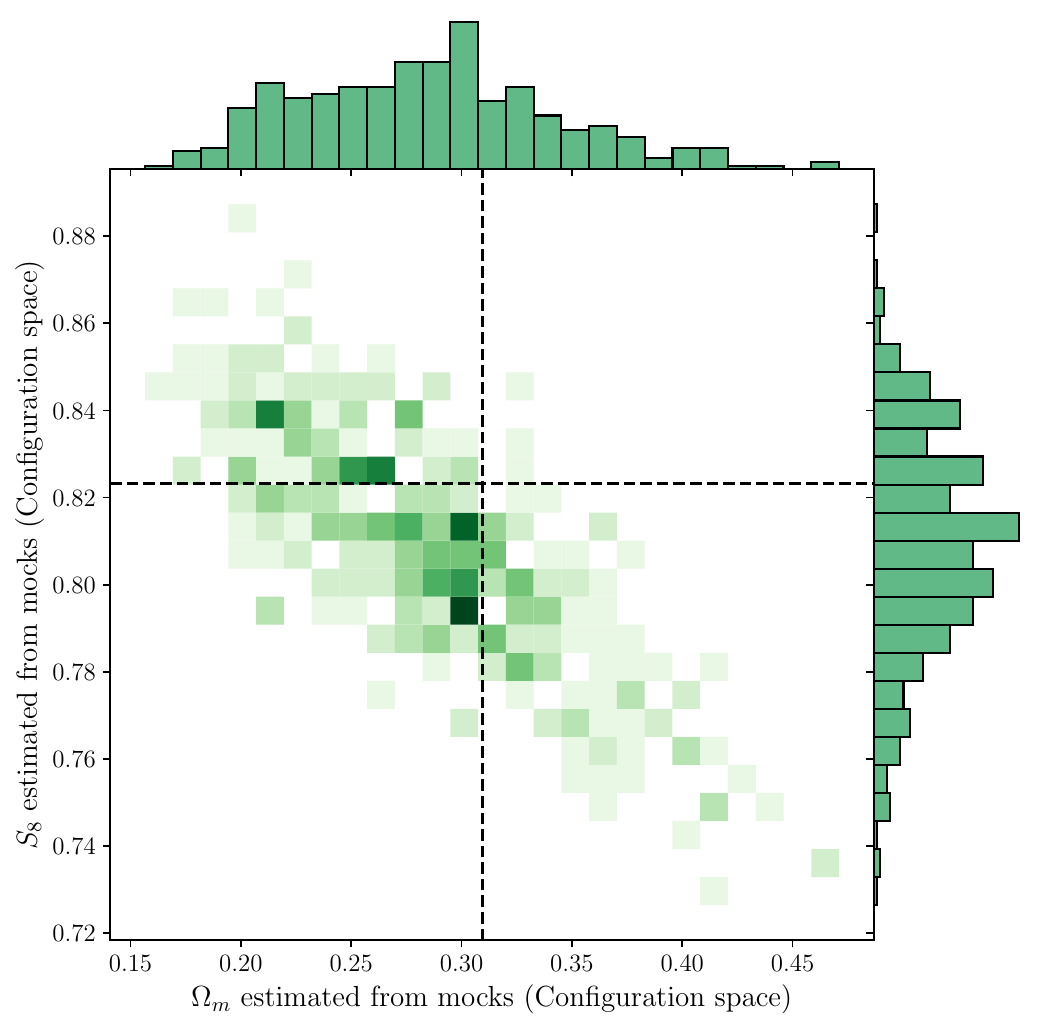}
    \caption{2D histogram of the recovered best-fit cosmological values in the $\Omega_\mathrm{m}-S_8$ plane, estimated from the 2D marginalised mode, for the configuration space analysis. The fiducial values have been marked out in dashed lines.}
    \label{fig:Om_s8_config}
\end{figure}

We also plot the histogram of the 2D marginalised modes of $\Omega_{\rm m}$ and $S_8$ in Fig. \ref{fig:Om_s8_config}, specifically those obtained in the configuration space mock pipeline analysis. 

\section{Additional inference results}

We present additional results from our inference.

\begin{figure}
    \centering
    \includegraphics[width=\linewidth]{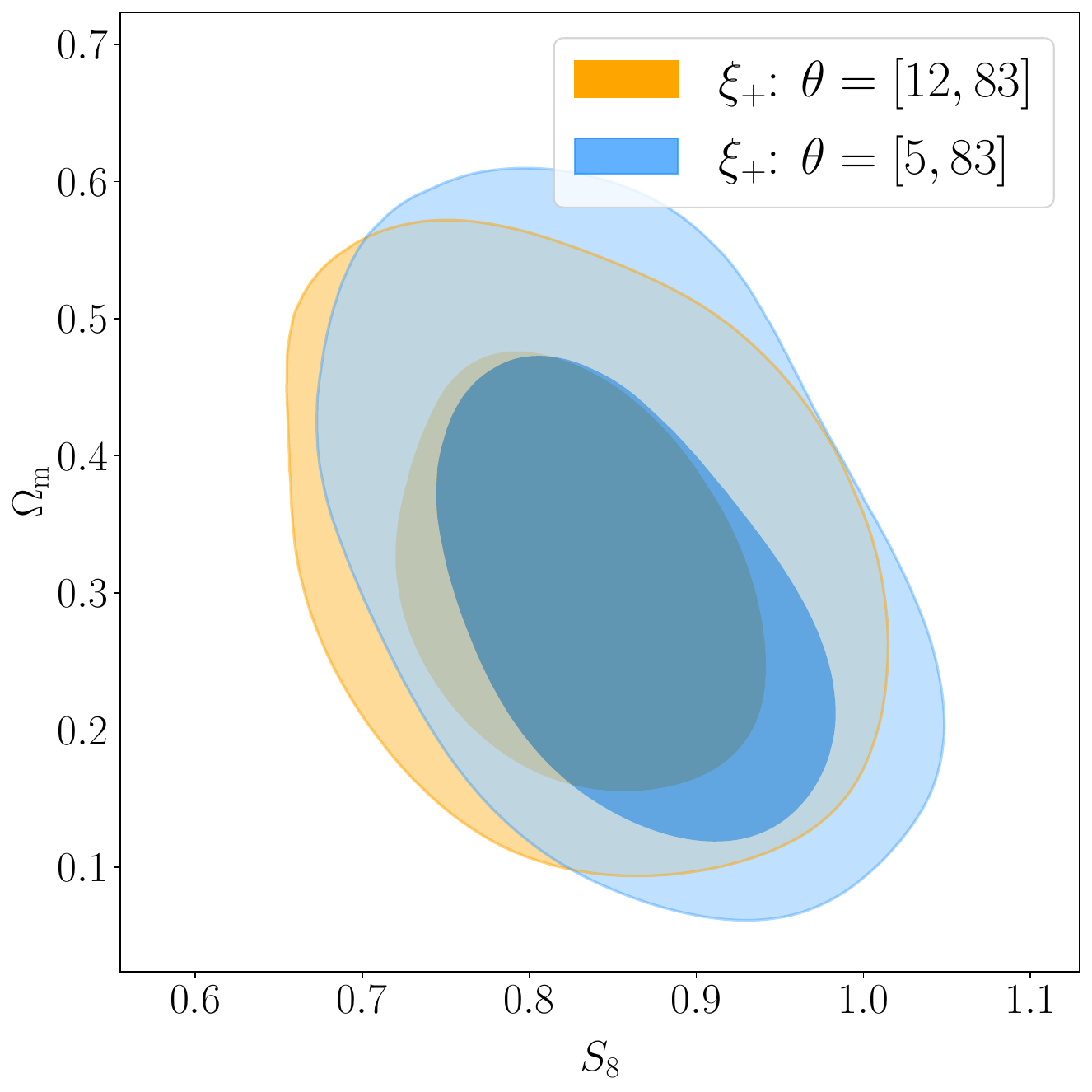}
    \caption{2D marginalised contour of $S_8$ and $\Omega_{\rm m}$ in the fiducial analysis when $\theta\in[12,83]$ arcmin (orange), and when $\theta\in[5,83]$ arcmin (blue).}
    \label{fig:scale_contour}
\end{figure}

\begin{figure}
    \centering
    \includegraphics[width=\linewidth]{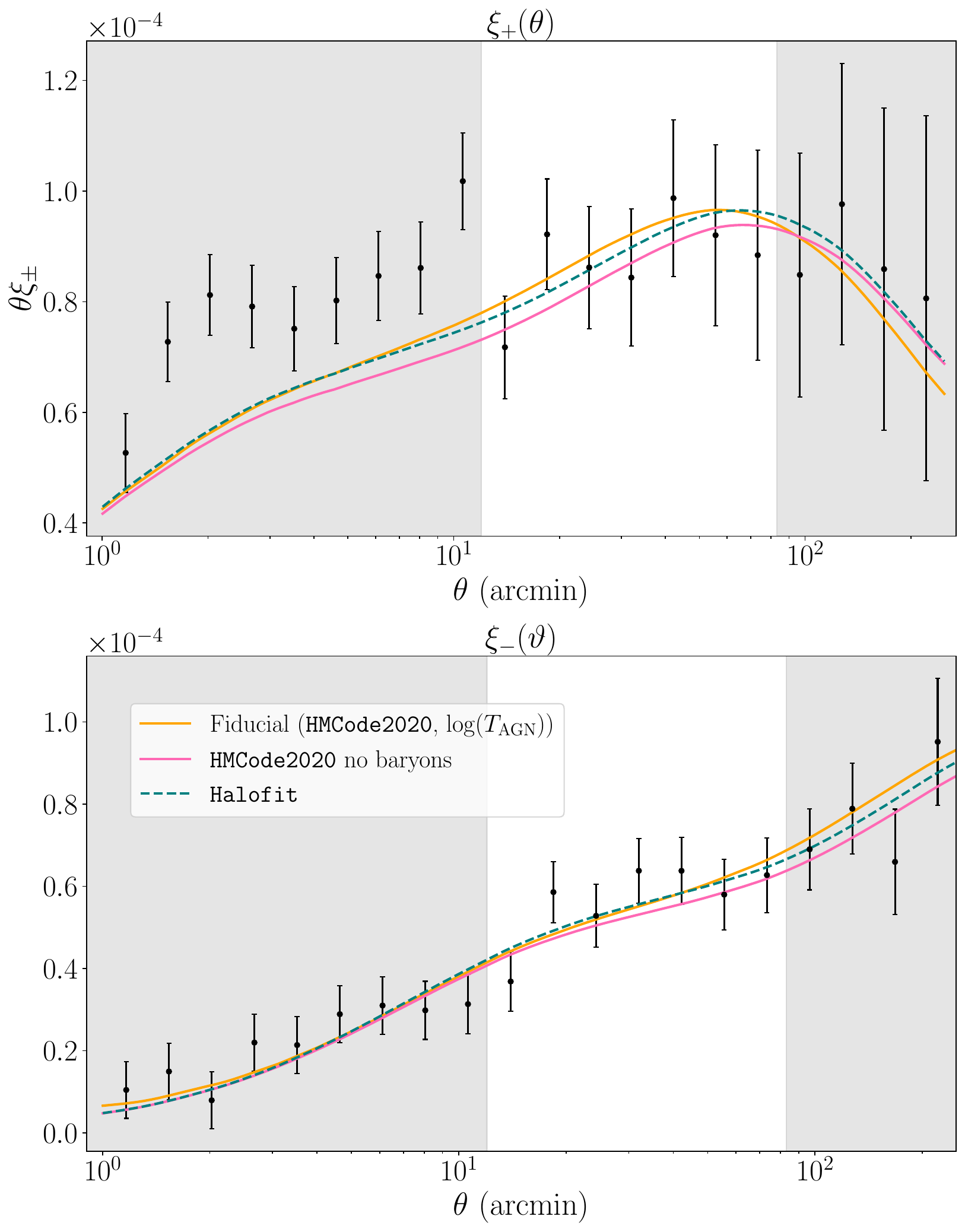}
    \caption{Plot of the best-fit $\xi_\pm$ data vectors for the three modelling choices of the nonlinear power spectrum: using \texttt{HMCode2020} including baryonic feedback (solid orange), without feedback (solid pink) or using \texttt{Halofit} instead (dotted teal).} 
    \label{fig:nonlin_xipm}
\end{figure}

\begin{figure*}
    \centering
    \includegraphics[width=\linewidth]{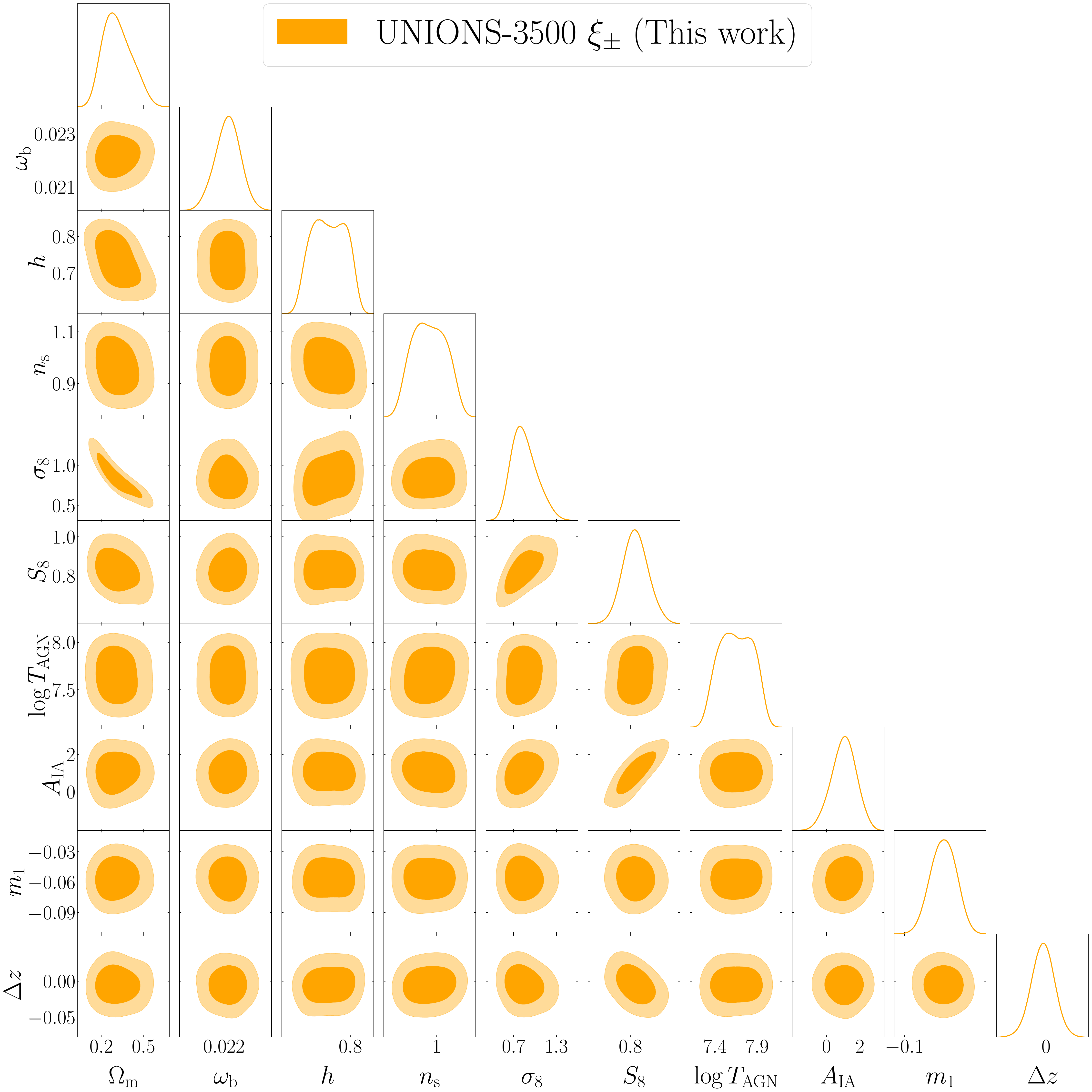}
    \caption{Full 2D marginalised posterior distributions for our fiducial setup run, for all parameters sampled (see Table \ref{tab:inference_priors}).}
    \label{fig:fid_contour_full}
\end{figure*}

\begin{sidewaystable*}
\centering
\renewcommand{\arraystretch}{1.5}
\begin{adjustbox}{center, max totalsize={\textheight}{\textwidth}}
\setlength{\tabcolsep}{4pt}
\begin{tabular}{l|cc|cc|cc|cc|cc|cc}
\hline
& \multicolumn{2}{c|}{$S_8$}
& \multicolumn{2}{c|}{$\xi_+$}
& \multicolumn{2}{c|}{$\xi_-$}
& \multicolumn{2}{c|}{$\xi_{\pm}$}
& \multicolumn{2}{c|}{$\tau_{0,2}$}
& \multicolumn{2}{c}{$\mathrm{Total}$} \\
\hline

& 2D MAP & Weighted mean
& $\chi^2/n$ & PTE
& $\chi^2/n$ & PTE
& $\chi^2/n$ & PTE
& $\chi^2/n$ & PTE
& $\chi^2/n$ & PTE \\
\hline

\textbf{Fiducial}
& $\mathbf{0.831^{+0.067}_{-0.078}}$ & $\mathbf{0.829}$
& \textbf{4.29} & \textbf{0.745} & \textbf{5.27} & \textbf{0.627} & \textbf{9.50/14} & \textbf{0.798}  &\textbf{65.77/40} & \textbf{0.00629} & \textbf{75.33/54} & \textbf{0.0291} \\

Including $\xi_+$ small scales 
& $0.886^{+0.049}_{-0.104}$ & 0.860
& 16.46/10 & 0.0872 
& 4.60/7 & 0.708 
& 22.85/17 & 0.154  
& 65.77/40 & 0.00628 
& 86.83/57 & 0.00663 \\

Flat $\alpha_{\rm PSF}$ and $\beta_{\rm PSF}$ priors 
& $0.825^{+0.071}_{-0.075}$ & 0.830
& 4.66/7 & 0.70 
& 5.47/7 & 0.603 
& 9.99/14 & 0.763  
&65.78/40 & 0.00627 
& 75.90/54 & 0.0263 \\

No $\xi^{\rm sys}_{\pm}$ 
& $0.824^{+0.063}_{-0.068}$ & 0.826
& 4.62/7 & 0.706 
& 5.27/7 & 0.626 
& 9.87/14 & 0.771  
& $-$ & $-$
& 9.87/14 & 0.771 \\

No object-wise leakage correction 
& $0.818\pm0.073$ & 0.820
& 4.44/7 & 0.728 
& 5.39/7 & 0.613 
& 9.61/14 & 0.790  
& 67.41/40 & 0.0043 
& 77.24/54 & 0.0207 \\

Flat $\Delta z$ prior 
& $0.808^{+0.121}_{-0.093}$ & 0.844
& 9.81/7 & 0.200 
& 17.67/7 & 0.0136 
& 24.28/14 & 0.0424  
& 65.77/40 & 0.00628 
& 93.25/54 & 0.000727 \\

No $\Delta z$ 
& $0.830^{+0.058}_{-0.071}$ & 0.825
& 5.17/7 & 0.639 
& 6.35/7 & 0.499 
& 11.17/14 & 0.672  
& 65.76/40 & 0.00629 
& 77.28/54 & 0.0205 \\

Flat $A_{\rm IA}$ prior 
& $0.895^{+0.172}_{-0.134}$ & 0.918
& 9.13/7 & 0.244 
& 14.67/7 & 0.0404 
& 21.31/14 & 0.094  
& 65.77/40 & 0.00628 
& 89.57/54 & 0.00168 \\

No $A_{\rm IA}$ 
& $0.770^{+0.029}_{-0.059}$ & 0.752
& 4.26/7 & 0.750 
& 5.27/7 & 0.627 
& 9.45/14 & 0.801  
& 65.77/40 & 0.00629 
& 75.30/54 & 0.0293 \\

No $m$ bias 
& $0.809^{+0.050}_{-0.079}$ & 0.796
& 4.69/7 & 0.697 
& 5.04/7 & 0.656 
& 9.62/14 & 0.789  
& 65.76/40 & 0.00629 
& 75.49/54 & 0.0283 \\

Unmasked covmat 
& $0.822^{+0.072}_{-0.070}$ & 0.823
& 4.78/7 & 0.686 
& 6.93/7 & 0.436 
& 11.39/14 & 0.655  
& 65.76/40 & 0.0063 
& 77.48/54 & 0.0198 \\

\texttt{Halofit} 
& $0.800^{+0.068}_{-0.074}$ & 0.806
& 4.58/7 & 0.711 
& 4.83/7 & 0.680 
& 9.40/14 & 0.804  
& 65.77/40 & 0.00628 
& 75.19/54 & 0.0299 \\

\texttt{HMCode} no baryons 
& $0.813^{+0.072}_{-0.068}$ & 0.822
& 4.84/7 & 0.680 
& 5.78/7 & 0.566 
& 10.46/14 & 0.728  
& 65.77/40 & 0.00628 
& 76.38/54 & 0.0242 \\

Nautilus sampler
& $0.826^{+0.068}_{-0.069}$ & 0.826
& 4.57/7 & 0.713 
&5.83/7 & 0.559
& 10.28/14 & 0.742  
& 65.76/40 & 0.0063 
& 76.16/54 & 0.0251 \\

UNIONS-3500 + CMB
& $0.855^{+0.020}_{-0.017}$ & 0.857
& 4.33/7 & 0.741 
& 4.92/7 & 0.67 
& 9.15/14 & 0.822  
&65.76/40 & 0.0063 
& 75.01/54 & 0.031 \\

UNIONS-3500 + CMB + BAO
& $0.849^{+0.015}_{-0.017}$ & 0.849
& 4.40/7 & 0.733 
& 4.99/7 & 0.661 
& 9.22/14 & 0.817  
&65.77/40 & 0.00628 
& 75.16/54 & 0.030 \\
\hline
\end{tabular}
\end{adjustbox}
\caption{Table presenting the best-fit values of $S_8$ and goodness-of-fit statistics (the $\chi^2/n$ and $p$ values of each component of the data vector: $\xi_\pm$, $\tau_{0,2}$ and their sum, where $n$ denotes the number of data points fitted) for different analysis choices and data sets. When quoting the $S_8$ best-fit values, we report both the 2D marginalised mode and its 68\% CI, as well as the weighted mean.}
\label{tab:best_fit_chi}
\end{sidewaystable*}
%%%%%%%%%%%%%%%%%%%%%%%%%%%%%%%%%%%%%%%%%%%%%%%%%%

% Don't change these lines
\bsp	% typesetting comment
\label{lastpage}
\end{document}

%% file: claims_macros.tex
% Auto-generated from claim evidence
% Regenerate: snakemake paper_macros
% See workflow/config/bmodes_paper.md for paper choices

% cosebis (SP_v1.4.6.3_leak_corr, n=6)

% pure_eb_data_vector (min across blinds per spec)
\newcommand{\ebthetaXipMin}{\num{12}}
\newcommand{\ebthetaXipMax}{\num{83}}
\newcommand{\ebthetaXimMin}{\num{12}}
\newcommand{\ebthetaXimMax}{\num{83}}

% pure_eb_covariance (block condition numbers)
\newcommand{\ebcovCondE}{\num{1.5e+05}}
\newcommand{\ebcovCondB}{\num{2.0e+05}}
\newcommand{\ebcovCondAmb}{\num{1.8e+10}}
\newcommand{\ebcovCondFull}{\num{2.7e+10}}
\newcommand{\ebcovNbins}{\num{20}}

% config_space_pte_matrices (all versions)

\newcommand{\configPteSixThreeXip}{\num{0.31}}
\newcommand{\configPteSixThreeXim}{\num{0.26}}
\newcommand{\configPteSixThreeCombined}{\num{0.18}}
\newcommand{\configPteSixThreeCosebis}{\num{0.78}}
\newcommand{\configPteSixThreeCosebisTwenty}{\num{0.94}}
\newcommand{\configPteSixThreeXipFull}{\num{0.45}}
\newcommand{\configPteSixThreeXimFull}{\num{0.16}}
\newcommand{\configPteSixThreeCombinedFull}{\num{0.40}}
\newcommand{\configPteSixThreeCosebisFull}{\num{1.37e-05}}
\newcommand{\configPteSixThreeCosebisTwentyFull}{\num{1.08e-04}}
\newcommand{\configPteFiveXip}{\num{0.51}}
\newcommand{\configPteFiveXim}{\num{0.005}}
\newcommand{\configPteFiveCombined}{\num{0.030}}
\newcommand{\configPteFiveCosebis}{\num{0.94}}
\newcommand{\configPteFiveCosebisTwenty}{\num{0.61}}
\newcommand{\configPteFiveXipFull}{\num{0.026}}
\newcommand{\configPteFiveXimFull}{\num{7.71e-04}}
\newcommand{\configPteFiveCombinedFull}{\num{0.004}}
\newcommand{\configPteFiveCosebisFull}{\num{7.37e-10}}
\newcommand{\configPteFiveCosebisTwentyFull}{\num{4.15e-07}}
\newcommand{\configPteEightXip}{\num{0.020}}
\newcommand{\configPteEightXim}{\num{0.047}}
\newcommand{\configPteEightCombined}{\num{0.014}}
\newcommand{\configPteEightCosebis}{\num{0.60}}
\newcommand{\configPteEightCosebisTwenty}{\num{0.81}}
\newcommand{\configPteEightXipFull}{\num{0.039}}
\newcommand{\configPteEightXimFull}{\num{0.020}}
\newcommand{\configPteEightCombinedFull}{\num{0.018}}
\newcommand{\configPteEightCosebisFull}{\num{4.94e-06}}
\newcommand{\configPteEightCosebisTwentyFull}{\num{2.50e-05}}
\newcommand{\configPteElevenThreeXip}{\num{0.78}}
\newcommand{\configPteElevenThreeXim}{\num{9.40e-04}}
\newcommand{\configPteElevenThreeCombined}{\num{0.014}}
\newcommand{\configPteElevenThreeCosebis}{\num{0.82}}
\newcommand{\configPteElevenThreeCosebisTwenty}{\num{0.83}}
\newcommand{\configPteElevenThreeXipFull}{\num{0.10}}
\newcommand{\configPteElevenThreeXimFull}{\num{2.94e-06}}
\newcommand{\configPteElevenThreeCombinedFull}{\num{9.07e-04}}
\newcommand{\configPteElevenThreeCosebisFull}{\num{1.70e-10}}
\newcommand{\configPteElevenThreeCosebisTwentyFull}{\num{1.16e-08}}

% harmonic_space_pte_matrices (all versions)

\newcommand{\clPteFiveFid}{\num{0.003}}
\newcommand{\clPteFiveFull}{\num{2.53e-04}}
\newcommand{\clPteSixThreeFid}{\num{0.30}}
\newcommand{\clPteSixThreeFull}{\num{0.13}}
\newcommand{\clPteEightFid}{\num{0.43}}
\newcommand{\clPteEightFull}{\num{0.11}}
\newcommand{\clPteElevenThreeFid}{\num{5.74e-10}}
\newcommand{\clPteElevenThreeFull}{\num{1.78e-12}}

% harmonic_config_cosebis_comparison_full (B-mode COSEBIS PTEs)

\newcommand{\harmCosebisPteSixThreeFull}{\num{1.61e-05}}
\newcommand{\harmCosebisChisqSixThreeFull}{\num{32.03}}
\newcommand{\harmCosebisPteElevenThreeFull}{\num{4.64e-18}}
\newcommand{\harmCosebisChisqElevenThreeFull}{\num{93.92}}
\newcommand{\harmCosebisPteFiveFull}{\num{6.41e-09}}
\newcommand{\harmCosebisChisqFiveFull}{\num{49.33}}
\newcommand{\harmCosebisPteEightFull}{\num{4.72e-05}}
\newcommand{\harmCosebisChisqEightFull}{\num{29.58}}
\newcommand{\cfgCosebisPteSixThreeFull}{\num{1.37e-05}}
\newcommand{\cfgCosebisChisqSixThreeFull}{\num{32.40}}
\newcommand{\cfgCosebisPteElevenThreeFull}{\num{1.70e-10}}
\newcommand{\cfgCosebisChisqElevenThreeFull}{\num{57.15}}
\newcommand{\cfgCosebisPteFiveFull}{\num{7.37e-10}}
\newcommand{\cfgCosebisChisqFiveFull}{\num{54.00}}
\newcommand{\cfgCosebisPteEightFull}{\num{4.94e-06}}
\newcommand{\cfgCosebisChisqEightFull}{\num{34.69}}

% harmonic_config_cosebis_comparison_fiducial (B-mode COSEBIS PTEs)

\newcommand{\harmCosebisPteSixThreeFid}{\num{0.60}}
\newcommand{\harmCosebisChisqSixThreeFid}{\num{4.59}}
\newcommand{\harmCosebisPteElevenThreeFid}{\num{0.85}}
\newcommand{\harmCosebisChisqElevenThreeFid}{\num{2.63}}
\newcommand{\harmCosebisPteFiveFid}{\num{0.73}}
\newcommand{\harmCosebisChisqFiveFid}{\num{3.59}}
\newcommand{\harmCosebisPteEightFid}{\num{0.40}}
\newcommand{\harmCosebisChisqEightFid}{\num{6.17}}
\newcommand{\cfgCosebisPteSixThreeFid}{\num{0.78}}
\newcommand{\cfgCosebisChisqSixThreeFid}{\num{3.25}}
\newcommand{\cfgCosebisPteElevenThreeFid}{\num{0.82}}
\newcommand{\cfgCosebisChisqElevenThreeFid}{\num{2.88}}
\newcommand{\cfgCosebisPteFiveFid}{\num{0.94}}
\newcommand{\cfgCosebisChisqFiveFid}{\num{1.79}}
\newcommand{\cfgCosebisPteEightFid}{\num{0.60}}
\newcommand{\cfgCosebisChisqEightFid}{\num{4.54}}

%% file: biblio.bib
@article{maccrann_dark_2021,
  title =        "Dark {Energy} {Survey} {Y3} results: blending shear
                 and redshift biases in image simulations",
  volume =       "509",
  ISSN =         "0035-8711, 1365-2966",
  shorttitle =   "Dark {Energy} {Survey} {Y3} results",
  URL =          "https://academic.oup.com/mnras/article/509/3/3371/6385771",
  DOI =          "10.1093/mnras/stab2870",
  language =     "en",
  number =       "3",
  urldate =      "2024-01-15",
  journal =      "MNRAS",
  author =       "N MacCrann and M R Becker and J McCullough and A Amon
                 and D Gruen and M Jarvis and A Choi and M A Troxel and
                 E Sheldon and B Yanny and K Herner and S Dodelson and J
                 Zuntz and K Eckert and R P Rollins and T N Varga and G
                 M Bernstein and R A Gruendl and I Harrison and W G
                 Hartley and I Sevilla-Noarbe and A Pieres and S L
                 Bridle and J Myles and A Alarcon and S Everett and C
                 Sánchez and E M Huff and F Tarsitano and M Gatti and L
                 F Secco and T M C Abbott and M Aguena and S Allam and J
                 Annis and D Bacon and E Bertin and D Brooks and D L
                 Burke and A Carnero Rosell and M Carrasco Kind and J
                 Carretero and M Costanzi and M Crocce and M E S Pereira
                 and J De Vicente and S Desai and H T Diehl and J P
                 Dietrich and P Doel and T F Eifler and I Ferrero and A
                 Ferté and B Flaugher and P Fosalba and J Frieman and J
                 García-Bellido and E Gaztanaga and D W Gerdes and T
                 Giannantonio and J Gschwend and G Gutierrez and S R
                 Hinton and D L Hollowood and K Honscheid and D J James
                 and O Lahav and M Lima and M A G Maia and M March and J
                 L Marshall and P Martini and P Melchior and F Menanteau
                 and R Miquel and J J Mohr and R Morgan and J Muir and R
                 L C Ogando and A Palmese and F Paz-Chinchón and A A
                 Plazas and M Rodriguez-Monroy and A Roodman and S
                 Samuroff and E Sanchez and V Scarpine and S Serrano and
                 M Smith and M Soares-Santos and E Suchyta and M E C
                 Swanson and G Tarle and D Thomas and C To and R D
                 Wilkinson and {(DES Collaboration)}",
  month =        nov,
  year =         "2021",
  pages =        "3371--3394",
}

@article{liKiDSLegacyCalibrationUnifying2023,
  title = {{{KiDS-Legacy}} Calibration: {{Unifying}} Shear and Redshift Calibration with the {{SKiLLS}} Multi-Band Image Simulations},
  shorttitle = {{{KiDS-Legacy}} Calibration},
  author = {Li, Shun-Sheng and Kuijken, Konrad and Hoekstra, Henk and Miller, Lance and Heymans, Catherine and Hildebrandt, Hendrik and {van den Busch}, Jan Luca and Wright, Angus H. and Yoon, Mijin and Bilicki, Maciej and Bravo, Mat{\'i}as and Lagos, Claudia del P.},
  year = {2023},
  month = feb,
  journal = {\aap},
  volume = {670},
  pages = {A100},
  issn = {0004-6361},
  doi = {10.1051/0004-6361/202245210},
  urldate = {2024-05-22}
}

@ARTICLE{2006astro.ph..9591A,
       author = {{Albrecht}, Andreas and {Bernstein}, Gary and {Cahn}, Robert and {Freedman}, Wendy L. and {Hewitt}, Jacqueline and {Hu}, Wayne and {Huth}, John and {Kamionkowski}, Marc and {Kolb}, Edward W. and {Knox}, Lloyd and {Mather}, John C. and {Staggs}, Suzanne and {Suntzeff}, Nicholas B.},
        title = "{Report of the Dark Energy Task Force}",
      journal = {arXiv e-prints},
         year = 2006,
        month = sep,
          eid = {astro-ph/0609591},
        pages = {astro-ph/0609591},
          doi = {10.48550/arXiv.astro-ph/0609591},
archivePrefix = {arXiv},
       eprint = {astro-ph/0609591},
 primaryClass = {astro-ph},
       adsurl = {https://ui.adsabs.harvard.edu/abs/2006astro.ph..9591A}
}

@ARTICLE{2000astro.ph..3338K,
       author = {{Kaiser}, Nick and {Wilson}, Gillian and {Luppino}, Gerard A.},
        title = "{Large-Scale Cosmic Shear Measurements}",
      journal = {arXiv e-prints},
         year = 2000,
        month = mar,
          eid = {astro-ph/0003338},
        pages = {astro-ph/0003338},
          doi = {10.48550/arXiv.astro-ph/0003338},
archivePrefix = {arXiv},
       eprint = {astro-ph/0003338},
 primaryClass = {astro-ph},
       adsurl = {https://ui.adsabs.harvard.edu/abs/2000astro.ph..3338K}
}

@ARTICLE{troxel2018maskingkids,
       author = {{Troxel}, M.~A. and {Krause}, E. and {Chang}, C. and {Eifler}, T.~F. and {Friedrich}, O. and {Gruen}, D. and {MacCrann}, N. and {Chen}, A. and {Davis}, C. and {DeRose}, J. and {Dodelson}, S. and {Gatti}, M. and {Hoyle}, B. and {Huterer}, D. and {Jarvis}, M. and {Lacasa}, F. and {Lemos}, P. and {Peiris}, H.~V. and {Prat}, J. and {Samuroff}, S. and {S{\'a}nchez}, C. and {Sheldon}, E. and {Vielzeuf}, P. and {Wang}, M. and {Zuntz}, J. and {Lahav}, O. and {Abdalla}, F.~B. and {Allam}, S. and {Annis}, J. and {Avila}, S. and {Bertin}, E. and {Brooks}, D. and {Burke}, D.~L. and {Carnero Rosell}, A. and {Carrasco Kind}, M. and {Carretero}, J. and {Crocce}, M. and {Cunha}, C.~E. and {D'Andrea}, C.~B. and {da Costa}, L.~N. and {De Vicente}, J. and {Diehl}, H.~T. and {Doel}, P. and {Evrard}, A.~E. and {Flaugher}, B. and {Fosalba}, P. and {Frieman}, J. and {Garc{\'\i}a-Bellido}, J. and {Gaztanaga}, E. and {Gerdes}, D.~W. and {Gruendl}, R.~A. and {Gschwend}, J. and {Gutierrez}, G. and {Hartley}, W.~G. and {Hollowood}, D.~L. and {Honscheid}, K. and {James}, D.~J. and {Kirk}, D. and {Kuehn}, K. and {Kuropatkin}, N. and {Li}, T.~S. and {Lima}, M. and {March}, M. and {Menanteau}, F. and {Miquel}, R. and {Mohr}, J.~J. and {Ogando}, R.~L.~C. and {Plazas}, A.~A. and {Roodman}, A. and {Sanchez}, E. and {Scarpine}, V. and {Schindler}, R. and {Sevilla-Noarbe}, I. and {Smith}, M. and {Soares-Santos}, M. and {Sobreira}, F. and {Suchyta}, E. and {Swanson}, M.~E.~C. and {Thomas}, D. and {Walker}, A.~R. and {Wechsler}, R.~H.},
        title = "{Survey geometry and the internal consistency of recent cosmic shear measurements}",
      journal = {\mnras},
         year = 2018,
        month = oct,
       volume = {479},
       number = {4},
        pages = {4998-5004},
          doi = {10.1093/mnras/sty1889},
archivePrefix = {arXiv},
       eprint = {1804.10663},
 primaryClass = {astro-ph.CO},
       adsurl = {https://ui.adsabs.harvard.edu/abs/2018MNRAS.479.4998T}
}

@ARTICLE{des_y6_cosmo,
       author = {{DES Collaboration} and {Abbott}, T.~M.~C. and {Adamow}, M. and {Aguena}, M. and {Alarcon}, A. and {Allam}, S.~S. and {Alves}, O. and {Amon}, A. and {Anbajagane}, D. and {Andrade-Oliveira}, F. and {Avila}, S. and {Bacon}, D. and {Baxter}, E.~J. and {Beas-Gonzalez}, J. and {Bechtol}, K. and {Becker}, M.~R. and {Bernstein}, G.~M. and {Bertin}, E. and {Blazek}, J. and {Bocquet}, S. and {Brooks}, D. and {Brout}, D. and {Camacho}, H. and {Camacho-Ciurana}, G. and {Camilleri}, R. and {Campailla}, G. and {Campos}, A. and {Carnero Rosell}, A. and {Carrasco Kind}, M. and {Carretero}, J. and {Carrilho}, P. and {Castander}, F.~J. and {Cawthon}, R. and {Chang}, C. and {Choi}, A. and {Coloma-Nadal}, J.~M. and {Costanzi}, M. and {Crocce}, M. and {d'Assignies}, W. and {da Costa}, L.~N. and {da Silva Pereira}, M.~E. and {Davis}, T.~M. and {De Vicente}, J. and {DeRose}, J. and {Diehl}, H.~T. and {Dodelson}, S. and {Doel}, P. and {Doux}, C. and {Drlica-Wagner}, A. and {Eifler}, T.~F. and {Elvin-Poole}, J. and {Estrada}, J. and {Everett}, S. and {Evrard}, A.~E. and {Fang}, J. and {Farahi}, A. and {Fert{\'e}}, A. and {Flaugher}, B. and {Fosalba}, P. and {Frieman}, J. and {Garc{\'\i}a-Bellido}, J. and {Gatti}, M. and {Gaztanaga}, E. and {Giannini}, G. and {Giles}, P. and {Glazebrook}, K. and {Gorsuch}, M. and {Gruen}, D. and {Gruendl}, R.~A. and {Gschwend}, J. and {Gutierrez}, G. and {Harrison}, I. and {Hartley}, W.~G. and {Henning}, E. and {Herner}, K. and {Hinton}, S.~R. and {Hollowood}, D.~L. and {Honscheid}, K. and {Huff}, E.~M. and {Huterer}, D. and {Jain}, B. and {James}, D.~J. and {Jarvis}, M. and {Jeffrey}, N. and {Jeltema}, T. and {Kacprzak}, T. and {Kent}, S. and {Kovacs}, A. and {Krause}, E. and {Kron}, R. and {Kuehn}, K. and {Lahav}, O. and {Lee}, S. and {Legnani}, E. and {Lidman}, C. and {Lin}, H. and {MacCrann}, N. and {Manera}, M. and {Manning}, T. and {Marshall}, J.~L. and {Mau}, S. and {McCullough}, J. and {Mena-Fern{\'a}ndez}, J. and {Menanteau}, F. and {Miquel}, R. and {Mohr}, J.~J. and {Muir}, J. and {Myles}, J. and {Nichol}, R.~C. and {Nord}, B. and {O'Donnell}, J.~H. and {Ogando}, R.~L.~C. and {Palmese}, A. and {Paterno}, M. and {Peoples}, J. and {Percival}, W.~J. and {Petravick}, D. and {Pieres}, A. and {Plazas Malag{\'o}n}, A.~A. and {Porredon}, A. and {Pourtsidou}, A. and {Prat}, J. and {Preston}, C. and {Raveri}, M. and {Riquelme}, W. and {Rodriguez-Monroy}, M. and {Rogozenski}, P. and {Romer}, A.~K. and {Roodman}, A. and {Rosenfeld}, R. and {Ross}, A.~J. and {Rozo}, E. and {Rykoff}, E.~S. and {Samuroff}, S. and {S{\'a}nchez}, C. and {Sanchez}, E. and {Sanchez Cid}, D. and {Schutt}, T. and {Sevilla-Noarbe}, I. and {Sheldon}, E. and {Sherman}, N. and {Shin}, T. and {Smith}, M. and {Soares-Santos}, M. and {Suchyta}, E. and {Swanson}, M.~E.~C. and {Tabbutt}, M. and {Tarle}, G. and {Thomas}, D. and {To}, C. and {Tong}, A. and {Toribio San Cipriano}, L. and {Troxel}, M.~A. and {Tsedrik}, M. and {Tucker}, D.~L. and {Vikram}, V. and {Walker}, A.~R. and {Weaverdyck}, N. and {Wechsler}, R.~H. and {Weinberg}, D.~H. and {Weller}, J. and {Wetzell}, V. and {Whyley}, A. and {Wilkinson}, R.~D. and {Wiseman}, P. and {Wu}, H.-Y. and {Yamamoto}, M. and {Yanny}, B. and {Yin}, B. and {Zacharegkas}, G. and {Zhang}, Y. and {Zuntz}, J.},
        title = "{Dark Energy Survey Year 6 Results: Cosmological Constraints from Galaxy Clustering and Weak Lensing}",
      journal = {arXiv e-prints},
         year = 2026,
        month = jan,
          eid = {arXiv:2601.14559},
        pages = {arXiv:2601.14559},
          doi = {10.48550/arXiv.2601.14559},
archivePrefix = {arXiv},
       eprint = {2601.14559},
 primaryClass = {astro-ph.CO},
       adsurl = {https://ui.adsabs.harvard.edu/abs/2026arXiv260114559D}
}

@ARTICLE{friedrich2018maskingdes,
       author = {{Friedrich}, O. and {Andrade-Oliveira}, F. and {Camacho}, H. and {Alves}, O. and {Rosenfeld}, R. and {Sanchez}, J. and {Fang}, X. and {Eifler}, T.~F. and {Krause}, E. and {Chang}, C. and {Omori}, Y. and {Amon}, A. and {Baxter}, E. and {Elvin-Poole}, J. and {Huterer}, D. and {Porredon}, A. and {Prat}, J. and {Terra}, V. and {Troja}, A. and {Alarcon}, A. and {Bechtol}, K. and {Bernstein}, G.~M. and {Buchs}, R. and {Campos}, A. and {Carnero Rosell}, A. and {Carrasco Kind}, M. and {Cawthon}, R. and {Choi}, A. and {Cordero}, J. and {Crocce}, M. and {Davis}, C. and {DeRose}, J. and {Diehl}, H.~T. and {Dodelson}, S. and {Doux}, C. and {Drlica-Wagner}, A. and {Elsner}, F. and {Everett}, S. and {Fosalba}, P. and {Gatti}, M. and {Giannini}, G. and {Gruen}, D. and {Gruendl}, R.~A. and {Harrison}, I. and {Hartley}, W.~G. and {Jain}, B. and {Jarvis}, M. and {MacCrann}, N. and {McCullough}, J. and {Muir}, J. and {Myles}, J. and {Pandey}, S. and {Raveri}, M. and {Roodman}, A. and {Rodriguez-Monroy}, M. and {Rykoff}, E.~S. and {Samuroff}, S. and {S{\'a}nchez}, C. and {Secco}, L.~F. and {Sevilla-Noarbe}, I. and {Sheldon}, E. and {Troxel}, M.~A. and {Weaverdyck}, N. and {Yanny}, B. and {Aguena}, M. and {Avila}, S. and {Bacon}, D. and {Bertin}, E. and {Bhargava}, S. and {Brooks}, D. and {Burke}, D.~L. and {Carretero}, J. and {Costanzi}, M. and {da Costa}, L.~N. and {Pereira}, M.~E.~S. and {De Vicente}, J. and {Desai}, S. and {Evrard}, A.~E. and {Ferrero}, I. and {Frieman}, J. and {Garc{\'\i}a-Bellido}, J. and {Gaztanaga}, E. and {Gerdes}, D.~W. and {Giannantonio}, T. and {Gschwend}, J. and {Gutierrez}, G. and {Hinton}, S.~R. and {Hollowood}, D.~L. and {Honscheid}, K. and {James}, D.~J. and {Kuehn}, K. and {Lahav}, O. and {Lima}, M. and {Maia}, M.~A.~G. and {Menanteau}, F. and {Miquel}, R. and {Morgan}, R. and {Palmese}, A. and {Paz-Chinch{\'o}n}, F. and {Plazas}, A.~A. and {Sanchez}, E. and {Scarpine}, V. and {Serrano}, S. and {Soares-Santos}, M. and {Smith}, M. and {Suchyta}, E. and {Tarle}, G. and {Thomas}, D. and {To}, C. and {Varga}, T.~N. and {Weller}, J. and {Wilkinson}, R.~D. and {Wilkinson}, R.~D. and {DES Collaboration}},
        title = "{Dark Energy Survey year 3 results: covariance modelling and its impact on parameter estimation and quality of fit}",
      journal = {\mnras},
         year = 2021,
        month = dec,
       volume = {508},
       number = {3},
        pages = {3125-3165},
          doi = {10.1093/mnras/stab2384},
archivePrefix = {arXiv},
       eprint = {2012.08568},
 primaryClass = {astro-ph.CO},
       adsurl = {https://ui.adsabs.harvard.edu/abs/2021MNRAS.508.3125F}
}

@article{sheldonNGMIXGaussianMixture2015,
  title = {{{NGMIX}}: {{Gaussian}} Mixture Models for {{2D}} Images},
  shorttitle = {{{NGMIX}}},
  author = {Sheldon, Erin},
  year = {2015},
  month = aug,
  journal = {Astrophysics Source Code Library},
  pages = {ascl:1508.008},
  urldate = {2024-11-15}
}

@article{huffMetacalibrationDirectSelfCalibration2017,
  title = {Metacalibration: {{Direct Self-Calibration}} of {{Biases}} in {{Shear Measurement}}},
  shorttitle = {Metacalibration},
  author = {Huff, Eric and Mandelbaum, Rachel},
  year = {2017},
  month = feb,
  journal = {arXiv e-prints},
  eid = {arXiv:1702.02600},
  eprint = {1702.02600},
  primaryclass = {astro-ph.CO},
  archivePrefix = {arXiv},
  doi = {10.48550/arXiv.1702.02600},
  urldate = {2025-09-12},
  archiveprefix = {arXiv}
}

@article{sheldonPracticalWeaklensingShear2017,
  title = {Practical {{Weak-lensing Shear Measurement}} with {{Metacalibration}}},
  author = {Sheldon, Erin S. and Huff, Eric M.},
  year = {2017},
  month = may,
  journal = {\apj},
  volume = {841},
  number = {1},
  pages = {24},
  publisher = {The American Astronomical Society},
  issn = {0004-637X},
  doi = {10.3847/1538-4357/aa704b},
  urldate = {2024-02-20},
  langid = {english}
}

@INPROCEEDINGS{bertinAutomatedMorphometrySExtractor2011a,
       author = {{Bertin}, E.},
        title = "{Automated Morphometry with SExtractor and PSFEx}",
    booktitle = {Astronomical Data Analysis Software and Systems XX},
         year = 2011,
       editor = {{Evans}, I.~N. and {Accomazzi}, A. and {Mink}, D.~J. and {Rots}, A.~H.},
       series = {Astronomical Society of the Pacific Conference Series},
       volume = {442},
        month = jul,
        pages = {435},
       adsurl = {https://ui.adsabs.harvard.edu/abs/2011ASPC..442..435B}
}

@article{farrensShapePipeModularWeaklensing2022a,
  title = {{{ShapePipe}}: {{A}} Modular Weak-Lensing Processing and Analysis Pipeline},
  shorttitle = {{{ShapePipe}}},
  author = {Farrens, S. and Guinot, A. and Kilbinger, M. and Liaudat, T. and Baumont, L. and Jimenez, X. and Peel, A. and Pujol, A. and Schmitz, M. and Starck, J.-L. and Vitorelli, A. Z.},
  year = {2022},
  month = aug,
  journal = {\aap},
  volume = {664},
  pages = {A141},
  publisher = {EDP Sciences},
  issn = {0004-6361, 1432-0746},
  doi = {10.1051/0004-6361/202243970},
  urldate = {2025-09-12},
  copyright = {{\copyright} S. Farrens et al. 2022},
  langid = {english}
}

@article{gwynUNIONSUltravioletNearInfrared2025,
       author = {{Gwyn}, Stephen and {McConnachie}, Alan W. and {Cuillandre}, Jean-Charles and {Chambers}, Kenneth C. and {Magnier}, Eugene A. and {de Boer}, Thomas and {Hudson}, Michael J. and {Oguri}, Masamune and {Furusawa}, Hisanori and {Hildebrandt}, Hendrik and {Carlberg}, Raymond and {Ellison}, Sara L. and {Furusawa}, Junko and {Gavazzi}, Rapha{\"e}l and {Ibata}, Rodrigo and {Mellier}, Yannick and {Osato}, Ken and {Aussel}, H. and {Baumont}, Lucie and {Bayer}, Manuel and {Boulade}, Olivier and {C{\^o}t{\'e}}, Patrick and {Chemaly}, David and {Daley}, Cail and {Duc}, Pierre-Alain and {Durret}, Florence and {Ellien}, A. and {Fabbro}, S{\'e}bastien and {Ferreira}, Leonardo and {Fitriana}, Itsna K. and {Le Floc'h}, Emeric and {Fudamoto}, Yoshinobu and {Gao}, Hua and {Goh}, L.~W.~K. and {Goto}, Tomotsugu and {Guerrini}, Sacha and {Guinot}, Axel and {H{\'e}nault-Brunet}, Vincent and {Hammer}, Francois and {Harikane}, Yuichi and {Hayashi}, Kohei and {Heesters}, Nick and {Ichikawa}, Kohei and {Kilbinger}, Martin and {Kuzma}, P.~B. and {Li}, Qinxun and {Liaudat}, Tob{\'\i}as I. and {Lin}, Chien-Cheng and {M{\"u}ller}, Oliver and {Martin}, Nicolas F. and {Matsuoka}, Yoshiki and {Medina}, Gustavo E. and {Miyatake}, Hironao and {Miyazaki}, Satoshi and {Mpetha}, Charlie T. and {Nagao}, Tohru and {Navarro}, Julio F. and {Niwano}, Masafumi and {Ogami}, Itsuki and {Okabe}, Nobuhiro and {Onoue}, Masafusa and {Paek}, Gregory S.~H. and {Parker}, Laura C. and {Patton}, David R. and {Peters}, Fabian Hervas and {Prunet}, Simon and {S{\'a}nchez-Janssen}, Rub{\'e}n and {Schultheis}, M. and {Sestito}, Federico and {Smith}, Simon E.~T. and {Starck}, J.-L. and {Starkenburg}, Else and {Stone}, Connor and {Storfer}, Christopher and {Suzuki}, Yoshihisa and {Erben}, T. and {Taibi}, Salvatore and {Thomas}, G.~F. and {Toba}, Yoshiki and {Uchiyama}, Hisakazu and {Valls-Gabaud}, David and {Venn}, Kim A. and {Van Waerbeke}, Ludovic and {Wainscoat}, Richard J. and {Wilkinson}, Scott and {Wittje}, Anna and {Yoshida}, Taketo and {Zhang}, TianFang and {Zhong}, Yuxing},
        title = "{UNIONS: The Ultraviolet Near-infrared Optical Northern Survey}",
      journal = {\aj},
         year = 2025,
        month = dec,
       volume = {170},
       number = {6},
          eid = {324},
        pages = {324},
          doi = {10.3847/1538-3881/ae03ab},
archivePrefix = {arXiv},
       eprint = {2503.13783},
 primaryClass = {astro-ph.GA},
       adsurl = {https://ui.adsabs.harvard.edu/abs/2025AJ....170..324G}
}

@Article{	  georgiou_dependence_2019,
  title		= {The dependence of intrinsic alignment of galaxies on
		  wavelength using {KiDS} and {GAMA}},
  volume	= {622},
  copyright	= {https://www.edpsciences.org/en/authors/copyright-and-licensing},
  issn		= {0004-6361, 1432-0746},
  url		= {https://www.aanda.org/10.1051/0004-6361/201834219},
  doi		= {10.1051/0004-6361/201834219},
  language	= {en},
  urldate	= {2024-05-24},
  journal	= {A\&A},
  author	= {Georgiou, Christos and Johnston, Harry and Hoekstra, Henk
		  and Viola, Massimo and Kuijken, Konrad and Joachimi,
		  Benjamin and Chisari, Nora Elisa and Farrow, Daniel J. and
		  Hildebrandt, Hendrik and Holwerda, Benne W. and Kannawadi,
		  Arun},
  month		= feb,
  year		= {2019},
  pages		= {A90}
}

@Article{	  singh_intrinsic_2016,
  title		= {Intrinsic alignments of {BOSS} {LOWZ} galaxies {II}:
		  {Impact} of shape measurement methods},
  volume	= {457},
  issn		= {0035-8711, 1365-2966},
  shorttitle	= {Intrinsic alignments of {BOSS} {LOWZ} galaxies {II}},
  url		= {http://arxiv.org/abs/1510.06752},
  doi		= {10.1093/mnras/stw144},
  language	= {en},
  number	= {3},
  urldate	= {2022-10-09},
  journal	= {MNRAS},
  author	= {Singh, Sukhdeep and Mandelbaum, Rachel},
  month		= apr,
  year		= {2016},
  note		= {arXiv:1510.06752 },
  pages		= {2301--2317}
}

@article{Krause_IA_2015,
    author = "Krause, Elisabeth and Eifler, Tim and Blazek, Jonathan",
    title = "{The impact of intrinsic alignment on current and future cosmic shear surveys}",
    eprint = "1506.08730",
    archivePrefix = "arXiv",
    primaryClass = "astro-ph.CO",
    doi = "10.1093/mnras/stv2615",
    journal = "\mnras",
    volume = "456",
    number = "1",
    pages = "207--222",
    year = "2016"
}

@Article{	  bridle_dark_2007,
  title		= {Dark energy constraints from cosmic shear power spectra:
		  impact of intrinsic alignments on photometric redshift
		  requirements},
  volume	= {9},
  issn		= {1367-2630},
  shorttitle	= {Dark energy constraints from cosmic shear power spectra},
  url		= {http://arxiv.org/abs/0705.0166},
  doi		= {10.1088/1367-2630/9/12/444},
  language	= {en},
  number	= {12},
  urldate	= {2024-03-08},
  journal	= {New Journal of Physics},
  author	= {Bridle, Sarah and King, Lindsay},
  month		= dec,
  year		= {2007},
  note		= {arXiv:0705.0166 },
  pages		= {444--444}
}

@article{Takahashi:2012em,
    author = "Takahashi, Ryuichi and Sato, Masanori and Nishimichi, Takahiro and Taruya, Atsushi and Oguri, Masamune",
    title = "{Revising the Halofit Model for the Nonlinear Matter Power Spectrum}",
    eprint = "1208.2701",
    archivePrefix = "arXiv",
    primaryClass = "astro-ph.CO",
    doi = "10.1088/0004-637X/761/2/152",
    journal = {\apj},
    volume = "761",
    pages = "152",
    year = "2012"
}

@Article{hirata_intrinsic_2004,
  title		= {Intrinsic alignment-lensing interference as a contaminant
		  of cosmic shear},
  volume	= {70},
  issn		= {1550-7998, 1550-2368},
  url		= {https://link.aps.org/doi/10.1103/PhysRevD.70.063526},
  doi		= {10.1103/PhysRevD.70.063526},
  language	= {en},
  number	= {6},
  urldate	= {2022-10-18},
  journal	= {Phys. Rev. D},
  author	= {Hirata, Christopher M. and Seljak, Uroš},
  month		= sep,
  year		= {2004},
  pages		= {063526}
}

@ARTICLE{Benitez_BPZ_2000,
       author = {{Ben{\'\i}tez}, Narciso},
        title = "{Bayesian Photometric Redshift Estimation}",
      journal = {\apj},
         year = 2000,
        month = jun,
       volume = {536},
       number = {2},
        pages = {571-583},
          doi = {10.1086/308947},
archivePrefix = {arXiv},
       eprint = {astro-ph/9811189},
 primaryClass = {astro-ph},
       adsurl = {https://ui.adsabs.harvard.edu/abs/2000ApJ...536..571B}
}

@misc{Lephare_2011,
       author = {{Arnouts}, S. and {Ilbert}, O.},
        title = "{LePHARE: Photometric Analysis for Redshift Estimate}",
 howpublished = {Astrophysics Source Code Library, record ascl:1108.009},
         year = 2011,
        month = aug,
          eid = {ascl:1108.009},
       adsurl = {https://ui.adsabs.harvard.edu/abs/2011ascl.soft08009A}
}

@Article{	  fortuna_halo_2021,
  title		= {The halo model as a versatile tool to predict intrinsic
		  alignments},
  volume	= {501},
  issn		= {0035-8711, 1365-2966},
  url		= {https://academic.oup.com/mnras/article/501/2/2983/6030045},
  doi		= {10.1093/mnras/staa3802},
  language	= {en},
  number	= {2},
  urldate	= {2022-02-22},
  journal	= {MNRAS},
  author	= {Fortuna, Maria Cristina and Hoekstra, Henk and Joachimi,
		  Benjamin and Johnston, Harry and Chisari, Nora Elisa and
		  Georgiou, Christos and Mahony, Constance},
  month		= jan,
  year		= {2021},
  pages		= {2983--3002}
}

@ARTICLE{getdist,
   author = "Lewis, Antony",
   title = "{GetDist: a Python package for analysing Monte Carlo samples}",
   eprint = "1910.13970",
   archivePrefix = "arXiv",
   primaryClass = "astro-ph.IM",
   doi = "10.1088/1475-7516/2025/08/025",
   journal = "JCAP",
   volume = "08",
   pages = "025",
   year = "2025"
}

@article{schneider.eifler.krause10,
       author = {{Schneider}, P. and {Eifler}, T. and {Krause}, E.},
        title = "{COSEBIs: Extracting the full E-/B-mode information from cosmic shear correlation functions}",
      journal = {\aap},
         year = 2010,
        month = sep,
       volume = {520},
          eid = {A116},
        pages = {A116},
          doi = {10.1051/0004-6361/201014235},
archivePrefix = {arXiv},
       eprint = {1002.2136},
 primaryClass = {astro-ph.CO},
       adsurl = {https://ui.adsabs.harvard.edu/abs/2010A&A...520A.116S}
}

@article{asgari.schneider.simon12,
       author = {{Asgari}, M. and {Schneider}, P. and {Simon}, P.},
        title = "{Cosmic shear tomography and efficient data compression using COSEBIs}",
      journal = {\aap},
         year = 2012,
        month = jun,
       volume = {542},
          eid = {A122},
        pages = {A122},
          doi = {10.1051/0004-6361/201218828},
archivePrefix = {arXiv},
       eprint = {1201.2669},
 primaryClass = {astro-ph.CO},
       adsurl = {https://ui.adsabs.harvard.edu/abs/2012A&A...542A.122A}
}

@ARTICLE{des_y6_cs,
       author = {{DES Collaboration} and {Abbott}, T.~M.~C. and {Aguena}, M. and {Alarcon}, A. and {Alves}, O. and {Amon}, A. and {Anbajagane}, D. and {Andrade-Oliveira}, F. and {d'Assignies}, W. and {Avila}, S. and {Bacon}, D. and {Beas-Gonzalez}, J. and {Bechtol}, K. and {Becker}, M.~R. and {Bernstein}, G.~M. and {Blazek}, J. and {Bocquet}, S. and {Brooks}, D. and {Camacho}, H. and {Camacho-Ciurana}, G. and {Camilleri}, R. and {Campailla}, G. and {Campos}, A. and {Carnero Rosell}, A. and {Carrasco Kind}, M. and {Carretero}, J. and {Castander}, F.~J. and {Cawthon}, R. and {Chang}, C. and {Choi}, A. and {Coloma-Nadal}, J.~M. and {Conselice}, C. and {da Costa}, L.~N. and {Costanzi}, M. and {Crocce}, M. and {Davis}, T.~M. and {De Vicente}, J. and {DePoy}, D.~L. and {DeRose}, J. and {Desai}, S. and {Diehl}, H.~T. and {Doel}, P. and {Doux}, C. and {Drlica-Wagner}, A. and {Eifler}, T.~F. and {Everett}, S. and {Evrard}, A.~E. and {Fert{\'e}}, A. and {Flaugher}, B. and {Fosalba}, P. and {Friedrich}, O. and {Frieman}, J. and {Garc{\'\i}a-Bellido}, J. and {Gatti}, M. and {Giannini}, G. and {Giles}, P. and {Glazebrook}, K. and {Gruen}, D. and {Gruendl}, R.~A. and {Gutierrez}, G. and {Harrison}, I. and {Hartley}, W.~G. and {Herner}, K. and {Hinton}, S.~R. and {Hollowood}, D.~L. and {Honscheid}, K. and {Huterer}, D. and {Jain}, B. and {James}, D.~J. and {Jarvis}, M. and {Jeffrey}, N. and {Jeltema}, T. and {Kacprzak}, T. and {Kent}, S. and {Krause}, E. and {Lahav}, O. and {Lee}, S. and {Legnani}, E. and {Lin}, H. and {Marshall}, J.~L. and {Mau}, S. and {Mena-Fern{\'a}ndez}, J. and {Menanteau}, F. and {Miquel}, R. and {Mohr}, J.~J. and {Muir}, J. and {Myles}, J. and {Nichol}, R.~C. and {Ogando}, R.~L.~C. and {Palmese}, A. and {Paterno}, M. and {Percival}, W.~J. and {Petravick}, D. and {Plazas Malag{\'o}n}, A.~A. and {Porredon}, A. and {Prat}, J. and {Preston}, C. and {Raveri}, M. and {Rodriguez-Monroy}, M. and {Romer}, A.~K. and {Roodman}, A. and {Rykoff}, E.~S. and {Samuroff}, S. and {S{\'a}nchez}, C. and {Sanchez}, E. and {Sanchez Cid}, D. and {Schutt}, T. and {Sevilla-Noarbe}, I. and {Sheldon}, E. and {Shin}, T. and {da Silva Pereira}, M.~E. and {Smith}, M. and {Soares-Santos}, M. and {Suchyta}, E. and {Swanson}, M.~E.~C. and {Tabbutt}, M. and {Tarle}, G. and {Thomas}, D. and {To}, C. and {Troxel}, M.~A. and {Vikram}, V. and {Vincenzi}, M. and {Weaverdyck}, N. and {Weller}, J. and {Wiseman}, P. and {Yamamoto}, M. and {Yanny}, B. and {Yin}, B. and {Zuntz}, J.},
        title = "{Dark Energy Survey Year 6 Results: Cosmological Constraints from Cosmic Shear}",
      journal = {arXiv e-prints},
         year = 2026,
        month = feb,
          eid = {arXiv:2602.10065},
        pages = {arXiv:2602.10065},
          doi = {10.48550/arXiv.2602.10065},
archivePrefix = {arXiv},
       eprint = {2602.10065},
 primaryClass = {astro-ph.CO},
       adsurl = {https://ui.adsabs.harvard.edu/abs/2026arXiv260210065D}
}

@ARTICLE{des-muir,
       author = {{Muir}, J. and {Bernstein}, G.~M. and {Huterer}, D. and {Elsner}, F. and {Krause}, E. and {Roodman}, A. and {Allam}, S. and {Annis}, J. and {Avila}, S. and {Bechtol}, K. and {Bertin}, E. and {Brooks}, D. and {Buckley-Geer}, E. and {Burke}, D.~L. and {Carnero Rosell}, A. and {Carrasco Kind}, M. and {Carretero}, J. and {Cawthon}, R. and {Costanzi}, M. and {da Costa}, L.~N. and {De Vicente}, J. and {Desai}, S. and {Dietrich}, J.~P. and {Doel}, P. and {Eifler}, T.~F. and {Everett}, S. and {Fosalba}, P. and {Frieman}, J. and {Garc{\'\i}a-Bellido}, J. and {Gerdes}, D.~W. and {Gruen}, D. and {Gruendl}, R.~A. and {Gschwend}, J. and {Hartley}, W.~G. and {Hollowood}, D.~L. and {James}, D.~J. and {Jarvis}, M. and {Kuehn}, K. and {Kuropatkin}, N. and {Lahav}, O. and {March}, M. and {Marshall}, J.~L. and {Melchior}, P. and {Menanteau}, F. and {Miquel}, R. and {Ogando}, R.~L.~C. and {Palmese}, A. and {Paz-Chinch{\'o}n}, F. and {Plazas}, A.~A. and {Romer}, A.~K. and {Sanchez}, E. and {Scarpine}, V. and {Schubnell}, M. and {Serrano}, S. and {Sevilla-Noarbe}, I. and {Smith}, M. and {Suchyta}, E. and {Tarle}, G. and {Thomas}, D. and {Troxel}, M.~A. and {Walker}, A.~R. and {Weller}, J. and {Wester}, W. and {Zuntz}, J. and {DES Collaboration}},
        title = "{Blinding multiprobe cosmological experiments}",
      journal = {\mnras},
         year = 2020,
        month = may,
       volume = {494},
       number = {3},
        pages = {4454-4470},
          doi = {10.1093/mnras/staa965},
archivePrefix = {arXiv},
       eprint = {1911.05929},
 primaryClass = {astro-ph.CO},
       adsurl = {https://ui.adsabs.harvard.edu/abs/2020MNRAS.494.4454M}
}

@ARTICLE{Li_Kids_2023,
       author = {{Li}, Shun-Sheng and {Hoekstra}, Henk and {Kuijken}, Konrad and {Asgari}, Marika and {Bilicki}, Maciej and {Giblin}, Benjamin and {Heymans}, Catherine and {Hildebrandt}, Hendrik and {Joachimi}, Benjamin and {Miller}, Lance and {van den Busch}, Jan Luca and {Wright}, Angus H. and {Kannawadi}, Arun and {Reischke}, Robert and {Shan}, HuanYuan},
        title = "{KiDS-1000: Cosmology with improved cosmic shear measurements}",
      journal = {\aap},
         year = 2023,
        month = nov,
       volume = {679},
          eid = {A133},
        pages = {A133},
          doi = {10.1051/0004-6361/202347236},
archivePrefix = {arXiv},
       eprint = {2306.11124},
 primaryClass = {astro-ph.CO},
       adsurl = {https://ui.adsabs.harvard.edu/abs/2023A&A...679A.133L}
}

@ARTICLE{kidslegacy_catalogue,
       author = {{Wright}, Angus H. and {Kuijken}, Konrad and {Hildebrandt}, Hendrik and {Radovich}, Mario and {Bilicki}, Maciej and {Dvornik}, Andrej and {Getman}, Fedor and {Heymans}, Catherine and {Hoekstra}, Henk and {Li}, Shun-Sheng and {Miller}, Lance and {Napolitano}, Nicola R. and {Xia}, Qianli and {Asgari}, Marika and {Brescia}, Massimo and {Buddelmeijer}, Hugo and {Burger}, Pierre and {Castignani}, Gianluca and {Cavuoti}, Stefano and {de Jong}, Jelte and {Edge}, Alastair and {Giblin}, Benjamin and {Giocoli}, Carlo and {Harnois-D{\'e}raps}, Joachim and {Jalan}, Priyanka and {Joachimi}, Benjamin and {John William}, Anjitha and {Joudaki}, Shahab and {Kannawadi}, Arun and {Kaur}, Gursharanjit and {La Barbera}, Francesco and {Linke}, Laila and {Mahony}, Constance and {Maturi}, Matteo and {Moscardini}, Lauro and {Nakoneczny}, Szymon J. and {Paolillo}, Maurizio and {Porth}, Lucas and {Puddu}, Emanuella and {Reischke}, Robert and {Schneider}, Peter and {Sereno}, Mauro and {Shan}, HuanYuan and {Sif{\'o}n}, Crist{\'o}bal and {St{\"o}lzner}, Benjamin and {Tr{\"o}ster}, Tilman and {Valentijn}, Edwin and {van den Busch}, Jan Luca and {Verdoes Kleijn}, Gijs and {Wittje}, Anna and {Yan}, Ziang and {Yao}, Ji and {Yoon}, Mijin and {Zhang}, Yun-Hao},
        title = "{The fifth data release of the Kilo Degree Survey: Multi-epoch optical/NIR imaging covering wide and legacy-calibration fields}",
      journal = {\aap},
         year = 2024,
        month = jun,
       volume = {686},
          eid = {A170},
        pages = {A170},
          doi = {10.1051/0004-6361/202346730},
archivePrefix = {arXiv},
       eprint = {2503.19439},
 primaryClass = {astro-ph.GA},
       adsurl = {https://ui.adsabs.harvard.edu/abs/2024A&A...686A.170W}
}

@article{Wright_kids_2025,
       author = {{Wright}, Angus H. and {St{\"o}lzner}, Benjamin and {Asgari}, Marika and {Bilicki}, Maciej and {Giblin}, Benjamin and {Heymans}, Catherine and {Hildebrandt}, Hendrik and {Hoekstra}, Henk and {Joachimi}, Benjamin and {Kuijken}, Konrad and {Li}, Shun-Sheng and {Reischke}, Robert and {von Wietersheim-Kramsta}, Maximilian and {Yoon}, Mijin and {Burger}, Pierre and {Chisari}, Nora Elisa and {de Jong}, Jelte and {Dvornik}, Andrej and {Georgiou}, Christos and {Harnois-D{\'e}raps}, Joachim and {Jalan}, Priyanka and {William}, Anjitha John and {Joudaki}, Shahab and {Lesci}, Giorgio Francesco and {Linke}, Laila and {Loureiro}, Arthur and {Mahony}, Constance and {Maturi}, Matteo and {Miller}, Lance and {Moscardini}, Lauro and {Napolitano}, Nicola R. and {Porth}, Lucas and {Radovich}, Mario and {Schneider}, Peter and {Tr{\"o}ster}, Tilman and {Wittje}, Anna and {Yan}, Ziang and {Zhang}, Yun-Hao},
        title = "{KiDS-Legacy: Cosmological constraints from cosmic shear with the complete Kilo-Degree Survey}",
      journal = {arXiv e-prints},
         year = 2025,
        month = mar,
          eid = {arXiv:2503.19441},
        pages = {arXiv:2503.19441},
          doi = {10.48550/arXiv.2503.19441},
archivePrefix = {arXiv},
       eprint = {2503.19441},
 primaryClass = {astro-ph.CO},
       adsurl = {https://ui.adsabs.harvard.edu/abs/2025arXiv250319441W}
}

@Article{	  blazek_beyond_2019,
  title		= {Beyond linear galaxy alignments},
  volume	= {100},
  issn		= {2470-0010, 2470-0029},
  url		= {https://link.aps.org/doi/10.1103/PhysRevD.100.103506},
  doi		= {10.1103/PhysRevD.100.103506},
  language	= {en},
  number	= {10},
  urldate	= {2023-02-24},
  journal	= {Phys. Rev. D},
  author	= {Blazek, Jonathan A. and MacCrann, Niall and Troxel,
		  M. A. and Fang, Xiao},
  month		= nov,
  year		= {2019},
  pages		= {103506}
}

@article{Hervas_Peters_IA_2024,
       author = {{Hervas Peters}, Fabian and {Kilbinger}, Martin and {Paviot}, Romain and {Baumont}, Lucie and {Russier}, Elisa and {Zhang}, Ziwen and {Murray}, Calum and {Pettorino}, Valeria and {de Boer}, Thomas and {Fabbro}, S{\'e}bastien and {Guerrini}, Sacha and {Hildebrandt}, Hendrik and {Hudson}, Michael J. and {Van Waerbeke}, Ludovic and {Wittje}, Anna},
        title = "{UNIONS: A direct measurement of intrinsic alignment with BOSS/eBOSS spectroscopy}",
      journal = {\aap},
         year = 2025,
        month = jul,
       volume = {699},
          eid = {A201},
        pages = {A201},
          doi = {10.1051/0004-6361/202453442},
archivePrefix = {arXiv},
       eprint = {2412.01790},
 primaryClass = {astro-ph.CO},
       adsurl = {https://ui.adsabs.harvard.edu/abs/2025A&A...699A.201H}
}

@article{Navarro-Girones_pau_2025,
       author = {{Navarro-Giron{\'e}s}, D. and {Gazta{\~n}aga}, M. Crocce E. and {Wittje}, A. and {Siudek}, M. and {Hoekstra}, H. and {Hildebrandt}, H. and {Joachimi}, B. and {Paviot}, R. and {Baugh}, C.~M. and {Carretero}, J. and {Casas}, R. and {Castander}, F.~J. and {Eriksen}, M. and {Fernandez}, E. and {Fosalba}, P. and {Garc{\'\i}a-Bellido}, J. and {Miquel}, R. and {Padilla}, C. and {Renard}, P. and {S{\'a}nchez}, E. and {Serrano}, S. and {Sevilla-Noarbe}, I. and {Tallada-Cresp{\'\i}}, P.},
        title = "{The PAU Survey: Measuring intrinsic galaxy alignments in deep wide fields as a function of colour, luminosity, stellar mass and redshift}",
      journal = {\mnras},
         year = 2025,
        month = sep,
          doi = {10.1093/mnras/staf1630},
archivePrefix = {arXiv},
       eprint = {2505.15470},
 primaryClass = {astro-ph.CO},
       adsurl = {https://ui.adsabs.harvard.edu/abs/2025MNRAS.tmp.1548N}
}

@ARTICLE{des_y6,
       author = {{Bechtol}, K. and {Sevilla-Noarbe}, I. and {Drlica-Wagner}, A. and {Yanny}, B. and {Gruendl}, R.~A. and {Sheldon}, E. and {Rykoff}, E.~S. and {De Vicente}, J. and {Adamow}, M. and {Anbajagane}, D. and {Becker}, M.~R. and {Bernstein}, G.~M. and {Carnero Rosell}, A. and {Gschwend}, J. and {Gorsuch}, M. and {Hartley}, W.~G. and {Jarvis}, M. and {Jeltema}, T. and {Kron}, R. and {Manning}, T.~A. and {O'Donnell}, J. and {Pieres}, A. and {Rodr{\'\i}guez-Monroy}, M. and {Sanchez Cid}, D. and {Tabbutt}, M. and {Toribio San Cipriano}, L. and {Tucker}, D.~L. and {Weaverdyck}, N. and {Yamamoto}, M. and {Abbott}, T.~M.~C. and {Aguena}, M. and {Alarc{\'o}n}, A. and {Allam}, S. and {Amon}, A. and {Andrade-Oliveira}, F. and {Avila}, S. and {Bernardinelli}, P.~H. and {Bertin}, E. and {Blazek}, J. and {Brooks}, D. and {Burke}, D.~L. and {Carretero}, J. and {Castander}, F.~J. and {Cawthon}, R. and {Chang}, C. and {Choi}, A. and {Conselice}, C. and {Costanzi}, M. and {Crocce}, M. and {da Costa}, L.~N. and {Davis}, T.~M. and {Desai}, S. and {Diehl}, H.~T. and {Dodelson}, S. and {Doel}, P. and {Doux}, C. and {Fert{\'e}}, A. and {Flaugher}, B. and {Fosalba}, P. and {Frieman}, J. and {Garc{\'\i}a-Bellido}, J. and {Gatti}, M. and {Gaztanaga}, E. and {Giannini}, G. and {Gruen}, D. and {Gutierrez}, G. and {Herner}, K. and {Hinton}, S.~R. and {Hollowood}, D.~L. and {Honscheid}, K. and {Huterer}, D. and {Jeffrey}, N. and {Krause}, E. and {Kuehn}, K. and {Lahav}, O. and {Lee}, S. and {Lidman}, C. and {Lima}, M. and {Lin}, H. and {Marshall}, J.~L. and {Mena-Fern{\'a}ndez}, J. and {Miquel}, R. and {Mohr}, J.~J. and {Muir}, J. and {Myles}, J. and {Ogando}, R.~L.~C. and {Palmese}, A. and {Plazas Malag{\'o}n}, A.~A. and {Porredon}, A. and {Prat}, J. and {Raveri}, M. and {Romer}, A.~K. and {Roodman}, A. and {Samuroff}, S. and {Sanchez}, E. and {Scarpine}, V. and {Smith}, M. and {Soares-Santos}, M. and {Suchyta}, E. and {Tarle}, G. and {Troxel}, M.~A. and {Vikram}, V. and {Walker}, A.~R. and {Weller}, J. and {Wiseman}, P. and {Zhang}, Y.},
        title = "{Dark Energy Survey Year 6 Results: Photometric Data Set for Cosmology}",
      journal = {arXiv e-prints},
         year = 2025,
        month = jan,
          eid = {arXiv:2501.05739},
        pages = {arXiv:2501.05739},
          doi = {10.48550/arXiv.2501.05739},
archivePrefix = {arXiv},
       eprint = {2501.05739},
 primaryClass = {astro-ph.CO},
       adsurl = {https://ui.adsabs.harvard.edu/abs/2025arXiv250105739B}
}

@Article{	  samuroff_dark_2023,
  title		= {The {Dark} {Energy} {Survey} {Year} 3 and {eBOSS}:
		  constraining galaxy intrinsic alignments across luminosity
		  and colour space},
  volume	= {524},
  issn		= {0035-8711},
  shorttitle	= {The {Dark} {Energy} {Survey} {Year} 3 and {eBOSS}},
  url		= {https://doi.org/10.1093/mnras/stad2013},
  doi		= {10.1093/mnras/stad2013},
  number	= {2},
  year      ={2023},
  urldate	= {2024-05-02},
  journal	= {MNRAS},
  author	= {Samuroff, S and Mandelbaum, R and Blazek, J and Campos, A
		  and MacCrann, N and Zacharegkas, G and Amon, A and Prat, J
		  and Singh, S and Elvin-Poole, J and Ross, A J and Alarcon,
		  A and Baxter, E and Bechtol, K and Becker, M R and
		  Bernstein, G M and Rosell, A Carnero and Kind, M Carrasco
		  and Cawthon, R and Chang, C and Chen, R and Choi, A and
		  Crocce, M and Davis, C and DeRose, J and Dodelson, S and
		  Doux, C and Drlica-Wagner, A and Eckert, K and Everett, S
		  and Ferté, A and Gatti, M and Giannini, G and Gruen, D and
		  Gruendl, R A and Harrison, I and Herner, K and Huff, E M
		  and Jarvis, M and Kuropatkin, N and Leget, P-F and Lemos, P
		  and McCullough, J and Myles, J and Navarro-Alsina, A and
		  Pandey, S and Porredon, A and Raveri, M and
		  Rodriguez-Monroy, M and Rollins, R P and Roodman, A and
		  Rossi, G and Rykoff, E S and Sánchez, C and Secco, L F and
		  Sevilla-Noarbe, I and Sheldon, E and Shin, T and Troxel, M
		  A and Tutusaus, I and Weaverdyck, N and Yanny, B and Yin, B
		  and Zhang, Y and Zuntz, J and Aguena, M and Alves, O and
		  Annis, J and Bacon, D and Bertin, E and Bocquet, S and
		  Brooks, D and Burke, D L and Carretero, J and Costanzi, M
		  and da Costa, L N and Pereira, M E S and De Vicente, J
		  and Desai, S and Diehl, H T and Dietrich, J P and Doel, P
		  and Ferrero, I and Flaugher, B and Frieman, J and
		  García-Bellido, J and Hinton, S R and Hollowood, D L and
		  Honscheid, K and James, D J and Kuehn, K and Lahav, O and
		  Marshall, J L and Melchior, P and Mena-Fernández, J and
		  Menanteau, F and Miquel, R and Newman, J and Palmese, A and
		  Pieres, A and Malagón, A A Plazas and Sanchez, E and
		  Scarpine, V and Smith, M and Suchyta, E and Swanson, M E C
		  and Tarle, G and To, C and {(DES Collaboration)}},
  month		= sep,
  pages		= {2195--2223}
}

@Article{	  mandelbaum_wigglez_2011,
  title		= {The {WiggleZ} {Dark} {Energy} {Survey}: direct constraints
		  on blue galaxy intrinsic alignments at intermediate
		  redshifts: {WiggleZ} intrinsic alignments},
  volume	= {410},
  issn		= {00358711},
  shorttitle	= {The {WiggleZ} {Dark} {Energy} {Survey}},
  url		= {https://academic.oup.com/mnras/article-lookup/doi/10.1111/j.1365-2966.2010.17485.x},
  doi		= {10.1111/j.1365-2966.2010.17485.x},
  language	= {en},
  number	= {2},
  urldate	= {2024-09-19},
  journal	= {MNRAS},
  author	= {Mandelbaum, Rachel and Blake, Chris and Bridle, Sarah and
		  Abdalla, Filipe B. and Brough, Sarah and Colless, Matthew
		  and Couch, Warrick and Croom, Scott and Davis, Tamara and
		  Drinkwater, Michael J. and Forster, Karl and Glazebrook,
		  Karl and Jelliffe, Ben and Jurek, Russell J. and Li, I-hui
		  and Madore, Barry and Martin, Chris and Pimbblet, Kevin and
		  Poole, Gregory B. and Pracy, Michael and Sharp, Rob and
		  Wisnioski, Emily and Woods, David and Wyder, Ted},
  month		= jan,
  year		= {2011},
  pages		= {844--859}
}

@Article{	  fortuna_kids-1000_2021,
  title		= {{KiDS}-1000: {Constraints} on the intrinsic alignment of
		  luminous red galaxies},
  volume	= {654},
  issn		= {0004-6361, 1432-0746},
  shorttitle	= {{KiDS}-1000},
  url		= {http://arxiv.org/abs/2109.02556},
  doi		= {10.1051/0004-6361/202140706},
  language	= {en},
  urldate	= {2023-03-15},
  journal	= {A\&A},
  author	= {Fortuna, Maria Cristina and Hoekstra, Henk and Johnston,
		  Harry and Vakili, Mohammadjavad and Kannawadi, Arun and
		  Georgiou, Christos and Joachimi, Benjamin and Wright, Angus
		  H. and Asgari, Marika and Bilicki, Maciej and Heymans,
		  Catherine and Hildebrandt, Hendrik and Kuijken, Konrad and
		  Von Wietersheim-Kramsta, Maximilian},
  month		= oct,
  year		= {2021},
  note		= {arXiv:2109.02556 },
  pages		= {A76}
}

@Article{	  johnston_kidsgama_2019,
  title		= {{KiDS}+{GAMA}: {Intrinsic} alignment model constraints for
		  current and future weak lensing cosmology},
  volume	= {624},
  issn		= {0004-6361},
  shorttitle	= {{KiDS}+{GAMA}},
  url		= {https://ui.adsabs.harvard.edu/abs/2019A&A...624A..30J/abstract},
  doi		= {10.1051/0004-6361/201834714},
  language	= {en},
  urldate	= {2022-10-31},
  journal	= {\aap},
  author	= {Johnston, Harry and Georgiou, Christos and Joachimi,
		  Benjamin and Hoekstra, Henk and Chisari, Nora Elisa and
		  Farrow, Daniel and Fortuna, Maria Cristina and Heymans,
		  Catherine and Joudaki, Shahab and Kuijken, Konrad and
		  Wright, Angus},
  month		= apr,
  year		= {2019},
  pages		= {A30}
}

@Article{	  singh_intrinsic_2015,
  title		= {Intrinsic alignments of {SDSS}-{III} {BOSS} {LOWZ} sample
		  galaxies},
  volume	= {450},
  issn		= {1365-2966, 0035-8711},
  url		= {http://arxiv.org/abs/1411.1755},
  doi		= {10.1093/mnras/stv778},
  language	= {en},
  number	= {2},
  urldate	= {2023-02-23},
  journal	= {MNRAS},
  author	= {Singh, Sukhdeep and Mandelbaum, Rachel and More, Surhud},
  month		= jun,
  year		= {2015},
  note		= {arXiv:1411.1755 },
  pages		= {2195--2216}
}

@Article{joachimi_constraints_2011,
  title		= {Constraints on intrinsic alignment contamination of weak
		  lensing surveys using the {MegaZ}-{LRG} sample},
  volume	= {527},
  issn		= {0004-6361, 1432-0746},
  url		= {http://arxiv.org/abs/1008.3491},
  doi		= {10.1051/0004-6361/201015621},
  urldate	= {2022-10-19},
  journal	= {A\&A},
  author	= {Joachimi, B. and Mandelbaum, R. and Abdalla, F. B. and
		  Bridle, S. L.},
  month		= mar,
  year		= {2011},
  note		= {arXiv:1008.3491 },
  pages		= {A26}
}

@article{Secco_DES_2022,
       author = {{Secco}, L.~F. and {Samuroff}, S. and {Krause}, E. and {Jain}, B. and {Blazek}, J. and {Raveri}, M. and {Campos}, A. and {Amon}, A. and {Chen}, A. and {Doux}, C. and {Choi}, A. and {Gruen}, D. and {Bernstein}, G.~M. and {Chang}, C. and {DeRose}, J. and {Myles}, J. and {Fert{\'e}}, A. and {Lemos}, P. and {Huterer}, D. and {Prat}, J. and {Troxel}, M.~A. and {MacCrann}, N. and {Liddle}, A.~R. and {Kacprzak}, T. and {Fang}, X. and {S{\'a}nchez}, C. and {Pandey}, S. and {Dodelson}, S. and {Chintalapati}, P. and {Hoffmann}, K. and {Alarcon}, A. and {Alves}, O. and {Andrade-Oliveira}, F. and {Baxter}, E.~J. and {Bechtol}, K. and {Becker}, M.~R. and {Brandao-Souza}, A. and {Camacho}, H. and {Carnero Rosell}, A. and {Carrasco Kind}, M. and {Cawthon}, R. and {Cordero}, J.~P. and {Crocce}, M. and {Davis}, C. and {Di Valentino}, E. and {Drlica-Wagner}, A. and {Eckert}, K. and {Eifler}, T.~F. and {Elidaiana}, M. and {Elsner}, F. and {Elvin-Poole}, J. and {Everett}, S. and {Fosalba}, P. and {Friedrich}, O. and {Gatti}, M. and {Giannini}, G. and {Gruendl}, R.~A. and {Harrison}, I. and {Hartley}, W.~G. and {Herner}, K. and {Huang}, H. and {Huff}, E.~M. and {Jarvis}, M. and {Jeffrey}, N. and {Kuropatkin}, N. and {Leget}, P. -F. and {Muir}, J. and {Mccullough}, J. and {Navarro Alsina}, A. and {Omori}, Y. and {Park}, Y. and {Porredon}, A. and {Rollins}, R. and {Roodman}, A. and {Rosenfeld}, R. and {Ross}, A.~J. and {Rykoff}, E.~S. and {Sanchez}, J. and {Sevilla-Noarbe}, I. and {Sheldon}, E.~S. and {Shin}, T. and {Troja}, A. and {Tutusaus}, I. and {Varga}, T.~N. and {Weaverdyck}, N. and {Wechsler}, R.~H. and {Yanny}, B. and {Yin}, B. and {Zhang}, Y. and {Zuntz}, J. and {Abbott}, T.~M.~C. and {Aguena}, M. and {Allam}, S. and {Annis}, J. and {Bacon}, D. and {Bertin}, E. and {Bhargava}, S. and {Bridle}, S.~L. and {Brooks}, D. and {Buckley-Geer}, E. and {Burke}, D.~L. and {Carretero}, J. and {Costanzi}, M. and {da Costa}, L.~N. and {De Vicente}, J. and {Diehl}, H.~T. and {Dietrich}, J.~P. and {Doel}, P. and {Ferrero}, I. and {Flaugher}, B. and {Frieman}, J. and {Garc{\'\i}a-Bellido}, J. and {Gaztanaga}, E. and {Gerdes}, D.~W. and {Giannantonio}, T. and {Gschwend}, J. and {Gutierrez}, G. and {Hinton}, S.~R. and {Hollowood}, D.~L. and {Honscheid}, K. and {Hoyle}, B. and {James}, D.~J. and {Jeltema}, T. and {Kuehn}, K. and {Lahav}, O. and {Lima}, M. and {Lin}, H. and {Maia}, M.~A.~G. and {Marshall}, J.~L. and {Martini}, P. and {Melchior}, P. and {Menanteau}, F. and {Miquel}, R. and {Mohr}, J.~J. and {Morgan}, R. and {Ogando}, R.~L.~C. and {Palmese}, A. and {Paz-Chinch{\'o}n}, F. and {Petravick}, D. and {Pieres}, A. and {Plazas Malag{\'o}n}, A.~A. and {Rodriguez-Monroy}, M. and {Romer}, A.~K. and {Sanchez}, E. and {Scarpine}, V. and {Schubnell}, M. and {Scolnic}, D. and {Serrano}, S. and {Smith}, M. and {Soares-Santos}, M. and {Suchyta}, E. and {Swanson}, M.~E.~C. and {Tarle}, G. and {Thomas}, D. and {To}, C. and {DES Collaboration}},
        title = "{Dark Energy Survey Year 3 results: Cosmology from cosmic shear and robustness to modeling uncertainty}",
      journal = {\prd},
         year = 2022,
        month = jan,
       volume = {105},
       number = {2},
          eid = {023515},
        pages = {023515},
          doi = {10.1103/PhysRevD.105.023515},
archivePrefix = {arXiv},
       eprint = {2105.13544},
 primaryClass = {astro-ph.CO},
       adsurl = {https://ui.adsabs.harvard.edu/abs/2022PhRvD.105b3515S}
}

@ARTICLE{hsc_clustering_redshift,
       author = {{Choppin de Janvry}, J. and {Dai}, B. and {Gontcho}, S. Gontcho A and {Seljak}, U. and {Zhang}, T.},
        title = "{Cosmic Shear constraints from HSC Year 3 with clustering calibration of the tomographic redshift distributions from DESI}",
      journal = {arXiv e-prints},
         year = 2025,
        month = nov,
          eid = {arXiv:2511.18134},
        pages = {arXiv:2511.18134},
          doi = {10.48550/arXiv.2511.18134},
archivePrefix = {arXiv},
       eprint = {2511.18134},
 primaryClass = {astro-ph.CO},
       adsurl = {https://ui.adsabs.harvard.edu/abs/2025arXiv251118134C}
}

@Article{asgari_kids-1000_2021,
  title		= {{KiDS}-1000 {Cosmology}: {Cosmic} shear constraints and
		  comparison between two point statistics},
  volume	= {645},
  issn		= {0004-6361, 1432-0746},
  shorttitle	= {{KiDS}-1000 {Cosmology}},
  url		= {http://arxiv.org/abs/2007.15633},
  doi		= {10.1051/0004-6361/202039070},
  language	= {en},
  urldate	= {2021-06-14},
  journal	= {A\&A},
  author	= {Asgari, Marika and Lin, Chieh-An and Joachimi, Benjamin
		  and Giblin, Benjamin and Heymans, Catherine and
		  Hildebrandt, Hendrik and Kannawadi, Arun and Stölzner,
		  Benjamin and Tröster, Tilman and Busch, Jan Luca van den
		  and Wright, Angus H. and Bilicki, Maciej and Blake, Chris
		  and de Jong, Jelte and Dvornik, Andrej and Erben, Thomas
		  and Getman, Fedor and Hoekstra, Henk and Köhlinger, Fabian
		  and Kuijken, Konrad and Miller, Lance and Radovich, Mario
		  and Schneider, Peter and Shan, HuanYuan and Valentijn,
		  Edwin},
  month		= jan,
  year		= {2021},
  note		= {arXiv: 2007.15633},
  pages		= {A104}
}

@ARTICLE{Joachimi_nulling_2010,
       author = {{Joachimi}, B. and {Schneider}, P.},
        title = "{Controlling intrinsic alignments in weak lensing statistics: The nulling and boosting techniques}",
      journal = {arXiv e-prints},
         year = 2010,
        month = sep,
          eid = {arXiv:1009.2024},
        pages = {arXiv:1009.2024},
          doi = {10.48550/arXiv.1009.2024},
archivePrefix = {arXiv},
       eprint = {1009.2024},
 primaryClass = {astro-ph.CO},
       adsurl = {https://ui.adsabs.harvard.edu/abs/2010arXiv1009.2024J}
}

@article{Bartelmann_1999yn,
    author = "Bartelmann, M. and Schneider, P.",
    title = "{Weak gravitational lensing}",
    eprint = "astro-ph/9912508",
    archivePrefix = "arXiv",
    doi = "10.1016/S0370-1573(00)00082-X",
    journal = "Phys. Rept.",
    volume = "340",
    pages = "291--472",
    year = "2001"
}

@article{Kilbinger_2014cea,
    author = "Kilbinger, M.",
    title = "{Cosmology with cosmic shear observations: a review}",
    eprint = "1411.0115",
    archivePrefix = "arXiv",
    primaryClass = "astro-ph.CO",
    doi = "10.1088/0034-4885/78/8/086901",
    journal = "Rept. Prog. Phys.",
    volume = "78",
    pages = "086901",
    year = "2015"
}

@article{Schneider:2002jd,
    author = "Schneider, Peter and van Waerbeke, Ludovic and Kilbinger, Martin and Mellier, Yannick",
    title = "{Analysis of two-point statistics of cosmic shear: I. estimators and covariances}",
    eprint = "astro-ph/0206182",
    archivePrefix = "arXiv",
    doi = "10.1051/0004-6361:20021341",
    journal = {\aap},
    volume = "396",
    pages = "1--20",
    year = "2002"
}

@ARTICLE{Kaiser1992,
       author = {{Kaiser}, Nick},
        title = "{Weak Gravitational Lensing of Distant Galaxies}",
      journal = {\apj},
         year = 1992,
        month = apr,
       volume = {388},
        pages = {272},
          doi = {10.1086/171151},
       adsurl = {https://ui.adsabs.harvard.edu/abs/1992ApJ...388..272K}
}

@article{Schoneberg:2024ifp,
    author = {Sch\"oneberg, Nils},
    title = "{The 2024 BBN baryon abundance update}",
    eprint = "2401.15054",
    archivePrefix = "arXiv",
    primaryClass = "astro-ph.CO",
    doi = "10.1088/1475-7516/2024/06/006",
    journal = "JCAP",
    volume = "06",
    pages = "006",
    year = "2024"
}

@article{Mead:2020vgs,
       author = {{Mead}, A.~J. and {Brieden}, S. and {Tr{\"o}ster}, T. and {Heymans}, C.},
        title = "{HMCODE-2020: improved modelling of non-linear cosmological power spectra with baryonic feedback}",
      journal = {\mnras},
         year = 2021,
        month = mar,
       volume = {502},
       number = {1},
        pages = {1401-1422},
          doi = {10.1093/mnras/stab082},
archivePrefix = {arXiv},
       eprint = {2009.01858},
 primaryClass = {astro-ph.CO},
       adsurl = {https://ui.adsabs.harvard.edu/abs/2021MNRAS.502.1401M}
}

@ARTICLE{decade,
       author = {{Anbajagane}, D. and {Chang}, C. and {Drlica-Wagner}, A. and {Tan}, C.~Y. and {Adamow}, M. and {Gruendl}, R.~A. and {Secco}, L.~F. and {Zhang}, Z. and {Becker}, M.~R. and {Ferguson}, P.~S. and {Chicoine}, N. and {Herron}, K. and {Alarcon}, A. and {Teixeira}, R. and {Suson}, D. and {Alsina}, A.~N. and {Amon}, A. and {Andrade-Oliveira}, F. and {Blazek}, J. and {Bom}, C.~R. and {Camacho}, H. and {Carballo-Bello}, J.~A. and {Carnero Rosell}, A. and {Cawthon}, R. and {Cerny}, W. and {Choi}, A. and {Choi}, Y. and {Dodelson}, S. and {Doux}, C. and {Eckert}, K. and {Elvin-Poole}, J. and {Esteves}, J. and {Gatti}, M. and {Giannini}, G. and {Gruen}, D. and {Hartley}, W.~G. and {Herner}, K. and {Huff}, E.~M. and {James}, D.~J. and {Jarvis}, M. and {Krause}, E. and {Kuropatkin}, N. and {Mart{\'\i}nez-V{\'a}zquez}, C.~E. and {Massana}, P. and {Mau}, S. and {McCullough}, J. and {Medina}, G.~E. and {Mutlu-Pakdil}, B. and {Myles}, J. and {Navabi}, M. and {No{\"e}l}, N.~E.~D. and {Pace}, A.~B. and {Porredon}, A. and {Prat}, J. and {Raveri}, M. and {Riley}, A.~H. and {Rykoff}, E.~S. and {Sakowska}, J.~D. and {Samuroff}, S. and {Sanchez-Cid}, D. and {Sand}, D.~J. and {Santana-Silva}, L. and {Sevilla-Noarbe}, I. and {Shin}, T. and {Soares-Santos}, M. and {Stringfellow}, G.~S. and {To}, C. and {Tong}, A. and {Troxel}, M.~A. and {Vivas}, A.~K. and {Yamamoto}, M. and {Yanny}, B. and {Yin}, B. and {Zhang}, Y. and {Zuntz}, J.},
            title = "{The DECADE cosmic shear project IV: cosmological constraints from 107 million galaxies across 5,400 deg$^2$ of the sky}",
    eprint = "2502.17677",
    archivePrefix = "arXiv",
    primaryClass = "astro-ph.CO",
    reportNumber = "FERMILAB-PUB-25-0066-LDRD-PPD",
    doi = "10.33232/001c.146161",
    journal = "Open J. Astrophys.",
    volume = "8",
    pages = "146161",
    year = "2025"
}

@ARTICLE{planck_data,
       author = {{Planck Collaboration}},
        title = "{Planck 2018 results. I. Overview and the cosmological legacy of Planck}",
      journal = {\aap},
         year = 2020,
        month = sep,
       volume = {641},
          eid = {A1},
        pages = {A1},
          doi = {10.1051/0004-6361/201833880},
archivePrefix = {arXiv},
       eprint = {1807.06205},
 primaryClass = {astro-ph.CO},
       adsurl = {https://ui.adsabs.harvard.edu/abs/2020A&A...641A...1P}
}

@ARTICLE{plike_lite,
       author = {{Prince}, Heather and {Dunkley}, Jo},
        title = "{Data compression in cosmology: A compressed likelihood for Planck data}",
      journal = {\prd},
         year = 2019,
        month = oct,
       volume = {100},
       number = {8},
          eid = {083502},
        pages = {083502},
          doi = {10.1103/PhysRevD.100.083502},
archivePrefix = {arXiv},
       eprint = {1909.05869},
 primaryClass = {astro-ph.CO},
       adsurl = {https://ui.adsabs.harvard.edu/abs/2019PhRvD.100h3502P}
}

@ARTICLE{desi_dr2_1,
       author = {{Abdul Karim}, M. and {Aguilar}, J. and {Ahlen}, S. and {Allende Prieto}, C. and {Alves}, O. and {Anand}, A. and {Andrade}, U. and {Armengaud}, E. and {Aviles}, A. and {Bailey}, S. and {Bault}, A. and {Behera}, J. and {BenZvi}, S. and {Bianchi}, D. and {Blake}, C. and {Brodzeller}, A. and {Brooks}, D. and {Buckley-Geer}, E. and {Burtin}, E. and {Calderon}, R. and {Canning}, R. and {Carnero Rosell}, A. and {Carrilho}, P. and {Casas}, L. and {Castander}, F.~J. and {Cereskaite}, R. and {Charles}, M. and {Chaussidon}, E. and {Chaves-Montero}, J. and {Chebat}, D. and {Claybaugh}, T. and {Cole}, S. and {Cooper}, A.~P. and {Cuceu}, A. and {Dawson}, K.~S. and {de Belsunce}, R. and {de la Macorra}, A. and {de Mattia}, A. and {Deiosso}, N. and {Della Costa}, J. and {Dey}, A. and {Dey}, B. and {Ding}, Z. and {Doel}, P. and {Edelstein}, J. and {Eisenstein}, D.~J. and {Elbers}, W. and {Fagrelius}, P. and {Fanning}, K. and {Ferraro}, S. and {Font-Ribera}, A. and {Forero-Romero}, J.~E. and {Garcia-Quintero}, C. and {Garrison}, L.~H. and {Gazta{\~n}aga}, E. and {Gil-Mar{\'\i}n}, H. and {Gontcho A Gontcho}, S. and {Gonzalez-Morales}, A.~X. and {Gordon}, C. and {Green}, D. and {Gutierrez}, G. and {Guy}, J. and {Hahn}, C. and {Herbold}, M. and {Herrera-Alcantar}, H.~K. and {Ho}, M. and {Ho}, M.-F. and {Honscheid}, K. and {Howlett}, C. and {Huterer}, D. and {Ishak}, M. and {Juneau}, S. and {Kara{\c{c}}ayl{\i}}, N.~G. and {Kehoe}, R. and {Kent}, S. and {Kirkby}, D. and {Kisner}, T. and {Kitaura}, F.-S. and {Koposov}, S.~E. and {Kremin}, A. and {Lahav}, O. and {Lamman}, C. and {Landriau}, M. and {Lang}, D. and {Lasker}, J. and {Le Goff}, J.~M. and {Le Guillou}, L. and {Leauthaud}, A. and {Levi}, M.~E. and {Li}, Q. and {Li}, T.~S. and {Lodha}, K. and {Lokken}, M. and {Magneville}, C. and {Manera}, M. and {Martini}, P. and {Matthewson}, W.~L. and {McDonald}, P. and {Meisner}, A. and {Mena-Fern{\'a}ndez}, J. and {Miquel}, R. and {Moustakas}, J. and {Mu{\~n}oz-Guti{\'e}rrez}, A. and {Mu{\~n}oz-Santos}, D. and {Myers}, A.~D. and {Newman}, J.~A. and {Niz}, G. and {Noriega}, H.~E. and {Paillas}, E. and {Palanque-Delabrouille}, N. and {Pan}, J. and {Percival}, W.~J. and {P{\'e}rez-R{\`a}fols}, I. and {Pieri}, M.~M. and {Poppett}, C. and {Prada}, F. and {Rabinowitz}, D. and {Raichoor}, A. and {Ram{\'\i}rez-P{\'e}rez}, C. and {Rashkovetskyi}, M. and {Ravoux}, C. and {Rich}, J. and {Rockosi}, C. and {Ross}, A.~J. and {Rossi}, G. and {Ruhlmann-Kleider}, V. and {Sanchez}, E. and {Sanders}, N. and {Satyavolu}, S. and {Schlegel}, D. and {Schubnell}, M. and {Seo}, H. and {Shafieloo}, A. and {Sharples}, R. and {Silber}, J. and {Sinigaglia}, F. and {Sprayberry}, D. and {Tan}, T. and {Tarl{\'e}}, G. and {Taylor}, P. and {Turner}, W. and {Valdes}, F. and {Vargas-Maga{\~n}a}, M. and {Walther}, M. and {Weaver}, B.~A. and {Wolfson}, M. and {Y{\`e}che}, C. and {Zarrouk}, P. and {Zhou}, R. and {Zou}, H. and {DESI Collaboration}},
        title = "{DESI DR2 results. I. Baryon acoustic oscillations from the Lyman alpha forest}",
      journal = {\prd},
         year = 2025,
        month = oct,
       volume = {112},
       number = {8},
          eid = {083514},
        pages = {083514},
          doi = {10.1103/2wwn-xjm5},
archivePrefix = {arXiv},
       eprint = {2503.14739},
 primaryClass = {astro-ph.CO},
       adsurl = {https://ui.adsabs.harvard.edu/abs/2025PhRvD.112h3514A}
}

@ARTICLE{desi_dr2_2,
       author = {{Abdul Karim}, M. and {Aguilar}, J. and {Ahlen}, S. and {Alam}, S. and {Allen}, L. and {Prieto}, C. Allende and {Alves}, O. and {Anand}, A. and {Andrade}, U. and {Armengaud}, E. and {Aviles}, A. and {Bailey}, S. and {Baltay}, C. and {Bansal}, P. and {Bault}, A. and {Behera}, J. and {BenZvi}, S. and {Bianchi}, D. and {Blake}, C. and {Brieden}, S. and {Brodzeller}, A. and {Brooks}, D. and {Buckley-Geer}, E. and {Burtin}, E. and {Calderon}, R. and {Canning}, R. and {Rosell}, A. Carnero and {Carrilho}, P. and {Casas}, L. and {Castander}, F.~J. and {Charles}, M. and {Chaussidon}, E. and {Chaves-Montero}, J. and {Chebat}, D. and {Chen}, X. and {Claybaugh}, T. and {Cole}, S. and {Cooper}, A.~P. and {Cuceu}, A. and {Dawson}, K.~S. and {de la Macorra}, A. and {de Mattia}, A. and {Deiosso}, N. and {Della Costa}, J. and {Demina}, R. and {Dey}, A. and {Dey}, B. and {Ding}, Z. and {Doel}, P. and {Edelstein}, J. and {Eisenstein}, D.~J. and {Elbers}, W. and {Fagrelius}, P. and {Fanning}, K. and {Fern{\'a}ndez-Garc{\'\i}a}, E. and {Ferraro}, S. and {Font-Ribera}, A. and {Forero-Romero}, J.~E. and {Frenk}, C.~S. and {Garcia-Quintero}, C. and {Garrison}, L.~H. and {Gazta{\~n}aga}, E. and {Gil-Mar{\'\i}n}, H. and {Gontcho A Gontcho}, S. and {Gonzalez}, D. and {Gonzalez-Morales}, A.~X. and {Gordon}, C. and {Green}, D. and {Gutierrez}, G. and {Guy}, J. and {Hadzhiyska}, B. and {Hahn}, C. and {He}, S. and {Herbold}, M. and {Herrera-Alcantar}, H.~K. and {Ho}, M.-F. and {Honscheid}, K. and {Howlett}, C. and {Huterer}, D. and {Ishak}, M. and {Juneau}, S. and {Kamble}, N.~V. and {Kara{\c{c}}ayl{\i}}, N.~G. and {Kehoe}, R. and {Kent}, S. and {Kim}, A.~G. and {Kirkby}, D. and {Kisner}, T. and {Koposov}, S.~E. and {Kremin}, A. and {Krolewski}, A. and {Lahav}, O. and {Lamman}, C. and {Landriau}, M. and {Lang}, D. and {Lasker}, J. and {Le Goff}, J.~M. and {Le Guillou}, L. and {Leauthaud}, A. and {Levi}, M.~E. and {Li}, Q. and {Li}, T.~S. and {Lodha}, K. and {Lokken}, M. and {Lozano-Rodr{\'\i}guez}, F. and {Magneville}, C. and {Manera}, M. and {Martini}, P. and {Matthewson}, W.~L. and {Meisner}, A. and {Mena-Fern{\'a}ndez}, J. and {Menegas}, A. and {Mergulh{\~a}o}, T. and {Miquel}, R. and {Moustakas}, J. and {Mu{\~n}oz-Guti{\'e}rrez}, A. and {Mu{\~n}oz-Santos}, D. and {Myers}, A.~D. and {Nadathur}, S. and {Naidoo}, K. and {Napolitano}, L. and {Newman}, J.~A. and {Niz}, G. and {Noriega}, H.~E. and {Paillas}, E. and {Palanque-Delabrouille}, N. and {Pan}, J. and {Peacock}, J.~A. and {Pellejero Ibanez}, M. and {Percival}, W.~J. and {P{\'e}rez-Fern{\'a}ndez}, A. and {P{\'e}rez-R{\`a}fols}, I. and {Pieri}, M.~M. and {Poppett}, C. and {Prada}, F. and {Rabinowitz}, D. and {Raichoor}, A. and {Ram{\'\i}rez-P{\'e}rez}, C. and {Rashkovetskyi}, M. and {Ravoux}, C. and {Rich}, J. and {Rocher}, A. and {Rockosi}, C. and {Rohlf}, J. and {Rom{\'a}n-Herrera}, J.~O. and {Ross}, A.~J. and {Rossi}, G. and {Ruggeri}, R. and {Ruhlmann-Kleider}, V. and {Samushia}, L. and {Sanchez}, E. and {Sanders}, N. and {Schlegel}, D. and {Schubnell}, M. and {Seo}, H. and {Shafieloo}, A. and {Sharples}, R. and {Silber}, J. and {Sinigaglia}, F. and {Sprayberry}, D. and {Tan}, T. and {Tarl{\'e}}, G. and {Taylor}, P. and {Turner}, W. and {Ure{\~n}a-L{\'o}pez}, L.~A. and {Vaisakh}, R. and {Valdes}, F. and {Valogiannis}, G. and {Vargas-Maga{\~n}a}, M. and {Verde}, L. and {Walther}, M. and {Weaver}, B.~A. and {Weinberg}, D.~H. and {White}, M. and {Wolfson}, M. and {Y{\`e}che}, C. and {Yu}, J. and {Zaborowski}, E.~A. and {Zarrouk}, P. and {Zhai}, Z. and {Zhang}, H. and {Zhao}, C. and {Zhao}, G.~B. and {Zhou}, R. and {Zou}, H. and {DESI Collaboration}},
        title = "{DESI DR2 results. II. Measurements of baryon acoustic oscillations and cosmological constraints}",
      journal = {\prd},
         year = 2025,
        month = oct,
       volume = {112},
       number = {8},
          eid = {083515},
        pages = {083515},
          doi = {10.1103/tr6y-kpc6},
archivePrefix = {arXiv},
       eprint = {2503.14738},
 primaryClass = {astro-ph.CO},
       adsurl = {https://ui.adsabs.harvard.edu/abs/2025PhRvD.112h3515A}
}

@ARTICLE{HSC-PSF,
       author = {{Zhang}, Tianqing and {Li}, Xiangchong and {Dalal}, Roohi and {Mandelbaum}, Rachel and {Strauss}, Michael A. and {Kannawadi}, Arun and {Miyatake}, Hironao and {Nicola}, Andrina and {Malag{\'o}n}, Andr{\'e}s A. Plazas and {Shirasaki}, Masato and {Sugiyama}, Sunao and {Takada}, Masahiro and {More}, Surhud},
        title = "{A general framework for removing point-spread function additive systematics in cosmological weak lensing analysis}",
      journal = {\mnras},
         year = 2023,
        month = oct,
       volume = {525},
       number = {2},
        pages = {2441-2471},
          doi = {10.1093/mnras/stad1801},
archivePrefix = {arXiv},
       eprint = {2212.03257},
 primaryClass = {astro-ph.CO},
       adsurl = {https://ui.adsabs.harvard.edu/abs/2023MNRAS.525.2441Z}
}

@ARTICLE{hsc-y3-2,
       author = {{Li}, Xiangchong and {Zhang}, Tianqing and {Sugiyama}, Sunao and {Dalal}, Roohi and {Terasawa}, Ryo and {Rau}, Markus M. and {Mandelbaum}, Rachel and {Takada}, Masahiro and {More}, Surhud and {Strauss}, Michael A. and {Miyatake}, Hironao and {Shirasaki}, Masato and {Hamana}, Takashi and {Oguri}, Masamune and {Luo}, Wentao and {Nishizawa}, Atsushi J. and {Takahashi}, Ryuichi and {Nicola}, Andrina and {Osato}, Ken and {Kannawadi}, Arun and {Sunayama}, Tomomi and {Armstrong}, Robert and {Bosch}, James and {Komiyama}, Yutaka and {Lupton}, Robert H. and {Lust}, Nate B. and {MacArthur}, Lauren A. and {Miyazaki}, Satoshi and {Murayama}, Hitoshi and {Nishimichi}, Takahiro and {Okura}, Yuki and {Price}, Paul A. and {Tait}, Philip J. and {Tanaka}, Masayuki and {Wang}, Shiang-Yu},
        title = "{Hyper Suprime-Cam Year 3 results: Cosmology from cosmic shear two-point correlation functions}",
      journal = {\prd},
         year = 2023,
        month = dec,
       volume = {108},
       number = {12},
          eid = {123518},
        pages = {123518},
          doi = {10.1103/PhysRevD.108.123518},
archivePrefix = {arXiv},
       eprint = {2304.00702},
 primaryClass = {astro-ph.CO},
       adsurl = {https://ui.adsabs.harvard.edu/abs/2023PhRvD.108l3518L}
}

@ARTICLE{hsc-y3,
       author = {{Dalal}, Roohi and {Li}, Xiangchong and {Nicola}, Andrina and {Zuntz}, Joe and {Strauss}, Michael A. and {Sugiyama}, Sunao and {Zhang}, Tianqing and {Rau}, Markus M. and {Mandelbaum}, Rachel and {Takada}, Masahiro and {More}, Surhud and {Miyatake}, Hironao and {Kannawadi}, Arun and {Shirasaki}, Masato and {Taniguchi}, Takanori and {Takahashi}, Ryuichi and {Osato}, Ken and {Hamana}, Takashi and {Oguri}, Masamune and {Nishizawa}, Atsushi J. and {Malag{\'o}n}, Andr{\'e}s A. Plazas and {Sunayama}, Tomomi and {Alonso}, David and {Slosar}, An{\v{z}}e and {Luo}, Wentao and {Armstrong}, Robert and {Bosch}, James and {Hsieh}, Bau-Ching and {Komiyama}, Yutaka and {Lupton}, Robert H. and {Lust}, Nate B. and {MacArthur}, Lauren A. and {Miyazaki}, Satoshi and {Murayama}, Hitoshi and {Nishimichi}, Takahiro and {Okura}, Yuki and {Price}, Paul A. and {Tait}, Philip J. and {Tanaka}, Masayuki and {Wang}, Shiang-Yu},
        title = "{Hyper Suprime-Cam Year 3 results: Cosmology from cosmic shear power spectra}",
      journal = {\prd},
         year = 2023,
        month = dec,
       volume = {108},
       number = {12},
          eid = {123519},
        pages = {123519},
          doi = {10.1103/PhysRevD.108.123519},
archivePrefix = {arXiv},
       eprint = {2304.00701},
 primaryClass = {astro-ph.CO},
       adsurl = {https://ui.adsabs.harvard.edu/abs/2023PhRvD.108l3519D}
}

@ARTICLE{2019PASJ...71...43H,
       author = {{Hikage}, Chiaki and {Oguri}, Masamune and {Hamana}, Takashi and {More}, Surhud and {Mandelbaum}, Rachel and {Takada}, Masahiro and {K{\"o}hlinger}, Fabian and {Miyatake}, Hironao and {Nishizawa}, Atsushi J. and {Aihara}, Hiroaki and {Armstrong}, Robert and {Bosch}, James and {Coupon}, Jean and {Ducout}, Anne and {Ho}, Paul and {Hsieh}, Bau-Ching and {Komiyama}, Yutaka and {Lanusse}, Fran{\c{c}}ois and {Leauthaud}, Alexie and {Lupton}, Robert H. and {Medezinski}, Elinor and {Mineo}, Sogo and {Miyama}, Shoken and {Miyazaki}, Satoshi and {Murata}, Ryoma and {Murayama}, Hitoshi and {Shirasaki}, Masato and {Sif{\'o}n}, Crist{\'o}bal and {Simet}, Melanie and {Speagle}, Joshua and {Spergel}, David N. and {Strauss}, Michael A. and {Sugiyama}, Naoshi and {Tanaka}, Masayuki and {Utsumi}, Yousuke and {Wang}, Shiang-Yu and {Yamada}, Yoshihiko},
        title = "{Cosmology from cosmic shear power spectra with Subaru Hyper Suprime-Cam first-year data}",
      journal = {\pasj},
         year = 2019,
        month = apr,
       volume = {71},
       number = {2},
          eid = {43},
        pages = {43},
          doi = {10.1093/pasj/psz010},
archivePrefix = {arXiv},
       eprint = {1809.09148},
 primaryClass = {astro-ph.CO},
       adsurl = {https://ui.adsabs.harvard.edu/abs/2019PASJ...71...43H}
}

@ARTICLE{2004MNRAS.352..338J,
       author = {{Jarvis}, M. and {Bernstein}, G. and {Jain}, B.},
        title = "{The skewness of the aperture mass statistic}",
      journal = {\mnras},
         year = 2004,
        month = jul,
       volume = {352},
       number = {1},
        pages = {338-352},
          doi = {10.1111/j.1365-2966.2004.07926.x},
archivePrefix = {arXiv},
       eprint = {astro-ph/0307393},
 primaryClass = {astro-ph},
       adsurl = {https://ui.adsabs.harvard.edu/abs/2004MNRAS.352..338J}
}

@article{PhysRevD.70.043009,
  title = {Joint galaxy-lensing observables and the dark energy},
  author = {Hu, W. and Jain, B.},
  journal = {Phys. Rev. D},
  volume = {70},
  issue = {4},
  pages = {043009},
  numpages = {16},
  year = {2004},
  month = Aug,
  publisher = {American Physical Society},
  doi = {10.1103/PhysRevD.70.043009},
  url = {https://link.aps.org/doi/10.1103/PhysRevD.70.043009}
}

@article{takada_jain_2009,
    author = {Takada, Masahiro and Jain, Bhuvnesh},
    title = {The impact of non-Gaussian errors on weak lensing surveys},
    journal = {\mnras},
    volume = {395},
    number = {4},
    pages = {2065-2086},
    year = {2009},
    month = {05},
    issn = {0035-8711},
    doi = {10.1111/j.1365-2966.2009.14504.x},
    url = {https://doi.org/10.1111/j.1365-2966.2009.14504.x},
    eprint = {https://academic.oup.com/mnras/article-pdf/395/4/2065/2929143/mnras0395-2065.pdf},
}

@article{Cooray:2002dia,
    author = "Cooray, Asantha and Sheth, Ravi K.",
    title = "{Halo Models of Large Scale Structure}",
    eprint = "astro-ph/0206508",
    archivePrefix = "arXiv",
    reportNumber = "FERMILAB-PUB-02-284-A",
    doi = "10.1016/S0370-1573(02)00276-4",
    journal = "Phys. Rept.",
    volume = "372",
    pages = "1--129",
    year = "2002"
}

@article{Lacasa_2016,
   title={Combining cluster number counts and galaxy clustering},
   volume={2016},
   ISSN={1475-7516},
   url={http://dx.doi.org/10.1088/1475-7516/2016/08/005},
   DOI={10.1088/1475-7516/2016/08/005},
   number={08},
   journal={JCAP},
   publisher={IOP Publishing},
   author={Lacasa, F. and Rosenfeld, R.},
   year={2016},
   month=aug, pages={005–005} }

@ARTICLE{Takada2013,
       author = {{Takada}, M. and {Hu}, W.},
        title = "{Power spectrum super-sample covariance}",
      journal = {Phys. Rev. D},
         year = 2013,
        month = jun,
       volume = {87},
       number = {12},
          eid = {123504},
        pages = {123504},
          doi = {10.1103/PhysRevD.87.123504},
archivePrefix = {arXiv},
       eprint = {1302.6994},
 primaryClass = {astro-ph.CO},
       adsurl = {https://ui.adsabs.harvard.edu/abs/2013PhRvD.87l3504T}
}

@ARTICLE{2017MNRAS.470.2100K,
       author = {{Krause}, E. and {Eifler}, T.},
        title = "{cosmolike - cosmological likelihood analyses for photometric galaxy surveys}",
      journal = {\mnras},
         year = 2017,
        month = sep,
       volume = {470},
       number = {2},
        pages = {2100-2112},
          doi = {10.1093/mnras/stx1261},
archivePrefix = {arXiv},
       eprint = {1601.05779},
 primaryClass = {astro-ph.CO},
       adsurl = {https://ui.adsabs.harvard.edu/abs/2017MNRAS.470.2100K}
}

@ARTICLE{onecovariance,
       author = {{Reischke}, Robert and {Unruh}, Sandra and {Asgari}, Marika and {Dvornik}, Andrej and {Hildebrandt}, Hendrik and {Joachimi}, Benjamin and {Porth}, Lucas and {von Wietersheim-Kramsta}, Maximilian and {van den Busch}, Jan Luca and {St{\"o}lzner}, Benjamin and {Wright}, Angus H. and {Yan}, Ziang and {Bilicki}, Maciej and {Burger}, Pierre and {Chisari}, Nora Elisa and {Harnois-D{\'e}raps}, Joachim and {Georgiou}, Christos and {Heymans}, Catherine and {Jalan}, Priyanka and {Joudaki}, Shahab and {Kuijken}, Konrad and {Li}, Shun-Sheng and {Linke}, Laila and {Mahony}, Constance and {Sciotti}, Davide and {Tr{\"o}ster}, Tilman and {Yoon}, Mijin},
        title = "{KiDS-Legacy: Covariance validation and the unified ONECOVARIANCE framework for projected large-scale structure observables}",
      journal = {\aap},
         year = 2025,
        month = jul,
       volume = {699},
          eid = {A124},
        pages = {A124},
          doi = {10.1051/0004-6361/202452592},
archivePrefix = {arXiv},
       eprint = {2410.06962},
 primaryClass = {astro-ph.CO},
       adsurl = {https://ui.adsabs.harvard.edu/abs/2025A&A...699A.124R}
}

@article{Kaiser:1996tp,
    author = "Kaiser, Nick",
    title = "{Weak lensing and cosmology}",
    eprint = "astro-ph/9610120",
    archivePrefix = "arXiv",
    reportNumber = "CITA-96-16",
    doi = "10.1086/305515",
    journal = {\apj},
    volume = "498",
    pages = "26",
    year = "1998"
}

@article{DES:2020daw,
    author = "Doux, C. and others",
    collaboration = "DES",
    title = "{Consistency of cosmic shear analyses in harmonic and real space}",
    eprint = "2011.06469",
    archivePrefix = "arXiv",
    primaryClass = "astro-ph.CO",
    reportNumber = "FERMILAB-PUB-20-550-AE",
    doi = "10.1093/mnras/stab661",
    journal = "\mnras",
    volume = "503",
    number = "3",
    pages = "3796--3817",
    year = "2021"
}

@article{desy3-cosmo,
  title = {Dark Energy Survey Year 3 results: Cosmological constraints from galaxy clustering and weak lensing},
  author = {Abbott, T. M. C. and Aguena, M. and Alarcon, A. and Allam, S. and Alves, O. and Amon, A. and Andrade-Oliveira, F. and Annis, J. and Avila, S. and Bacon, D. and Baxter, E. and Bechtol, K. and Becker, M. R. and Bernstein, G. M. and Bhargava, S. and Birrer, S. and Blazek, J. and Brandao-Souza, A. and Bridle, S. L. and Brooks, D. and Buckley-Geer, E. and Burke, D. L. and Camacho, H. and Campos, A. and Carnero Rosell, A. and Carrasco Kind, M. and Carretero, J. and Castander, F. J. and Cawthon, R. and Chang, C. and Chen, A. and Chen, R. and Choi, A. and Conselice, C. and Cordero, J. and Costanzi, M. and Crocce, M. and da Costa, L. N. and da Silva Pereira, M. E. and Davis, C. and Davis, T. M. and De Vicente, J. and DeRose, J. and Desai, S. and Di Valentino, E. and Diehl, H. T. and Dietrich, J. P. and Dodelson, S. and Doel, P. and Doux, C. and Drlica-Wagner, A. and Eckert, K. and Eifler, T. F. and Elsner, F. and Elvin-Poole, J. and Everett, S. and Evrard, A. E. and Fang, X. and Farahi, A. and Fernandez, E. and Ferrero, I. and Fert\'e, A. and Fosalba, P. and Friedrich, O. and Frieman, J. and Garc\'{\i}a-Bellido, J. and Gatti, M. and Gaztanaga, E. and Gerdes, D. W. and Giannantonio, T. and Giannini, G. and Gruen, D. and Gruendl, R. A. and Gschwend, J. and Gutierrez, G. and Harrison, I. and Hartley, W. G. and Herner, K. and Hinton, S. R. and Hollowood, D. L. and Honscheid, K. and Hoyle, B. and Huff, E. M. and Huterer, D. and Jain, B. and James, D. J. and Jarvis, M. and Jeffrey, N. and Jeltema, T. and Kovacs, A. and Krause, E. and Kron, R. and Kuehn, K. and Kuropatkin, N. and Lahav, O. and Leget, P.-F. and Lemos, P. and Liddle, A. R. and Lidman, C. and Lima, M. and Lin, H. and MacCrann, N. and Maia, M. A. G. and Marshall, J. L. and Martini, P. and McCullough, J. and Melchior, P. and Mena-Fern\'andez, J. and Menanteau, F. and Miquel, R. and Mohr, J. J. and Morgan, R. and Muir, J. and Myles, J. and Nadathur, S. and Navarro-Alsina, A. and Nichol, R. C. and Ogando, R. L. C. and Omori, Y. and Palmese, A. and Pandey, S. and Park, Y. and Paz-Chinch\'on, F. and Petravick, D. and Pieres, A. and Plazas Malag\'on, A. A. and Porredon, A. and Prat, J. and Raveri, M. and Rodriguez-Monroy, M. and Rollins, R. P. and Romer, A. K. and Roodman, A. and Rosenfeld, R. and Ross, A. J. and Rykoff, E. S. and Samuroff, S. and S\'anchez, C. and Sanchez, E. and Sanchez, J. and Sanchez Cid, D. and Scarpine, V. and Schubnell, M. and Scolnic, D. and Secco, L. F. and Serrano, S. and Sevilla-Noarbe, I. and Sheldon, E. and Shin, T. and Smith, M. and Soares-Santos, M. and Suchyta, E. and Swanson, M. E. C. and Tabbutt, M. and Tarle, G. and Thomas, D. and To, C. and Troja, A. and Troxel, M. A. and Tucker, D. L. and Tutusaus, I. and Varga, T. N. and Walker, A. R. and Weaverdyck, N. and Wechsler, R. and Weller, J. and Yanny, B. and Yin, B. and Zhang, Y. and Zuntz, J.},
  collaboration = {DES Collaboration},
  journal = {Phys. Rev. D},
  volume = {105},
  issue = {2},
  pages = {023520},
  numpages = {42},
  year = {2022},
  month = {Jan},
  publisher = {American Physical Society},
  doi = {10.1103/PhysRevD.105.023520},
  url = {https://link.aps.org/doi/10.1103/PhysRevD.105.023520}
}

@ARTICLE{euclid-cloe,
       author = {{Euclid Collaboration: Ca{\~n}as-Herrera}, G. and {Goh}, L.~W.~K. and {Blot}, L. and {Bonici}, M. and {Camera}, S. and {Cardone}, V.~F. and {Carrilho}, P. and {Casas}, S. and {Davini}, S. and {Di Domizio}, S. and {Farrens}, S. and {Gouyou Beauchamps}, S. and {Ili{\'c}}, S. and {Joudaki}, S. and {Keil}, F. and {Le Brun}, A.~M.~C. and {Martinelli}, M. and {Moretti}, C. and {Pettorino}, V. and {Pezzotta}, A. and {Sakr}, Z. and {S{\'a}nchez}, A.~G. and {Sciotti}, D. and {Tanidis}, K. and {Tutusaus}, I. and {Ajani}, V. and {Crocce}, M. and {Fumagalli}, A. and {Giocoli}, C. and {Legrand}, L. and {Lembo}, M. and {Lesci}, G.~F. and {Navarro Girones}, D. and {Nouri-Zonoz}, A. and {Pamuk}, S. and {Pourtsidou}, A. and {Tsedrik}, M. and {Bel}, J. and {Carbone}, C. and {Claramunt Gonzalez}, J. and {Duncan}, C.~A.~J. and {Kilbinger}, M. and {Porredon}, A. and {Sapone}, D. and {Sellentin}, E. and {Taylor}, P.~L. and {Tessore}, N. and {Altieri}, B. and {Amara}, A. and {Amendola}, L. and {Andreon}, S. and {Auricchio}, N. and {Baccigalupi}, C. and {Baldi}, M. and {Bardelli}, S. and {Bender}, R. and {Biviano}, A. and {Bonino}, D. and {Branchini}, E. and {Brescia}, M. and {Brinchmann}, J. and {Capobianco}, V. and {Carretero}, J. and {Castellano}, M. and {Castignani}, G. and {Cavuoti}, S. and {Chambers}, K.~C. and {Cimatti}, A. and {Colodro-Conde}, C. and {Congedo}, G. and {Conselice}, C.~J. and {Conversi}, L. and {Copin}, Y. and {Courbin}, F. and {Courtois}, H.~M. and {Cropper}, M. and {Da Silva}, A. and {Degaudenzi}, H. and {de la Torre}, S. and {De Lucia}, G. and {Di Giorgio}, A.~M. and {Dole}, H. and {Dubath}, F. and {Dupac}, X. and {Dusini}, S. and {Escoffier}, S. and {Farina}, M. and {Faustini}, F. and {Ferriol}, S. and {Finelli}, F. and {Fosalba}, P. and {Fotopoulou}, S. and {Fourmanoit}, N. and {Frailis}, M. and {Franceschi}, E. and {Galeotta}, S. and {George}, K. and {Gillard}, W. and {Gillis}, B. and {G{\'o}mez-Alvarez}, P. and {Gracia-Carpio}, J. and {Granett}, B.~R. and {Grazian}, A. and {Grupp}, F. and {Guzzo}, L. and {Haugan}, S.~V.~H. and {Hoekstra}, H. and {Holmes}, W. and {Hook}, I. and {Hormuth}, F. and {Hornstrup}, A. and {Hudelot}, P. and {Jahnke}, K. and {Jhabvala}, M. and {Joachimi}, B. and {Keih{\"a}nen}, E. and {Kermiche}, S. and {Kiessling}, A. and {Kubik}, B. and {Kuijken}, K. and {K{\"u}mmel}, M. and {Kunz}, M. and {Kurki-Suonio}, H. and {Lahav}, O. and {Laureijs}, R. and {Ligori}, S. and {Lilje}, P.~B. and {Lindholm}, V. and {Lloro}, I. and {Mainetti}, G. and {Maino}, D. and {Maiorano}, E. and {Mansutti}, O. and {Marcin}, S. and {Marggraf}, O. and {Markovic}, K. and {Martinet}, N. and {Marulli}, F. and {Massey}, R. and {McCracken}, H.~J. and {Medinaceli}, E. and {Melchior}, M. and {Mellier}, Y. and {Meneghetti}, M. and {Merlin}, E. and {Meylan}, G. and {Mora}, A. and {Moresco}, M. and {Moscardini}, L. and {Neissner}, C. and {Niemi}, S. -M. and {Nightingale}, J.~W. and {Padilla}, C. and {Paltani}, S. and {Pasian}, F. and {Pedersen}, K. and {Percival}, W.~J. and {Pires}, S. and {Polenta}, G. and {Poncet}, M. and {Popa}, L.~A. and {Pozzetti}, L. and {Raison}, F. and {Rebolo}, R. and {Renzi}, A. and {Rhodes}, J. and {Riccio}, G. and {Romelli}, E. and {Roncarelli}, M. and {Saglia}, R. and {Sartoris}, B. and {Schewtschenko}, J.~A. and {Schneider}, P. and {Schrabback}, T. and {Secroun}, A. and {Sefusatti}, E. and {Seidel}, G. and {Seiffert}, M. and {Serrano}, S. and {Simon}, P. and {Sirignano}, C. and {Sirri}, G. and {Spurio Mancini}, A. and {Stanco}, L. and {Steinwagner}, J. and {Tallada-Cresp{\'\i}}, P. and {Tavagnacco}, D. and {Taylor}, A.~N. and {Tereno}, I. and {Toft}, S. and {Toledo-Moreo}, R. and {Torradeflot}, F. and {Valenziano}, L. and {Valiviita}, J. and {Vassallo}, T. and {Verdoes Kleijn}, G. and {Veropalumbo}, A. and {Wang}, Y. and {Weller}, J.},
        title = "{Euclid preparation. Cosmology Likelihood for Observables in Euclid (CLOE). 3. Inference and Forecasts}",
      journal = {arXiv e-prints},
         year = 2025,
        month = oct,
          eid = {arXiv:2510.09153},
        pages = {arXiv:2510.09153},
          doi = {10.48550/arXiv.2510.09153},
archivePrefix = {arXiv},
       eprint = {2510.09153},
 primaryClass = {astro-ph.CO},
       adsurl = {https://ui.adsabs.harvard.edu/abs/2025arXiv251009153E}
}

@article{Fang__2020,
   title={2D-FFTLog: efficient computation of real-space covariance matrices for galaxy clustering and weak lensing},
   volume={497},
   ISSN={1365-2966},
   url={http://dx.doi.org/10.1093/mnras/staa1726},
   DOI={10.1093/mnras/staa1726},
   number={3},
   journal={\mnras},
   publisher={Oxford University Press (OUP)},
   author={Fang, X. and Eifler, T. and Krause, E.},
   year={2020},
   month=jun, pages={2699–2714} }

@article{Kilo-DegreeSurvey:2023gfr,
    author = "Abbott, T. M. C. and others",
    collaboration = "Kilo-Degree Survey, DES",
    title = "{DES Y3 + KiDS-1000: Consistent cosmology combining cosmic shear surveys}",
    eprint = "2305.17173",
    archivePrefix = "arXiv",
    primaryClass = "astro-ph.CO",
    reportNumber = "FERMILAB-PUB-23-267-PPD, DES-2023-0769",
    doi = "10.21105/astro.2305.17173",
    journal = "Open J. Astrophys.",
    volume = "6",
    pages = "2305.17173",
    year = "2023"
}

@ARTICLE{glass,
       author = {{Tessore}, Nicolas and {Loureiro}, Arthur and {Joachimi}, Benjamin and {von Wietersheim-Kramsta}, Maximilian and {Jeffrey}, Niall},
        title = "{GLASS: Generator for Large Scale Structure}",
      journal = {Open J. Astrophys.},
         year = 2023,
        month = mar,
       volume = {6},
          eid = {11},
        pages = {11},
          doi = {10.21105/astro.2302.01942},
archivePrefix = {arXiv},
       eprint = {2302.01942},
 primaryClass = {astro-ph.CO},
       adsurl = {https://ui.adsabs.harvard.edu/abs/2023OJAp....6E..11T}
}

@article{Lewis:1999bs,
      author         = "Lewis, Antony and Challinor, Anthony and Lasenby,
                        Anthony",
      title          = "{Efficient computation of CMB anisotropies in closed FRW
                        models}",
      journal        = "\apj",
      volume         = "538",
      year           = "2000",
      pages          = "473-476",
      doi            = "10.1086/309179",
      eprint         = "astro-ph/9911177",
      archivePrefix  = "arXiv",
      primaryClass   = "astro-ph",
      SLACcitation   = "%%CITATION = ASTRO-PH/9911177;%%"
}

@article{Zuntz:2014csq,
    author = "Zuntz, Joe and Paterno, Marc and Jennings, Elise and Rudd, Douglas and Manzotti, Alessandro and Dodelson, Scott and Bridle, Sarah and Sehrish, Saba and Kowalkowski, James",
    title = "{CosmoSIS: modular cosmological parameter estimation}",
    eprint = "1409.3409",
    archivePrefix = "arXiv",
    primaryClass = "astro-ph.CO",
    reportNumber = "FERMILAB-PUB-14-408-A",
    doi = "10.1016/j.ascom.2015.05.005",
    journal = "Astron. Comput.",
    volume = "12",
    pages = "45--59",
    year = "2015"
}

@ARTICLE{2010MNRAS.404..350R,
   author = {{Rowe}, B.},
    title = "{Improving PSF modelling for weak gravitational lensing using new methods in model selection}",
  journal = {\mnras},
archivePrefix = "arXiv",
   eprint = {0904.3056},
 primaryClass = "astro-ph.CO",
     year = 2010,
    OPTmonth = may,
   volume = 404,
    pages = {350-366},
      doi = {10.1111/j.1365-2966.2010.16277.x},
   adsurl = {http://adsabs.harvard.edu/abs/2010MNRAS.404..350R}
}

@ARTICLE{2016MNRAS.460.2245J,
   author = {{Jarvis}, M. and {Sheldon}, E. and {Zuntz}, J. and {Kacprzak}, T. and 
	{Bridle}, S.~L. and {Amara}, A. and {Armstrong}, R. and {Becker}, M.~R. and 
	{Bernstein}, G.~M. and {Bonnett}, C. and {Chang}, C. and {Das}, R. and 
	{Dietrich}, J.~P. and {Drlica-Wagner}, A. and {Eifler}, T.~F. and 
	{Gangkofner}, C. and {Gruen}, D. and {Hirsch}, M. and {Huff}, E.~M. and 
	{Jain}, B. and {Kent}, S. and {Kirk}, D. and {MacCrann}, N. and 
	{Melchior}, P. and {Plazas}, A.~A. and {Refregier}, A. and {Rowe}, B. and 
	{Rykoff}, E.~S. and {Samuroff}, S. and {S{\'a}nchez}, C. and 
	{Suchyta}, E. and {Troxel}, M.~A. and {Vikram}, V. and {Abbott}, T. and 
	{Abdalla}, F.~B. and {Allam}, S. and {Annis}, J. and {Benoit-L{\'e}vy}, A. and 
	{Bertin}, E. and {Brooks}, D. and {Buckley-Geer}, E. and {Burke}, D.~L. and 
	{Capozzi}, D. and {Carnero Rosell}, A. and {Carrasco Kind}, M. and 
	{Carretero}, J. and {Castander}, F.~J. and {Clampitt}, J. and 
	{Crocce}, M. and {Cunha}, C.~E. and {D'Andrea}, C.~B. and {da Costa}, L.~N. and 
	{DePoy}, D.~L. and {Desai}, S. and {Diehl}, H.~T. and {Doel}, P. and 
	{Fausti Neto}, A. and {Flaugher}, B. and {Fosalba}, P. and {Frieman}, J. and 
	{Gaztanaga}, E. and {Gerdes}, D.~W. and {Gruendl}, R.~A. and 
	{Gutierrez}, G. and {Honscheid}, K. and {James}, D.~J. and {Kuehn}, K. and 
	{Kuropatkin}, N. and {Lahav}, O. and {Li}, T.~S. and {Lima}, M. and 
	{March}, M. and {Martini}, P. and {Miquel}, R. and {Mohr}, J.~J. and 
	{Neilsen}, E. and {Nord}, B. and {Ogando}, R. and {Reil}, K. and 
	{Romer}, A.~K. and {Roodman}, A. and {Sako}, M. and {Sanchez}, E. and 
	{Scarpine}, V. and {Schubnell}, M. and {Sevilla-Noarbe}, I. and 
	{Smith}, R.~C. and {Soares-Santos}, M. and {Sobreira}, F. and 
	{Swanson}, M.~E.~C. and {Tarle}, G. and {Thaler}, J. and {Thomas}, D. and 
	{Walker}, A.~R. and {Wechsler}, R.~H.},
    title = "{The DES Science Verification weak lensing shear catalogues}",
  journal = {\mnras},
archivePrefix = "arXiv",
   eprint = {1507.05603},
 primaryClass = "astro-ph.IM",
     year = 2016,
    month = aug,
   volume = 460,
    pages = {2245-2281},
      doi = {10.1093/mnras/stw990},
   adsurl = {http://cdsads.u-strasbg.fr/abs/2016MNRAS.460.2245J}
}

@ARTICLE{lsst,
       author = {{Ivezi{\'c}}, {\v{Z}}eljko and {Kahn}, Steven M. and {Tyson}, J. Anthony and {Abel}, Bob and {Acosta}, Emily and {Allsman}, Robyn and {Alonso}, David and {AlSayyad}, Yusra and {Anderson}, Scott F. and {Andrew}, John and {Angel}, James Roger P. and {Angeli}, George Z. and {Ansari}, Reza and {Antilogus}, Pierre and {Araujo}, Constanza and {Armstrong}, Robert and {Arndt}, Kirk T. and {Astier}, Pierre and {Aubourg}, {\'E}ric and {Auza}, Nicole and {Axelrod}, Tim S. and {Bard}, Deborah J. and {Barr}, Jeff D. and {Barrau}, Aurelian and {Bartlett}, James G. and {Bauer}, Amanda E. and {Bauman}, Brian J. and {Baumont}, Sylvain and {Bechtol}, Ellen and {Bechtol}, Keith and {Becker}, Andrew C. and {Becla}, Jacek and {Beldica}, Cristina and {Bellavia}, Steve and {Bianco}, Federica B. and {Biswas}, Rahul and {Blanc}, Guillaume and {Blazek}, Jonathan and {Blandford}, Roger D. and {Bloom}, Josh S. and {Bogart}, Joanne and {Bond}, Tim W. and {Booth}, Michael T. and {Borgland}, Anders W. and {Borne}, Kirk and {Bosch}, James F. and {Boutigny}, Dominique and {Brackett}, Craig A. and {Bradshaw}, Andrew and {Brandt}, William Nielsen and {Brown}, Michael E. and {Bullock}, James S. and {Burchat}, Patricia and {Burke}, David L. and {Cagnoli}, Gianpietro and {Calabrese}, Daniel and {Callahan}, Shawn and {Callen}, Alice L. and {Carlin}, Jeffrey L. and {Carlson}, Erin L. and {Chandrasekharan}, Srinivasan and {Charles-Emerson}, Glenaver and {Chesley}, Steve and {Cheu}, Elliott C. and {Chiang}, Hsin-Fang and {Chiang}, James and {Chirino}, Carol and {Chow}, Derek and {Ciardi}, David R. and {Claver}, Charles F. and {Cohen-Tanugi}, Johann and {Cockrum}, Joseph J. and {Coles}, Rebecca and {Connolly}, Andrew J. and {Cook}, Kem H. and {Cooray}, Asantha and {Covey}, Kevin R. and {Cribbs}, Chris and {Cui}, Wei and {Cutri}, Roc and {Daly}, Philip N. and {Daniel}, Scott F. and {Daruich}, Felipe and {Daubard}, Guillaume and {Daues}, Greg and {Dawson}, William and {Delgado}, Francisco and {Dellapenna}, Alfred and {de Peyster}, Robert and {de Val-Borro}, Miguel and {Digel}, Seth W. and {Doherty}, Peter and {Dubois}, Richard and {Dubois-Felsmann}, Gregory P. and {Durech}, Josef and {Economou}, Frossie and {Eifler}, Tim and {Eracleous}, Michael and {Emmons}, Benjamin L. and {Fausti Neto}, Angelo and {Ferguson}, Henry and {Figueroa}, Enrique and {Fisher-Levine}, Merlin and {Focke}, Warren and {Foss}, Michael D. and {Frank}, James and {Freemon}, Michael D. and {Gangler}, Emmanuel and {Gawiser}, Eric and {Geary}, John C. and {Gee}, Perry and {Geha}, Marla and {Gessner}, Charles J.~B. and {Gibson}, Robert R. and {Gilmore}, D. Kirk and {Glanzman}, Thomas and {Glick}, William and {Goldina}, Tatiana and {Goldstein}, Daniel A. and {Goodenow}, Iain and {Graham}, Melissa L. and {Gressler}, William J. and {Gris}, Philippe and {Guy}, Leanne P. and {Guyonnet}, Augustin and {Haller}, Gunther and {Harris}, Ron and {Hascall}, Patrick A. and {Haupt}, Justine and {Hernandez}, Fabio and {Herrmann}, Sven and {Hileman}, Edward and {Hoblitt}, Joshua and {Hodgson}, John A. and {Hogan}, Craig and {Howard}, James D. and {Huang}, Dajun and {Huffer}, Michael E. and {Ingraham}, Patrick and {Innes}, Walter R. and {Jacoby}, Suzanne H. and {Jain}, Bhuvnesh and {Jammes}, Fabrice and {Jee}, M. James and {Jenness}, Tim and {Jernigan}, Garrett and {Jevremovi{\'c}}, Darko and {Johns}, Kenneth and {Johnson}, Anthony S. and {Johnson}, Margaret W.~G. and {Jones}, R. Lynne and {Juramy-Gilles}, Claire and {Juri{\'c}}, Mario and {Kalirai}, Jason S. and {Kallivayalil}, Nitya J. and {Kalmbach}, Bryce and {Kantor}, Jeffrey P. and {Karst}, Pierre and {Kasliwal}, Mansi M. and {Kelly}, Heather and {Kessler}, Richard and {Kinnison}, Veronica and {Kirkby}, David and {Knox}, Lloyd and {Kotov}, Ivan V. and {Krabbendam}, Victor L. and {Krughoff}, K. Simon and {Kub{\'a}nek}, Petr and {Kuczewski}, John and {Kulkarni}, Shri and {Ku}, John and {Kurita}, Nadine R. and {Lage}, Craig S. and {Lambert}, Ron and {Lange}, Travis and {Langton}, J. Brian and {Le Guillou}, Laurent and {Levine}, Deborah and {Liang}, Ming and {Lim}, Kian-Tat and {Lintott}, Chris J. and {Long}, Kevin E. and {Lopez}, Margaux and {Lotz}, Paul J. and {Lupton}, Robert H. and {Lust}, Nate B. and {MacArthur}, Lauren A. and {Mahabal}, Ashish and {Mandelbaum}, Rachel and {Markiewicz}, Thomas W. and {Marsh}, Darren S. and {Marshall}, Philip J. and {Marshall}, Stuart and {May}, Morgan and {McKercher}, Robert and {McQueen}, Michelle and {Meyers}, Joshua and {Migliore}, Myriam and {Miller}, Michelle and {Mills}, David J.},
        title = "{LSST: From Science Drivers to Reference Design and Anticipated Data Products}",
      journal = {\apj},
         year = 2019,
        month = mar,
       volume = {873},
       number = {2},
          eid = {111},
        pages = {111},
          doi = {10.3847/1538-4357/ab042c},
archivePrefix = {arXiv},
       eprint = {0805.2366},
 primaryClass = {astro-ph},
       adsurl = {https://ui.adsabs.harvard.edu/abs/2019ApJ...873..111I}
}

@ARTICLE{cosmoverse,
       author = {{Di Valentino}, Eleonora and {Said}, Jackson Levi and {Riess}, Adam and {Pollo}, Agnieszka and {Poulin}, Vivian and {G{\'o}mez-Valent}, Adri{\`a} and {Weltman}, Amanda and {Palmese}, Antonella and {Huang}, Caroline D. and {van de Bruck}, Carsten and {Saraf}, Chandra Shekhar and {Kuo}, Cheng-Yu and {Uhlemann}, Cora and {Grand{\'o}n}, Daniela and {Paz}, Dante and {Eckert}, Dominique and {Teixeira}, Elsa M. and {Saridakis}, Emmanuel N. and {Colg{\'a}in}, Eoin {\'O}. and {Beutler}, Florian and {Niedermann}, Florian and {Bajardi}, Francesco and {Barenboim}, Gabriela and {Gubitosi}, Giulia and {Musella}, Ilaria and {Banik}, Indranil and {Szapudi}, Istvan and {Singal}, Jack and {Cases}, Jaume Haro and {Chluba}, Jens and {Torrado}, Jes{\'u}s and {Mifsud}, Jurgen and {Jedamzik}, Karsten and {Said}, Khaled and {Dialektopoulos}, Konstantinos and {Herold}, Laura and {Perivolaropoulos}, Leandros and {Zu}, Lei and {Galbany}, Llu{\'\i}s and {Breuval}, Louise and {Visinelli}, Luca and {Escamilla}, Luis A. and {Anchordoqui}, Luis A. and {Sheikh-Jabbari}, M.~M. and {Lembo}, Margherita and {Dainotti}, Maria Giovanna and {Vincenzi}, Maria and {Asgari}, Marika and {Gerbino}, Martina and {Forconi}, Matteo and {Cantiello}, Michele and {Moresco}, Michele and {Benetti}, Micol and {Sch{\"o}neberg}, Nils and {Akarsu}, {\"O}zg{\"u}r and {Nunes}, Rafael C. and {Bernardo}, Reginald Christian and {Ch{\'a}vez}, Ricardo and {Anderson}, Richard I. and {Watkins}, Richard and {Capozziello}, Salvatore and {Li}, Siyang and {Vagnozzi}, Sunny and {Pan}, Supriya and {Treu}, Tommaso and {Irsic}, Vid and {Handley}, Will and {Giar{\`e}}, William and {Murakami}, Yukei and {Banihashemi}, Abdolali and {Poudou}, Ad{\`e}le and {Heavens}, Alan and {Kogut}, Alan and {Domi}, Alba and {Lenart}, Aleksander {\L}ukasz and {Melchiorri}, Alessandro and {Vadal{\`a}}, Alessandro and {Amon}, Alexandra and {Rivera}, Alexander Bonilla and {Reeves}, Alexander and {Zhuk}, Alexander and {Bonanno}, Alfio and {{\"O}vg{\"u}n}, Ali and {Pisani}, Alice and {Talebian}, Alireza and {Abebe}, Amare and {Aboubrahim}, Amin and {Gonz{\'a}lez Mor{\'a}n}, Ana Luisa and {Kov{\'a}cs}, Andr{\'a}s and {Lymperis}, Andreas and {Papatriantafyllou}, Andreas and {Liddle}, Andrew R. and {Paliathanasis}, Andronikos and {Borowiec}, Andrzej and {Yadav}, Anil Kumar and {Yadav}, Anita and {Sen}, Anjan Ananda and {William}, Anjitha John and {Davis}, Anne Christine and {Shajib}, Anowar J. and {Walters}, Anthony and {Lonappan}, Anto Idicherian and {Chudaykin}, Anton and {Capodagli}, Antonio and {da Silva}, Antonio and {De Felice}, Antonio and {Racioppi}, Antonio and {Oficial}, Araceli Soler and {Montiel}, Ariadna and {Favale}, Arianna and {Bernui}, Armando and {Velasco}, Arrianne Crystal and {Heinesen}, Asta and {Bakopoulos}, Athanasios and {Chatzistavrakidis}, Athanasios and {Khanpour}, Bahman and {Sathyaprakash}, Bangalore S. and {Zgirski}, Bartek and {L'Huillier}, Benjamin and {Famaey}, Benoit and {Jain}, Bhuvnesh and {Zhang}, Bing and {Karmakar}, Biswajit and {Dragovich}, Branko and {Thomas}, Brooks and {Correa}, Carlos and {Boiza}, Carlos G. and {Marques}, Catarina and {Escamilla-Rivera}, Celia and {Tzerefos}, Charalampos and {Zhang}, Chi and {De Leo}, Chiara and {Pfeifer}, Christian and {Lee}, Christine and {Venter}, Christo and {Gomes}, Cl{\'a}udio and {Roque De bom}, Clecio and {Moreno-Pulido}, Cristian and {Iosifidis}, Damianos and {Grin}, Dan and {Blixt}, Daniel and {Scolnic}, Dan and {Oriti}, Daniele and {Dobrycheva}, Daria and {Bettoni}, Dario and {Benisty}, David and {Fern{\'a}ndez-Arenas}, David and {Wiltshire}, David L. and {Sanchez Cid}, David and {Tamayo}, David and {Valls-Gabaud}, David and {Pedrotti}, Davide and {Wang}, Deng and {Staicova}, Denitsa and {Totolou}, Despoina and {Rubiera-Garcia}, Diego and {Milakovi{\'c}}, Dinko and {Pesce}, Dominic W. and {Sluse}, Dominique and {Borka}, Du{\v{s}}ko and {Yusofi}, Ebrahim and {Giusarma}, Elena and {Terlevich}, Elena and {Tomasetti}, Elena and {Vagenas}, Elias C. and {Fazzari}, Elisa and {Ferreira}, Elisa G.~M. and {Barakovic}, Elvis and {Dimastrogiovanni}, Emanuela and {Holm}, Emil Brinch and {Mottola}, Emil and {{\"O}z{\"u}lker}, Emre and {Specogna}, Enrico and {Brocato}, Enzo and {Jensko}, Erik and {Enriquez}, Erika Antonette and {Bhatia}, Esha and {Bresolin}, Fabio and {Avila}, Felipe and {Bouch{\`e}}, Filippo and {Bombacigno}, Flavio and {Anagnostopoulos}, Fotios K. and {Pace}, Francesco and {Sorrenti}, Francesco and {Lobo}, Francisco S.~N. and {Courbin}, Fr{\'e}d{\'e}ric and {Hansen}, Frode K. and {Sloan}, Greg and {Farrugia}, Gabriel and {Lynch}, Gabriel and {Garcia-Arroyo}, Gabriela and {Raimondo}, Gabriella and {Lambiase}, Gaetano and {Anand}, Gagandeep S. and {Poulot}, Gaspard and {Leon}, Genly and {Kouniatalis}, Gerasimos and {Nardini}, Germano and {Cs{\"o}rnyei}, G{\'e}za and {Galloni}, Giacomo},
        title = "{The CosmoVerse White Paper: Addressing observational tensions in cosmology with systematics and fundamental physics}",
      journal = {Physics of the Dark Universe},
         year = 2025,
        month = sep,
       volume = {49},
          eid = {101965},
        pages = {101965},
          doi = {10.1016/j.dark.2025.101965},
archivePrefix = {arXiv},
       eprint = {2504.01669},
 primaryClass = {astro-ph.CO},
       adsurl = {https://ui.adsabs.harvard.edu/abs/2025PDU....4901965D}
}

@ARTICLE{euclid_overview,
       author = {{Euclid Collaboration: Mellier}, Y. and {Abdurro'uf} and {Acevedo Barroso}, J.~A. and {Ach{\'u}carro}, A. and {Adamek}, J. and {Adam}, R. and {Addison}, G.~E. and {Aghanim}, N. and {Aguena}, M. and {Ajani}, V. and {Akrami}, Y. and {Al-Bahlawan}, A. and {Alavi}, A. and {Albuquerque}, I.~S. and {Alestas}, G. and {Alguero}, G. and {Allaoui}, A. and {Allen}, S.~W. and {Allevato}, V. and {Alonso-Tetilla}, A.~V. and {Altieri}, B. and {Alvarez-Candal}, A. and {Alvi}, S. and {Amara}, A. and {Amendola}, L. and {Amiaux}, J. and {Andika}, I.~T. and {Andreon}, S. and {Andrews}, A. and {Angora}, G. and {Angulo}, R.~E. and {Annibali}, F. and {Anselmi}, A. and {Anselmi}, S. and {Arcari}, S. and {Archidiacono}, M. and {Aric{\`o}}, G. and {Arnaud}, M. and {Arnouts}, S. and {Asgari}, M. and {Asorey}, J. and {Atayde}, L. and {Atek}, H. and {Atrio-Barandela}, F. and {Aubert}, M. and {Aubourg}, E. and {Auphan}, T. and {Auricchio}, N. and {Aussel}, B. and {Aussel}, H. and {Avelino}, P.~P. and {Avgoustidis}, A. and {Avila}, S. and {Awan}, S. and {Azzollini}, R. and {Baccigalupi}, C. and {Bachelet}, E. and {Bacon}, D. and {Baes}, M. and {Bagley}, M.~B. and {Bahr-Kalus}, B. and {Balaguera-Antolinez}, A. and {Balbinot}, E. and {Balcells}, M. and {Baldi}, M. and {Baldry}, I. and {Balestra}, A. and {Ballardini}, M. and {Ballester}, O. and {Balogh}, M. and {Ba{\~n}ados}, E. and {Barbier}, R. and {Bardelli}, S. and {Baron}, M. and {Barreiro}, T. and {Barrena}, R. and {Barriere}, J.-C. and {Barros}, B.~J. and {Barthelemy}, A. and {Bartolo}, N. and {Basset}, A. and {Battaglia}, P. and {Battisti}, A.~J. and {Baugh}, C.~M. and {Baumont}, L. and {Bazzanini}, L. and {Beaulieu}, J.-P. and {Beckmann}, V. and {Belikov}, A.~N. and {Bel}, J. and {Bellagamba}, F. and {Bella}, M. and {Bellini}, E. and {Benabed}, K. and {Bender}, R. and {Benevento}, G. and {Bennett}, C.~L. and {Benson}, K. and {Bergamini}, P. and {Bermejo-Climent}, J.~R. and {Bernardeau}, F. and {Bertacca}, D. and {Berthe}, M. and {Berthier}, J. and {Bethermin}, M. and {Beutler}, F. and {Bevillon}, C. and {Bhargava}, S. and {Bhatawdekar}, R. and {Bianchi}, D. and {Bisigello}, L. and {Biviano}, A. and {Blake}, R.~P. and {Blanchard}, A. and {Blazek}, J. and {Blot}, L. and {Bosco}, A. and {Bodendorf}, C. and {Boenke}, T. and {B{\"o}hringer}, H. and {Boldrini}, P. and {Bolzonella}, M. and {Bonchi}, A. and {Bonici}, M. and {Bonino}, D. and {Bonino}, L. and {Bonvin}, C. and {Bon}, W. and {Booth}, J.~T. and {Borgani}, S. and {Borlaff}, A.~S. and {Borsato}, E. and {Bose}, B. and {Botticella}, M.~T. and {Boucaud}, A. and {Bouche}, F. and {Boucher}, J.~S. and {Boutigny}, D. and {Bouvard}, T. and {Bouwens}, R. and {Bouy}, H. and {Bowler}, R.~A.~A. and {Bozza}, V. and {Bozzo}, E. and {Branchini}, E. and {Brando}, G. and {Brau-Nogue}, S. and {Brekke}, P. and {Bremer}, M.~N. and {Brescia}, M. and {Breton}, M.-A. and {Brinchmann}, J. and {Brinckmann}, T. and {Brockley-Blatt}, C. and {Brodwin}, M. and {Brouard}, L. and {Brown}, M.~L. and {Bruton}, S. and {Bucko}, J. and {Buddelmeijer}, H. and {Buenadicha}, G. and {Buitrago}, F. and {Burger}, P. and {Burigana}, C. and {Busillo}, V. and {Busonero}, D. and {Cabanac}, R. and {Cabayol-Garcia}, L. and {Cagliari}, M.~S. and {Caillat}, A. and {Caillat}, L. and {Calabrese}, M. and {Calabro}, A. and {Calderone}, G. and {Calura}, F. and {Camacho Quevedo}, B. and {Camera}, S. and {Campos}, L. and {Ca{\~n}as-Herrera}, G. and {Candini}, G.~P. and {Cantiello}, M. and {Capobianco}, V. and {Cappellaro}, E. and {Cappelluti}, N. and {Cappi}, A. and {Caputi}, K.~I. and {Cara}, C. and {Carbone}, C. and {Cardone}, V.~F. and {Carella}, E. and {Carlberg}, R.~G. and {Carle}, M. and {Carminati}, L. and {Caro}, F. and {Carrasco}, J.~M. and {Carretero}, J. and {Carrilho}, P. and {Carron Duque}, J. and {Carry}, B.},
        title = "{Euclid: I. Overview of the Euclid mission}",
      journal = {\aap},
         year = 2025,
        month = may,
       volume = {697},
          eid = {A1},
        pages = {A1},
          doi = {10.1051/0004-6361/202450810},
archivePrefix = {arXiv},
       eprint = {2405.13491},
 primaryClass = {astro-ph.CO},
       adsurl = {https://ui.adsabs.harvard.edu/abs/2025A&A...697A...1E}
}

@ARTICLE{hsc_dr3,
       author = {{Li}, Xiangchong and {Miyatake}, Hironao and {Luo}, Wentao and {More}, Surhud and {Oguri}, Masamune and {Hamana}, Takashi and {Mandelbaum}, Rachel and {Shirasaki}, Masato and {Takada}, Masahiro and {Armstrong}, Robert and {Kannawadi}, Arun and {Takita}, Satoshi and {Miyazaki}, Satoshi and {Nishizawa}, Atsushi J. and {Plazas Malagon}, Andres A. and {Strauss}, Michael A. and {Tanaka}, Masayuki and {Yoshida}, Naoki},
        title = "{The three-year shear catalog of the Subaru Hyper Suprime-Cam SSP Survey}",
      journal = {\pasj},
         year = 2022,
        month = apr,
       volume = {74},
       number = {2},
        pages = {421-459},
          doi = {10.1093/pasj/psac006},
archivePrefix = {arXiv},
       eprint = {2107.00136},
 primaryClass = {astro-ph.CO},
       adsurl = {https://ui.adsabs.harvard.edu/abs/2022PASJ...74..421L}
}

@ARTICLE{2020PASJ...72...16H,
       author = {{Hamana}, Takashi and {Shirasaki}, Masato and {Miyazaki}, Satoshi and {Hikage}, Chiaki and {Oguri}, Masamune and {More}, Surhud and {Armstrong}, Robert and {Leauthaud}, Alexie and {Mandelbaum}, Rachel and {Miyatake}, Hironao and {Nishizawa}, Atsushi J. and {Simet}, Melanie and {Takada}, Masahiro and {Aihara}, Hiroaki and {Bosch}, James and {Komiyama}, Yutaka and {Lupton}, Robert and {Murayama}, Hitoshi and {Strauss}, Michael A. and {Tanaka}, Masayuki},
        title = "{Cosmological constraints from cosmic shear two-point correlation functions with HSC survey first-year data}",
      journal = {\pasj},
         year = 2020,
        month = feb,
       volume = {72},
       number = {1},
          eid = {16},
        pages = {16},
          doi = {10.1093/pasj/psz138},
archivePrefix = {arXiv},
       eprint = {1906.06041},
 primaryClass = {astro-ph.CO},
       adsurl = {https://ui.adsabs.harvard.edu/abs/2020PASJ...72...16H}
}

@article{Giblin21,
   title={KiDS-1000 catalogue: Weak gravitational lensing shear measurements},
   volume={645},
   ISSN={1432-0746},
   url={http://dx.doi.org/10.1051/0004-6361/202038850},
   DOI={10.1051/0004-6361/202038850},
   journal={\aap},
   publisher={EDP Sciences},
   author={Giblin, Benjamin and Heymans, Catherine and Asgari, Marika and Hildebrandt, Hendrik and Hoekstra, Henk and Joachimi, Benjamin and Kannawadi, Arun and Kuijken, Konrad and Lin, Chieh-An and Miller, Lance and Tröster, Tilman and van den Busch, Jan Luca and Wright, Angus H. and Bilicki, Maciej and Blake, Chris and de Jong, Jelte and Dvornik, Andrej and Erben, Thomas and Getman, Fedor and Napolitano, Nicola R. and Schneider, Peter and Shan, HuanYuan and Valentijn, Edwin},
   year={2021},
   month=jan, pages={A105}
}

@ARTICLE{prymordial,
       author = {{Burns}, Anne-Katherine and {Tait}, Tim M.~P. and {Valli}, Mauro},
        title = "{PRyMordial: the first three minutes, within and beyond the standard model}",
      journal = {European Physical Journal C},
         year = 2024,
        month = jan,
       volume = {84},
       number = {1},
          eid = {86},
        pages = {86},
          doi = {10.1140/epjc/s10052-024-12442-0},
archivePrefix = {arXiv},
       eprint = {2307.07061},
 primaryClass = {hep-ph},
       adsurl = {https://ui.adsabs.harvard.edu/abs/2024EPJC...84...86B}
}

@ARTICLE{shapepipe_axel,
       author = {{Guinot}, Axel and {Kilbinger}, Martin and {Farrens}, Samuel and {Peel}, Austin and {Pujol}, Arnau and {Schmitz}, Morgan and {Starck}, Jean-Luc and {Erben}, Thomas and {Gavazzi}, Raphael and {Gwyn}, Stephen and {Hudson}, Michael J. and {Hildebrandt}, Hendrik and {Tobias}, Liaudat and {Miller}, Lance and {Spitzer}, Isaac and {Van Waerbeke}, Ludovic and {Cuillandre}, Jean-Charles and {Fabbro}, S{\'e}bastien and {McConnachie}, Alan and {Mellier}, Yannick},
        title = "{ShapePipe: A new shape measurement pipeline and weak-lensing application to UNIONS/CFIS data}",
      journal = {\aap},
         year = 2022,
        month = oct,
       volume = {666},
          eid = {A162},
        pages = {A162},
          doi = {10.1051/0004-6361/202141847},
archivePrefix = {arXiv},
       eprint = {2204.04798},
 primaryClass = {astro-ph.CO},
       adsurl = {https://ui.adsabs.harvard.edu/abs/2022A&A...666A.162G}
}

@article{guerriniGalaxyPointSpread2025,
  title = {Galaxy--Point Spread Function Correlations as a Probe of Weak-Lensing Systematics with {{UNIONS}} Data},
  author = {Guerrini, Sacha and Kilbinger, Martin and Leterme, Hubert and Guinot, Axel and Wang, Jingwei and Peters, Fabian Hervas and Hildebrandt, Hendrik and Hudson, Michael J. and McConnachie, Alan},
  year = 2025,
  month = aug,
  journal = {A\&A},
  volume = {700},
  pages = {A215},
  publisher = {EDP Sciences},
  issn = {0004-6361, 1432-0746},
  doi = {10.1051/0004-6361/202453512},
  urldate = {2025-10-27},
  copyright = {\copyright{} The Authors 2025},
  langid = {english}
}

@article{Polychord, author = "Handley, W. and Hobson, M. P. and Lasenby, A. N.", title = "{POLYCHORD: nested sampling for cosmology}", journal = "\mnras", volume = "450", number = "1", pages = "L61-L65", year = "2015", doi = "10.1093/mnrasl/slv047" }

@article{Polychord2, author = "Handley, W. and Hobson, M. P. and Lasenby, A. N.", title = "{POLYCHORD: next-generation nested sampling}", journal = "\mnras", volume = "453", number = "4", pages = "4384-4398", year = "2015", doi = "10.1093/mnras/stv1911" }

@ARTICLE{heymans2012cfhtlens,
       author = {{Heymans}, Catherine and {Van Waerbeke}, Ludovic and {Miller}, Lance and {Erben}, Thomas and {Hildebrandt}, Hendrik and {Hoekstra}, Henk and {Kitching}, Thomas D. and {Mellier}, Yannick and {Simon}, Patrick and {Bonnett}, Christopher and {Coupon}, Jean and {Fu}, Liping and {Harnois D{\'e}raps}, Joachim and {Hudson}, Michael J. and {Kilbinger}, Martin and {Kuijken}, Koenraad and {Rowe}, Barnaby and {Schrabback}, Tim and {Semboloni}, Elisabetta and {van Uitert}, Edo and {Vafaei}, Sanaz and {Velander}, Malin},
        title = "{CFHTLenS: the Canada-France-Hawaii Telescope Lensing Survey}",
      journal = {\mnras},
         year = 2012,
        month = nov,
       volume = {427},
       number = {1},
        pages = {146-166},
          doi = {10.1111/j.1365-2966.2012.21952.x},
archivePrefix = {arXiv},
       eprint = {1210.0032},
 primaryClass = {astro-ph.CO},
       adsurl = {https://ui.adsabs.harvard.edu/abs/2012MNRAS.427..146H}
}

@ARTICLE{erben2013cfhtlens,
       author = {{Erben}, T. and {Hildebrandt}, H. and {Miller}, L. and {van Waerbeke}, L. and {Heymans}, C. and {Hoekstra}, H. and {Kitching}, T.~D. and {Mellier}, Y. and {Benjamin}, J. and {Blake}, C. and {Bonnett}, C. and {Cordes}, O. and {Coupon}, J. and {Fu}, L. and {Gavazzi}, R. and {Gillis}, B. and {Grocutt}, E. and {Gwyn}, S.~D.~J. and {Holhjem}, K. and {Hudson}, M.~J. and {Kilbinger}, M. and {Kuijken}, K. and {Milkeraitis}, M. and {Rowe}, B.~T.~P. and {Schrabback}, T. and {Semboloni}, E. and {Simon}, P. and {Smit}, M. and {Toader}, O. and {Vafaei}, S. and {van Uitert}, E. and {Velander}, M.},
        title = "{CFHTLenS: the Canada-France-Hawaii Telescope Lensing Survey - imaging data and catalogue products}",
      journal = {\mnras},
         year = 2013,
        month = aug,
       volume = {433},
       number = {3},
        pages = {2545-2563},
          doi = {10.1093/mnras/stt928},
archivePrefix = {arXiv},
       eprint = {1210.8156},
 primaryClass = {astro-ph.CO},
       adsurl = {https://ui.adsabs.harvard.edu/abs/2013MNRAS.433.2545E}
}

@ARTICLE{hildebrandt2012cfhtlens,
       author = {{Hildebrandt}, H. and {Erben}, T. and {Kuijken}, K. and {van Waerbeke}, L. and {Heymans}, C. and {Coupon}, J. and {Benjamin}, J. and {Bonnett}, C. and {Fu}, L. and {Hoekstra}, H. and {Kitching}, T.~D. and {Mellier}, Y. and {Miller}, L. and {Velander}, M. and {Hudson}, M.~J. and {Rowe}, B.~T.~P. and {Schrabback}, T. and {Semboloni}, E. and {Ben{\'\i}tez}, N.},
        title = "{CFHTLenS: improving the quality of photometric redshifts with precision photometry}",
      journal = {\mnras},
         year = 2012,
        month = apr,
       volume = {421},
       number = {3},
        pages = {2355-2367},
          doi = {10.1111/j.1365-2966.2012.20468.x},
archivePrefix = {arXiv},
       eprint = {1111.4434},
 primaryClass = {astro-ph.CO},
       adsurl = {https://ui.adsabs.harvard.edu/abs/2012MNRAS.421.2355H}
}

@ARTICLE{newman2013deep2,
       author = {{Newman}, Jeffrey A. and {Cooper}, Michael C. and {Davis}, Marc and {Faber}, S.~M. and {Coil}, Alison L. and {Guhathakurta}, Puragra and {Koo}, David C. and {Phillips}, Andrew C. and {Conroy}, Charlie and {Dutton}, Aaron A. and {Finkbeiner}, Douglas P. and {Gerke}, Brian F. and {Rosario}, David J. and {Weiner}, Benjamin J. and {Willmer}, C.~N.~A. and {Yan}, Renbin and {Harker}, Justin J. and {Kassin}, Susan A. and {Konidaris}, N.~P. and {Lai}, Kamson and {Madgwick}, Darren S. and {Noeske}, K.~G. and {Wirth}, Gregory D. and {Connolly}, A.~J. and {Kaiser}, N. and {Kirby}, Evan N. and {Lemaux}, Brian C. and {Lin}, Lihwai and {Lotz}, Jennifer M. and {Luppino}, G.~A. and {Marinoni}, C. and {Matthews}, Daniel J. and {Metevier}, Anne and {Schiavon}, Ricardo P.},
        title = "{The DEEP2 Galaxy Redshift Survey: Design, Observations, Data Reduction, and Redshifts}",
      journal = {\apjs},
         year = 2013,
        month = sep,
       volume = {208},
       number = {1},
          eid = {5},
        pages = {5},
          doi = {10.1088/0067-0049/208/1/5},
archivePrefix = {arXiv},
       eprint = {1203.3192},
 primaryClass = {astro-ph.CO},
       adsurl = {https://ui.adsabs.harvard.edu/abs/2013ApJS..208....5N}
}

@ARTICLE{2008A&A...484...67P,
       author = {{Paulin-Henriksson}, S. and {Amara}, A. and {Voigt}, L. and {Refregier}, A. and {Bridle}, S.~L.},
        title = "{Point spread function calibration requirements for dark energy from cosmic shear}",
      journal = {\aap},
         year = 2008,
        month = jun,
       volume = {484},
       number = {1},
        pages = {67-77},
          doi = {10.1051/0004-6361:20079150},
archivePrefix = {arXiv},
       eprint = {0711.4886},
 primaryClass = {astro-ph},
       adsurl = {https://ui.adsabs.harvard.edu/abs/2008A&A...484...67P}
}

@ARTICLE{2021A&A...646A.129J,
       author = {{Joachimi}, B. and {Lin}, C.-A. and {Asgari}, M. and {Tr{\"o}ster}, T. and {Heymans}, C. and {Hildebrandt}, H. and {K{\"o}hlinger}, F. and {S{\'a}nchez}, A.~G. and {Wright}, A.~H. and {Bilicki}, M. and {Blake}, C. and {van den Busch}, J.~L. and {Crocce}, M. and {Dvornik}, A. and {Erben}, T. and {Getman}, F. and {Giblin}, B. and {Hoekstra}, H. and {Kannawadi}, A. and {Kuijken}, K. and {Napolitano}, N.~R. and {Schneider}, P. and {Scoccimarro}, R. and {Sellentin}, E. and {Shan}, H.~Y. and {von Wietersheim-Kramsta}, M. and {Zuntz}, J.},
        title = "{KiDS-1000 methodology: Modelling and inference for joint weak gravitational lensing and spectroscopic galaxy clustering analysis}",
      journal = {\aap},
         year = 2021,
        month = feb,
       volume = {646},
          eid = {A129},
        pages = {A129},
          doi = {10.1051/0004-6361/202038831},
archivePrefix = {arXiv},
       eprint = {2007.01844},
 primaryClass = {astro-ph.CO},
       adsurl = {https://ui.adsabs.harvard.edu/abs/2021A&A...646A.129J}
}

@ARTICLE{lefevre2005vvds,
       author = {{Le F{\`e}vre}, O. and {Vettolani}, G. and {Garilli}, B. and {Tresse}, L. and {Bottini}, D. and {Le Brun}, V. and {Maccagni}, D. and {Picat}, J.~P. and {Scaramella}, R. and {Scodeggio}, M. and {Zanichelli}, A. and {Adami}, C. and {Arnaboldi}, M. and {Arnouts}, S. and {Bardelli}, S. and {Bolzonella}, M. and {Cappi}, A. and {Charlot}, S. and {Ciliegi}, P. and {Contini}, T. and {Foucaud}, S. and {Franzetti}, P. and {Gavignaud}, I. and {Guzzo}, L. and {Ilbert}, O. and {Iovino}, A. and {McCracken}, H.~J. and {Marano}, B. and {Marinoni}, C. and {Mathez}, G. and {Mazure}, A. and {Meneux}, B. and {Merighi}, R. and {Paltani}, S. and {Pell{\`o}}, R. and {Pollo}, A. and {Pozzetti}, L. and {Radovich}, M. and {Zamorani}, G. and {Zucca}, E. and {Bondi}, M. and {Bongiorno}, A. and {Busarello}, G. and {Lamareille}, F. and {Mellier}, Y. and {Merluzzi}, P. and {Ripepi}, V. and {Rizzo}, D.},
        title = "{The VIMOS VLT deep survey. First epoch VVDS-deep survey: 11 564 spectra with 17.5 {\ensuremath{\leq}} IAB {\ensuremath{\leq}} 24, and the redshift distribution over 0 {\ensuremath{\leq}} z {\ensuremath{\leq}} 5}",
      journal = {\aap},
         year = 2005,
        month = sep,
       volume = {439},
       number = {3},
        pages = {845-862},
          doi = {10.1051/0004-6361:20041960},
archivePrefix = {arXiv},
       eprint = {astro-ph/0409133},
 primaryClass = {astro-ph},
       adsurl = {https://ui.adsabs.harvard.edu/abs/2005A&A...439..845L}
}

@ARTICLE{scodeggio2018vipers,
       author = {{Scodeggio}, M. and {Guzzo}, L. and {Garilli}, B. and {Granett}, B.~R. and {Bolzonella}, M. and {de la Torre}, S. and {Abbas}, U. and {Adami}, C. and {Arnouts}, S. and {Bottini}, D. and {Cappi}, A. and {Coupon}, J. and {Cucciati}, O. and {Davidzon}, I. and {Franzetti}, P. and {Fritz}, A. and {Iovino}, A. and {Krywult}, J. and {Le Brun}, V. and {Le F{\`e}vre}, O. and {Maccagni}, D. and {Ma{\l}ek}, K. and {Marchetti}, A. and {Marulli}, F. and {Polletta}, M. and {Pollo}, A. and {Tasca}, L.~A.~M. and {Tojeiro}, R. and {Vergani}, D. and {Zanichelli}, A. and {Bel}, J. and {Branchini}, E. and {De Lucia}, G. and {Ilbert}, O. and {McCracken}, H.~J. and {Moutard}, T. and {Peacock}, J.~A. and {Zamorani}, G. and {Burden}, A. and {Fumana}, M. and {Jullo}, E. and {Marinoni}, C. and {Mellier}, Y. and {Moscardini}, L. and {Percival}, W.~J.},
        title = "{The VIMOS Public Extragalactic Redshift Survey (VIPERS). Full spectroscopic data and auxiliary information release (PDR-2)}",
      journal = {\aap},
         year = 2018,
        month = jan,
       volume = {609},
          eid = {A84},
        pages = {A84},
          doi = {10.1051/0004-6361/201630114},
archivePrefix = {arXiv},
       eprint = {1611.07048},
 primaryClass = {astro-ph.GA},
       adsurl = {https://ui.adsabs.harvard.edu/abs/2018A&A...609A..84S}
}

@article{kohonen1982som,
  title={Self-organized formation of topologically correct feature maps},
  author={Kohonen, Teuvo},
  journal={Biological cybernetics},
  volume={43},
  number={1},
  pages={59--69},
  year={1982},
  publisher={Springer}
}

@ARTICLE{masters2015c3r2,
       author = {{Masters}, Daniel and {Capak}, Peter and {Stern}, Daniel and {Ilbert}, Olivier and {Salvato}, Mara and {Schmidt}, Samuel and {Longo}, Giuseppe and {Rhodes}, Jason and {Paltani}, Stephane and {Mobasher}, Bahram and {Hoekstra}, Henk and {Hildebrandt}, Hendrik and {Coupon}, Jean and {Steinhardt}, Charles and {Speagle}, Josh and {Faisst}, Andreas and {Kalinich}, Adam and {Brodwin}, Mark and {Brescia}, Massimo and {Cavuoti}, Stefano},
        title = "{Mapping the Galaxy Color-Redshift Relation: Optimal Photometric Redshift Calibration Strategies for Cosmology Surveys}",
      journal = {\apj},
         year = 2015,
        month = nov,
       volume = {813},
       number = {1},
          eid = {53},
        pages = {53},
          doi = {10.1088/0004-637X/813/1/53},
archivePrefix = {arXiv},
       eprint = {1509.03318},
 primaryClass = {astro-ph.CO},
       adsurl = {https://ui.adsabs.harvard.edu/abs/2015ApJ...813...53M}
}

@ARTICLE{wright2020som,
       author = {{Wright}, Angus H. and {Hildebrandt}, Hendrik and {van den Busch}, Jan Luca and {Heymans}, Catherine},
        title = "{Photometric redshift calibration with self-organising maps}",
      journal = {\aap},
         year = 2020,
        month = may,
       volume = {637},
          eid = {A100},
        pages = {A100},
          doi = {10.1051/0004-6361/201936782},
archivePrefix = {arXiv},
       eprint = {1909.09632},
 primaryClass = {astro-ph.CO},
       adsurl = {https://ui.adsabs.harvard.edu/abs/2020A&A...637A.100W}
}

@article{Bacon2000,
       author = {{Bacon}, David J. and {Refregier}, Alexandre R. and {Ellis}, Richard S.},
        title = "{Detection of weak gravitational lensing by large-scale structure}",
      journal = {\mnras},
         year = 2000,
        month = oct,
       volume = {318},
       number = {2},
        pages = {625-640},
          doi = {10.1046/j.1365-8711.2000.03851.x},
archivePrefix = {arXiv},
       eprint = {astro-ph/0003008},
 primaryClass = {astro-ph},
       adsurl = {https://ui.adsabs.harvard.edu/abs/2000MNRAS.318..625B}
}

@article{vanWaerbeke2000,
       author = {{Van Waerbeke}, L. and {Mellier}, Y. and {Erben}, T. and {Cuillandre}, J.~C. and {Bernardeau}, F. and {Maoli}, R. and {Bertin}, E. and {McCracken}, H.~J. and {Le F{\`e}vre}, O. and {Fort}, B. and {Dantel-Fort}, M. and {Jain}, B. and {Schneider}, P.},
        title = "{Detection of correlated galaxy ellipticities from CFHT data: first evidence for gravitational lensing by large-scale structures}",
      journal = {\aap},
         year = 2000,
        month = jun,
       volume = {358},
        pages = {30-44},
          doi = {10.48550/arXiv.astro-ph/0002500},
archivePrefix = {arXiv},
       eprint = {astro-ph/0002500},
 primaryClass = {astro-ph},
       adsurl = {https://ui.adsabs.harvard.edu/abs/2000A&A...358...30V}
}

@article{Wittman2000,
       author = {{Wittman}, David M. and {Tyson}, J. Anthony and {Kirkman}, David and {Dell'Antonio}, Ian and {Bernstein}, Gary},
        title = "{Detection of weak gravitational lensing distortions of distant galaxies by cosmic dark matter at large scales}",
      journal = {\nat},
         year = 2000,
        month = may,
       volume = {405},
       number = {6783},
        pages = {143-148},
          doi = {10.1038/35012001},
archivePrefix = {arXiv},
       eprint = {astro-ph/0003014},
 primaryClass = {astro-ph},
       adsurl = {https://ui.adsabs.harvard.edu/abs/2000Natur.405..143W}
}

@ARTICLE{nautilus,
       author = {{Lange}, Johannes U.},
        title = "{NAUTILUS: boosting Bayesian importance nested sampling with deep learning}",
      journal = {\mnras},
         year = 2023,
        month = oct,
       volume = {525},
       number = {2},
        pages = {3181-3194},
          doi = {10.1093/mnras/stad2441},
archivePrefix = {arXiv},
       eprint = {2306.16923},
 primaryClass = {astro-ph.IM},
       adsurl = {https://ui.adsabs.harvard.edu/abs/2023MNRAS.525.3181L}
}

@ARTICLE{polychord-test,
       author = {{Lemos}, P. and {Weaverdyck}, N. and {Rollins}, R.~P. and {Muir}, J. and {Fert{\'e}}, A. and {Liddle}, A.~R. and {Campos}, A. and {Huterer}, D. and {Raveri}, M. and {Zuntz}, J. and {Di Valentino}, E. and {Fang}, X. and {Hartley}, W.~G. and {Aguena}, M. and {Allam}, S. and {Annis}, J. and {Bertin}, E. and {Bocquet}, S. and {Brooks}, D. and {Burke}, D.~L. and {Carnero Rosell}, A. and {Carrasco Kind}, M. and {Carretero}, J. and {Castander}, F.~J. and {Choi}, A. and {Costanzi}, M. and {Crocce}, M. and {da Costa}, L.~N. and {Pereira}, M.~E.~S. and {Dietrich}, J.~P. and {Everett}, S. and {Ferrero}, I. and {Frieman}, J. and {Garc{\'\i}a-Bellido}, J. and {Gatti}, M. and {Gaztanaga}, E. and {Gerdes}, D.~W. and {Gruen}, D. and {Gruendl}, R.~A. and {Gschwend}, J. and {Gutierrez}, G. and {Hinton}, S.~R. and {Hollowood}, D.~L. and {Honscheid}, K. and {James}, D.~J. and {Kuehn}, K. and {Kuropatkin}, N. and {Lima}, M. and {March}, M. and {Melchior}, P. and {Menanteau}, F. and {Miquel}, R. and {Morgan}, R. and {Palmese}, A. and {Paz-Chinch{\'o}n}, F. and {Pieres}, A. and {Malag{\'o}n}, A.~A. Plazas and {Porredon}, A. and {Sanchez}, E. and {Scarpine}, V. and {Schubnell}, M. and {Serrano}, S. and {Sevilla-Noarbe}, I. and {Smith}, M. and {Suchyta}, E. and {Swanson}, M.~E.~C. and {Tarle}, G. and {Thomas}, D. and {To}, C. and {Varga}, T.~N. and {Weller}, J. and {DES Collaboration}},
        title = "{Robust sampling for weak lensing and clustering analyses with the Dark Energy Survey}",
      journal = {\mnras},
         year = 2023,
        month = may,
       volume = {521},
       number = {1},
        pages = {1184-1199},
          doi = {10.1093/mnras/stac2786},
archivePrefix = {arXiv},
       eprint = {2202.08233},
 primaryClass = {astro-ph.CO},
       adsurl = {https://ui.adsabs.harvard.edu/abs/2023MNRAS.521.1184L}
}

@article{gatti2021shapecatalogue,
   title={Dark energy survey year 3 results: weak lensing shape catalogue},
   volume={504},
   ISSN={1365-2966},
   url={http://dx.doi.org/10.1093/mnras/stab918},
   DOI={10.1093/mnras/stab918},
   number={3},
   journal={\mnras},
   publisher={Oxford University Press (OUP)},
   author={Gatti, M and Sheldon, E and Amon, A and Becker, M and Troxel, M and Choi, A and Doux, C and MacCrann, N and Navarro-Alsina, A and Harrison, I and Gruen, D and Bernstein, G and Jarvis, M and Secco, L F and Ferté, A and Shin, T and McCullough, J and Rollins, R P and Chen, R and Chang, C and Pandey, S and Tutusaus, I and Prat, J and Elvin-Poole, J and Sanchez, C and Plazas, A A and Roodman, A and Zuntz, J and Abbott, T M C and Aguena, M and Allam, S and Annis, J and Avila, S and Bacon, D and Bertin, E and Bhargava, S and Brooks, D and Burke, D L and Carnero Rosell, A and Carrasco Kind, M and Carretero, J and Castander, F J and Conselice, C and Costanzi, M and Crocce, M and da Costa, L N and Davis, T M and De Vicente, J and Desai, S and Diehl, H T and Dietrich, J P and Doel, P and Drlica-Wagner, A and Eckert, K and Everett, S and Ferrero, I and Frieman, J and García-Bellido, J and Gerdes, D W and Giannantonio, T and Gruendl, R A and Gschwend, J and Gutierrez, G and Hartley, W G and Hinton, S R and Hollowood, D L and Honscheid, K and Hoyle, B and Huff, E M and Huterer, D and Jain, B and James, D J and Jeltema, T and Krause, E and Kron, R and Kuropatkin, N and Lima, M and Maia, M A G and Marshall, J L and Miquel, R and Morgan, R and Myles, J and Palmese, A and Paz-Chinchón, F and Rykoff, E S and Samuroff, S and Sanchez, E and Scarpine, V and Schubnell, M and Serrano, S and Sevilla-Noarbe, I and Smith, M and Soares-Santos, M and Suchyta, E and Swanson, M E C and Tarle, G and Thomas, D and To, C and Tucker, D L and Varga, T N and Wechsler, R H and Weller, J and Wester, W and Wilkinson, R D},
   year={2021},
   month=apr, pages={4312–4336} }

@article{carretero2015mice2,
  title = {An Algorithm to Build Mock Galaxy Catalogues Using {{MICE}} Simulations},
  author = {Carretero, J. and Castander, F. J. and Gaztanaga, E. and Crocce, M. and Fosalba, P.},
  year = 2015,
  month = feb,
  journal = {MNRAS},
  volume = {447},
  number = {1},
  eprint = {1411.3286},
  pages = {646--670},
  issn = {1365-2966, 0035-8711},
  doi = {10.1093/mnras/stu2402},
  urldate = {2021-05-07},
  archiveprefix = {arXiv}
}

@article{crocce2015mice2,
  title = {The {{MICE Grand Challenge}} Lightcone Simulation -- {{II}}. {{Halo}} and Galaxy Catalogues},
  author = {Crocce, M. and Castander, F. J. and Gazta{\~n}aga, E. and Fosalba, P. and Carretero, J.},
  year = 2015,
  month = oct,
  journal = {MRNAS},
  volume = {453},
  number = {2},
  pages = {1513--1530},
  issn = {0035-8711, 1365-2966},
  doi = {10.1093/mnras/stv1708},
  urldate = {2023-12-06},
  langid = {english}
}

@article{fosalba2015mice2a,
  title = {The {{MICE Grand Challenge Lightcone Simulation I}}: {{Dark}} Matter Clustering},
  shorttitle = {The {{MICE Grand Challenge Lightcone Simulation I}}},
  author = {Fosalba, P. and Crocce, M. and Gaztanaga, E. and Castander, F. J.},
  year = 2015,
  month = apr,
  journal = {MNRAS},
  volume = {448},
  number = {4},
  eprint = {1312.1707},
  pages = {2987--3000},
  issn = {1365-2966, 0035-8711},
  doi = {10.1093/mnras/stv138},
  urldate = {2021-05-07},
  archiveprefix = {arXiv}
}

@article{fosalba2015mice2b,
  title = {The {{MICE Grand Challenge Lightcone Simulation III}}: {{Galaxy}} Lensing Mocks from All-Sky Lensing Maps},
  shorttitle = {The {{MICE Grand Challenge Lightcone Simulation III}}},
  author = {Fosalba, P. and Gaztanaga, E. and Castander, F. J. and Crocce, M.},
  year = 2015,
  month = feb,
  journal = {MNRAS},
  volume = {447},
  number = {2},
  eprint = {1312.2947},
  pages = {1319--1332},
  issn = {1365-2966, 0035-8711},
  doi = {10.1093/mnras/stu2464},
  urldate = {2021-05-07},
  archiveprefix = {arXiv}
}

@article{vandenbusch2020kidsredshiftcalibration,
  title = {Testing {{KiDS}} Cross-Correlation Redshifts with Simulations},
  author = {{van den Busch}, J. L. and Hildebrandt, H. and Wright, A. H. and Morrison, C. B. and Blake, C. and Joachimi, B. and Erben, T. and Heymans, C. and Kuijken, K. and Taylor, E. N.},
  year = 2020,
  month = oct,
  journal = {A\&A},
  volume = {642},
  eprint = {2007.01846},
  pages = {A200},
  issn = {0004-6361, 1432-0746},
  doi = {10.1051/0004-6361/202038835},
  urldate = {2021-05-04},
  archiveprefix = {arXiv}
}

@article{wright2025legacynz,
  title = {{{KiDS-Legacy}}: {{Redshift}} Distributions and Their Calibration},
  shorttitle = {{{KiDS-Legacy}}},
  author = {Wright, Angus H. and Hildebrandt, Hendrik and van den Busch, Jan Luca and Bilicki, Maciej and Heymans, Catherine and Joachimi, Benjamin and Mahony, Constance and Reischke, Robert and St{\"o}lzner, Benjamin and Wittje, Anna and Asgari, Marika and Chisari, Nora Elisa and Dvornik, Andrej and Georgiou, Christos and Giblin, Benjamin and Hoekstra, Henk and Jalan, Priyanka and William, Anjitha John and Joudaki, Shahab and Kuijken, Konrad and Lesci, Giorgio Francesco and Li, Shun-Sheng and Linke, Laila and Loureiro, Arthur and Maturi, Matteo and Moscardin, Lauro and Porth, Lucas and Radovich, Mario and Tr{\"o}ster, Tilman and von {Wietersheim-Kramsta}, Maximilian and Yan, Ziang and Yoon, Mijin and Zhang, Yun-Hao},
  year = 2025,
  month = nov,
  journal = {A\&A},
  volume = {703},
  eprint = {2503.19440},
  primaryclass = {astro-ph},
  pages = {A144},
  issn = {0004-6361, 1432-0746},
  doi = {10.1051/0004-6361/202554909},
  urldate = {2025-11-14},
  archiveprefix = {arXiv}
}

@ARTICLE{planck2018,
       author = {{Planck Collaboration}},
        title = "{Planck 2018 results. VI. Cosmological parameters}",
      journal = {\aap},
         year = 2020,
        month = sep,
       volume = {641},
          eid = {A6},
        pages = {A6},
          doi = {10.1051/0004-6361/201833910},
archivePrefix = {arXiv},
       eprint = {1807.06209},
 primaryClass = {astro-ph.CO},
       adsurl = {https://ui.adsabs.harvard.edu/abs/2020A&A...641A...6P}
}

@article{kilbinger.etal25,
    author = {Hervas Peters, Fabian and others},
    title = {{UNIONS-3500 Weak Lensing: I. Shear catalogue construction}},
    journal = {in preparation},
    year = {2026}
}

@article{guerrini.etal25b,
    author = {Guerrini, Sacha and others},
    title = {{UNIONS-3500 Weak Lensing: IV. Cosmological constraints in harmonic space}},
    journal = {in preparation},
    year = {2026}
}

@article{hervaspaters.etal25,
    author = {Hervas Peters, F. and others},
    title = {{UNIONS-3500 Weak Lensing: V. Image simulations and validation}},
    journal = {in preparation},
    year = {2026}
}
